\documentclass[table]{article}

\usepackage{arxiv}

\usepackage[utf8]{inputenc} 
\usepackage[T1]{fontenc}    
\usepackage{graphicx}
\usepackage{hyperref}       
\usepackage{url}            
\usepackage{booktabs}       
\usepackage{amsfonts}       
\usepackage{nicefrac}       
\usepackage{microtype}      
\usepackage{lipsum}		
\usepackage{graphicx}
\usepackage[square, sort, numbers]{natbib}
\usepackage{doi}
\usepackage{subcaption}
\usepackage[utf8]{inputenc}
\usepackage{amsmath}
\usepackage{cleveref}
\usepackage{svg}
\usepackage{multirow}
\usepackage{array}
\usepackage{svg-extract}

\definecolor{linDarkGray}{HTML}{2c2e34}
\definecolor{linLightGray}{HTML}{85878b}
\definecolor{tblGray}{HTML}{F2F2F2}

\newcolumntype{P}[1]{>{\centering\arraybackslash}m{#1}}
\newcolumntype{M}[1]{>{\arraybackslash}m{#1}}

\title{ExoFabric: A Re-moldable Textile System for Creating Customizable Soft Goods and Wearable Applications}

\author{ \href{https://orcid.org/0009-0005-9975-3882}{\includegraphics[scale=0.06]{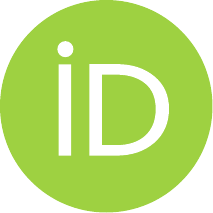}\hspace{1mm}Rosalie Lin} \\
	Future Technologies R\&D Group\\
    Accenture Labs\\
	San Francisco, California, 94105, USA \\
	\texttt{rosalie.lin@accenture.com} \\
	\And
	\href{https://orcid.org/0000-0001-9828-0570}{\includegraphics[scale=0.06]{orcid.pdf}\hspace{1mm}Aditi Maheshwari} \\
	Future Technologies R\&D Group\\
    Accenture Labs\\
	San Francisco, California, 94105, USA \\
	\texttt{aditi.maheshwari@accenture.com} \\
	\And
	\href{https://orcid.org/0000-0002-0360-1085}{\includegraphics[scale=0.06]{orcid.pdf}\hspace{1mm}Jung Wook Park} \\
	Future Technologies R\&D Group\\
    Accenture Labs\\
	San Francisco, California, 94105, USA \\
	\texttt{jungwook.park@accenture.com} \\
    \And
	\href{https://orcid.org/0000-0001-7460-2467}{\includegraphics[scale=0.06]{orcid.pdf}\hspace{1mm}Andreea Danielescu} \\
	Future Technologies R\&D Group\\
	Accenture Labs\\
	San Francisco, California, 94105, USA \\
	\texttt{andreea.danielescu@accenture.com} \\
}

\renewcommand{\shorttitle}{ExoFabric: A Re-moldable Textile System}

\hypersetup{
pdftitle={ExoFabric: A Re-moldable Textile System for Creating Customizable Soft Goods and Wearable Applications},
pdfsubject={cs.HC},
pdfauthor={Rosalie Lin, Aditi Maheshwari, Jung Wook Park, Andreea Danielescu},
pdfkeywords={ExoFabric, Re-moldable fabric, Thermoplastic threads, Customization, Stiffness, Geometric formability, Stretchability},
}

\begin{document}

\maketitle

\begin{abstract}
Fabric has been a fundamental part of human life for thousands of years, providing comfort, protection, and aesthetic expression. While modern advancements have enhanced fabric's functionality, it remains static and unchangeable, failing to adapt to our evolving body shapes and preferences. This lack of adaptability can lead to unsustainable practices, as consumers often buy more items to meet their changing needs. In this paper, we propose ExoFabric, a re-moldable fabric system for customized soft goods applications. We created ExoFabric by embedding thermoplastic threads into fabric through computerized embroidery to allow for tunability between rigid plastic and conformable fabric. We defined a library of design primitives to enable geometric formability, stiffness, and stretchability by identifying suitable fabrics, threads, embroidery parameters, and machine limitations. To facilitate practical applications, we demonstrated practical methods for linking parameters to application requirements, showcasing form-fitting wearables, structural support, and shape-changeable furniture for repeatable or one-time customization. 
\end{abstract}



\section{Introduction}
Fabric is ubiquitous, found in everyday objects from wearables and furniture to automobile interiors, providing our lives with comfort, protection, and aesthetic expression. It is one of the earliest human technologies, dating back tens of thousands of years, used for shelter, clothing, and cultural attire. Fast forward to modern times, the Industrial Revolution enabled mass production and a globalized supply chain in the textile industry. This industry has continually advanced, contributing to new functions such as moisture-wicking \cite{de_sousa_effects_2014}, sensing \cite{cai_flexible_2017}, and biodegradability \cite{ribul_mechanical_2021}, as well as novel manufacturing techniques like 3D knitting \cite{narayanan_automatic_2018}, 3D printing \cite{forman_defextiles_2020}, and even spray coating \cite{gong_fabrication_2018}. Despite its widespread use and rich advancements, once created, it typically remains static and unchangeable for most consumers. In contrast, our body shapes, behaviors, and preferences evolve over time, and the inability to adjust fabric to accommodate these changes prevents users from customizing it to fit various shapes and needs. This lack of adaptability may lead to unsustainable practices, such as purchasing and manufacturing more items to satisfy ever-changing needs.

To address this, researchers in HCI and related fields have proposed creating shape-changing and actuated textile interfaces by combining actuators with textiles' softness and conformability. Oftentimes, rigid and phase-change materials are introduced to the system to perform tasks such as ``holding'' and ``changing'' shapes. However, creating complex fabric interfaces often requires new materials and processes, leading to incompatibility with standard textile machinery and design practices, which inhibits their adoption by industry practitioners. For example, while pneumatic or mechanical actuators can generate high strain, they introduce bulkiness, noise, and rigidity, which may hinder the user experience \cite{pardomuan_vabricbeads_2024, yehoshua_wald_magnetform_2021}. Thermoplastic, another commonly used shape-changing material, has inspired works such as self-transforming, form-fitting, or reconfigurable interfaces \cite{yamaoka_rapid_2020,wang_thermofit_2022, del_valle_punchprint_2023}. Although thermoplastic is lightweight and does not require a tethered stimuli source (e.g., a pneumatic system needs an attached air pump to allow shape change, but thermoplastic can use an external heat source), it is usually made with standalone 3D-printed scaffolds, printed on fabric, or integrated with yarns through manual threading, which again diminishes wearability and fabrication efficiency. This work goes beyond previous works by using thermoplastic in ``thread'' form to enable easy integration with hobbyist or industrial textile machinery. 

The goal of this work is to explore re-moldability in textiles, offering customizability and ease of fabrication to meet individual needs. We introduce \textbf{\textit{ExoFabric}}, a re-moldable textile system that enables repeatable, instant, and macro-scale shape changes for versatile soft-good applications, including form-fitting wearables, reconfigurable softgoods, and structural daily objects. ExoFabric is both shape-retainable and shape-reversible, meaning it can retain a particular shape yet be reshaped with the application of heat and force, allowing users to mold and lock the shape through direct hands-on manipulation, with potentially endless physical adjustments. Unlike other shape-changing textile mechanisms that often embed rigid components \cite{yehoshua_wald_magnetform_2021} require digitally controlled systems \cite{kilic_afsar_omnifiber_2021}, or involve complex integration methods \cite{han_parametric_2023}, ExoFabric is lightweight, electronics-free, and highly compatible with standard textile design and fabrication practices. It is a homogeneous fabric-thermoplastic composite that can be made as rigid as plastic or as conformable and lightweight as fabric by leveraging thermoplastic fiber and digital embroidery techniques to deposit thermoplastic thread onto fabric, achieving a resolution close to the original fabric mesh. It is fully constructed with commercially available fabrics, thermoplastic fibers, and textile machinery, aiming to facilitate adoption among a broader range of practitioners. ExoFabric is created through a one-step fabrication process, with no need for assembly between parts or switching between machines, making it easy for designers to quickly iterate ideas both in terms of aesthetics and functionality. In introducing a new process for creating re-moldable textiles, we conduct a comprehensive evaluation to characterize the material properties and validate the design and fabrication pipeline. This approach provides valuable insights for meeting specific application requirements.

In this paper, we present the design space and fabrication pipeline for creating re-moldable fabrics by incorporating thermoplastic fibers within fabrics through digital embroidery. Our aim is to provide researchers, designers, and engineers in the HCI community and beyond—such as those in textile, fashion, and soft goods design and development—a new way to utilize textiles. We identify fibers, fabrics, and embroidery patterns through a diverse lens, considering factors such as functionality (stiffness, formability, and stretchability), interaction workflow (heating and molding), fabrication advantages (quality control, process time, etc.), aesthetic choices, sourcing availability, and scalability from DIY to mass production. Additionally, we characterize ExoFabric’s mechanical properties by conducting compression (stiffness) and tensile (stretchability) tests to establish a baseline understanding of the relationship between fiber layout and the fabric's intrinsic properties. Lastly, we demonstrate applications including a structural finger splint, a customizable bra or prosthesis, and a shape-changing lampshade. In summary, the main contributions of this paper are:
\begin{itemize}
    \item Design and fabrication pipeline for ExoFabric: understanding underlying working principles, material selection, fabrication tools, relationships between input parameters and desired outcomes, activation processes, and application domains
    \item Characterization of ExoFabric in terms of stiffness, geometrical formability, and stretchability, highlighting how different fibers, fabrics, embroidery parameters and fundamental geometries affect the mechanical stability of ExoFabric
    \item Three demonstrations showcasing different application requirements, including 1) a customizable finger splint, 2) a form-fitting bra, and 3) a shape changeable lamp shade, validating accessibility, customizability, and interactivity through an operationalized approach
\end{itemize}

\section{Related Works}

\begin{figure}[htbp]
    \centering
    \includegraphics[width=1\linewidth]{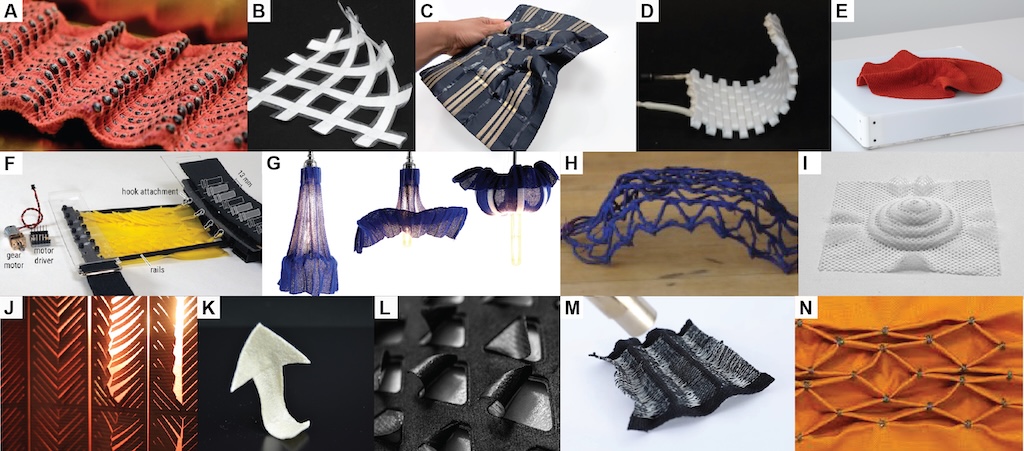}
    \caption{An overview of shape-change textiles mechanism. A-D: pneumatic driven \cite{kilic_afsar_omnifiber_2021, ou_jamsheets_2014,  sanchez_smart_2020, pardomuan_vabricbeads_2024}; E-F: magnetic \cite{yehoshua_wald_magnetform_2021} and motor  \cite{han_parametric_2023} driven ; G-N: material driven \cite{albaugh_digital_2019,glazko_shaping_2024, sun_fabricfit_2020,noauthor_active_nodate,rivera_hydrogel-textile_2020,yao_biologic_2015,forman_fiberobo_2023,haynes_flextiles_2024} }
    \label{fig:An overview of shape-change textiles}
\end{figure}

\subsection{Re-moldable Materials in HCI}

 Re-moldable objects such as clay and fabric can change their shape through human mechanical manipulation. Re-moldable materials are widely investigated in the HCI community because they can be reconfigured to cater to a wide range of geometries and functions \cite{boem_non-rigid_2019} while minimizing the fabrication of multiple objects, facilitating both creativity and sustainability. Clay's moldability makes it ideal for free-form music control interfaces \cite{tahiroglu_soundflex_2014, troiano_deformable_2015} and serves as a landscape topology display for projecting spatial information \cite{piper_illuminating_2002}. Similarly, thermoplastics, which are able to reshape under the application of heat, have inspired the creation of self-transformed textures and objects typically called '4D printing' in the HCI domain \cite{narumi_inkjet_2023, qin_exoform_2021, yamaoka_rapid_2020, sun_fabricfit_2020, hong_thermoformed_2021, sun_4dtexture_2020}, form-fitting orthoses \cite{sun_fabricfit_2020, wang_thermofit_2022}, or scaffolds that allow yarns to be woven or embroidered on \cite{deshpande_escapeloom_2021, del_valle_punchprint_2023}. Chocolate and wax, in light of sustainability and unmaking efforts, have also been used for transient and reconfigurable electronics due to their re-moldability \cite{cheng_functional_2023}. Our work expands the materials used in re-moldable interfaces to textiles, introducing re-moldable textiles that blend fabric and thermoplastic qualities to form composites that are as rigid as plastic and as breathable and lightweight as fabric. 

\subsection {Rigid Fabrics through Textile Engineering}

Rigid textiles, such as carbon fiber composites, are a type of textile that focus on mechanical strength and are used in areas like automobiles, the military, and outdoor gear, where high strength and lightweight properties are required simultaneously. They are also used in fashion, furniture, and sportswear, where structural support and shaping are needed. Given the hierarchical nature of textile construction, the mechanical properties can be formed anisotropically through 1D fibers and yarns, 2D textile structures, or 3D geometric forms. We focus on fabric-level construction techniques such as weaving, knitting, and braiding to identify guidelines for tuning the mechanics, which may be general or unique to each technique.
In weaving, the angle between warp and weft yarns \cite{liang_multi-scale_2019,mitchell_investigation_2016}, or the pattern of how weft yarns pass over warp yarns, can affect mechanical properties \cite{kumar_6_2018}. In knitting, different stitch types can be programmed to influence global stiffness \cite{singal_programming_2024}. Even within the same stitch types, stitch density and knit loop length can affect mechanical properties. Similar to weaving, braids are produced by interlacing two or more sets of yarns in a circular fashion to form hollow tubes. They are commonly produced with various angles to allow for a wide range of mechanical properties \cite{chen_mechanical_2019}, as seen in McKibben actuators \cite{felt_contraction_2016}. Sewing or embroidery is often seen as a post-processing method rather than textile construction, as it requires a base layer for interlacing threads. Its mechanical properties can be affected by stitch pattern, angle, and density \cite{poniecka_mechanical_2022}. For example, a strain sensor can be created through a zigzag stitch \cite{colli_alfaro_design_2023}, while a bending structure would prefer a running stitch to ensure continuous strength in the same direction. In our work, we drew inspiration from these textile techniques and identified parameters for tuning mechanical properties that are compatible with digital embroidery procedures while satisfying our targeted features, including rigidity, stretchability, and shape-holding.

\subsection{Shape-change Textile Interfaces}
Shape-changing fabric interfaces have opened up opportunities in wearables and smart homes due to their conformability to organic-shaped surfaces. Some enhance aesthetic expression in art and fashion, while others introduce new functions to fabrics, ranging from thermal, humidity, and light regulation \cite{yao_biologic_2015, noauthor_active_nodate}, to form-fitting garments \cite{sun_fabricfit_2020}, structural objects \cite{pardomuan_vabricbeads_2024}, and shape-changing devices \cite{hong_thermoformed_2021}.

To create these shape-changing fabric interfaces, different actuation methods have been developed to stimulate morphing effects. These can be categorized as pneumatic, mechanical, and material-based actuation methods, each of which has a different actuation speed, stiffness range, and form factor \cite{manti_stiffening_2016}. Pneumatically actuated interfaces often appear as soft composites constructed by expandable tubes with constrained parts, flowing with compressed air or liquid \cite{kilic_afsar_omnifiber_2021,ou_jamsheets_2014,sanchez_smart_2020,pardomuan_vabricbeads_2024}. Although they can perform relatively high strain under compact volume, they require pumps, air-tight pipes, and are often tethered to a digitally controlled system, introducing complexity in design and end-use. Mechanically actuated interfaces, often including motors \cite{han_parametric_2023}, or magnets \cite{yehoshua_wald_magnetform_2021}, among others, and offer high-loading actuation. However, they also bring bulkiness, noise, thickness, and rigidity to the otherwise soft interface, which may compromise wearability. Material-based interfaces, often actuated by sources such as heat, light, moisture, and pH, can either be triggered by physically or digitally enabled sources \cite {forman_fiberobo_2023, noauthor_active_nodate, yao_biologic_2015,rivera_hydrogel-textile_2020, sun_fabricfit_2020, haynes_flextiles_2024, glazko_shaping_2024}. Shape-change often occurs when the material contracts or expands under certain stimuli, while leveraging anisotropic or bi-layer structure to achieve a more complex shape-change system. Although this approach might require prior knowledge in material science, the ultimate outcome tends to be cleaner, quieter, and electronic-free.

Our work builds upon material-based interfaces, adopting shape memory material in fiber form to produce re-moldable fabric through state-of-the-art textile procedures. It allows users to mold and lock the shape through hands-on manipulation when heat is applied. We aim to provide an accessible remolding experience to the user without compromising the quality of traditional textiles such as comfort, breathability, washability, and aesthetics.

\section{ExoFabric Working Principle and Design Space}
\label{sec:exofabric_working_principle}
As outlined in the introduction, ExoFabric is a re-moldable, shape-retainable textile system designed for versatile, large-scale shape transformations, offering a lightweight, electronics-free solution compatible with conventional textile practices. Our aim is to add re-moldability, geometrical formability, and shape retainability to fabrics without hindering intended fabric qualities such as stretchability and breathability. This system can be engineered by incorporating thermoplastic materials into fabrics through multiple assembly techniques. To comprehensively define ExoFabric, this section introduces the underlying working principles, selection of materials, fabrication tools, relationships between input parameters and desired outcome, activation processes, and application domains. 

\subsection{Materials} 


\begin{figure}[htbp]
    \centering
    \begin{subfigure}[t]{0.43\linewidth}
        \centering
        \includegraphics[width=\linewidth]{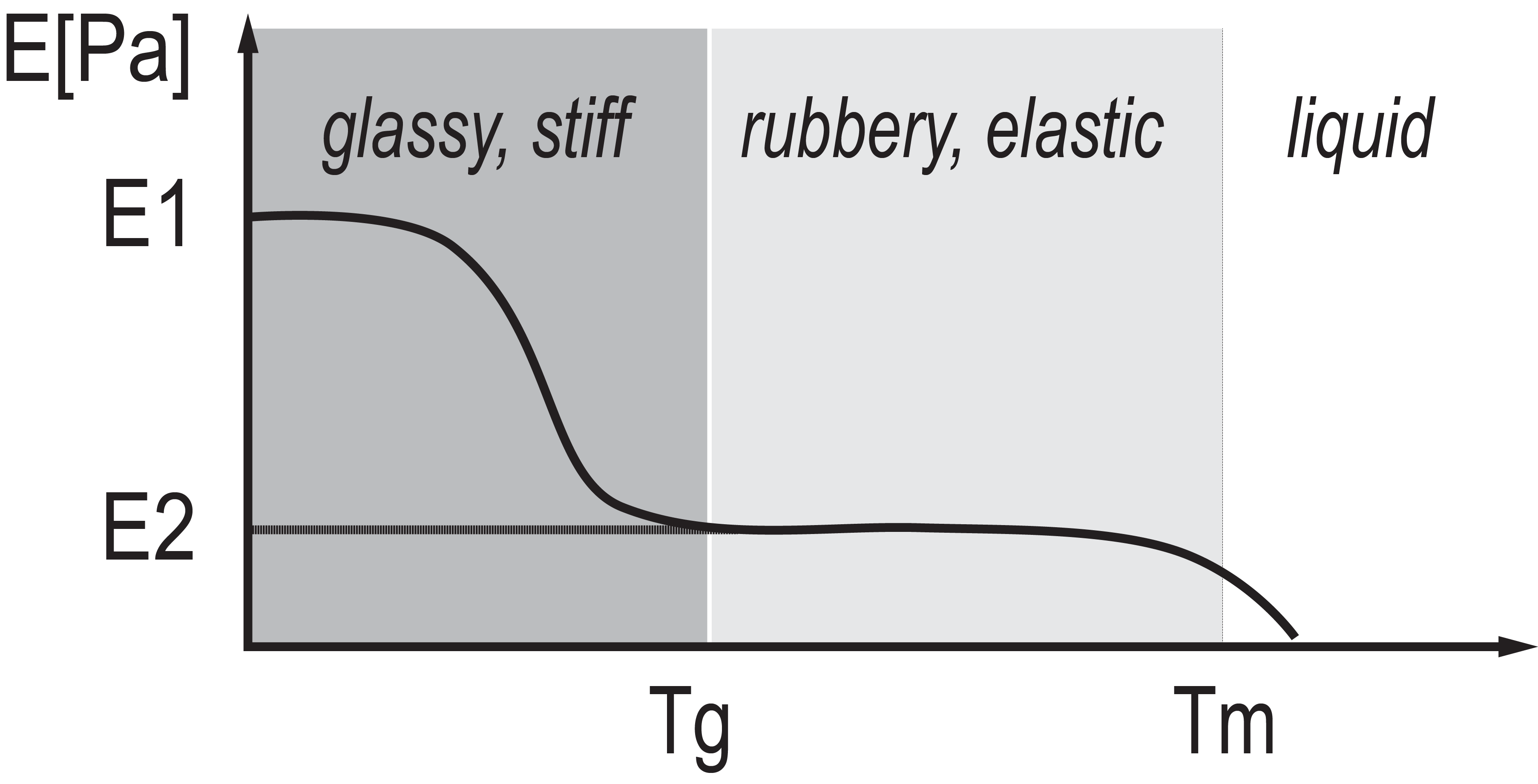}
        \caption{Thermoplastic Stiffness Transition}
        \label{fig:working_principle_transition}
    \end{subfigure}
    \hfill
    \begin{subfigure}[t]{0.55\linewidth}
        \centering
        \includegraphics[width=\linewidth]{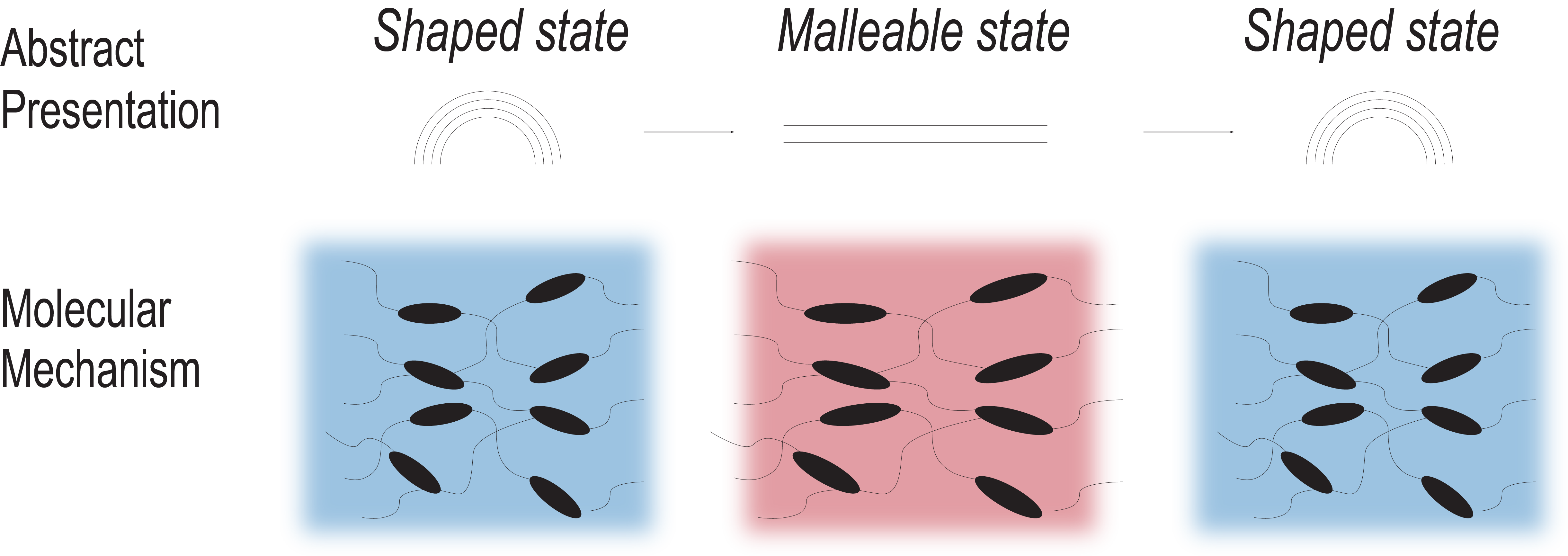}
        \caption{Molecular Mechanism of Shape-Memory Effect}
        \label{fig:working_principle_smeffect}
    \end{subfigure}
    \caption{Working principle of re-moldability through thermoplastic. (a) Thermoplastic stiffness transition states. (b) Shape-memory effect at molecular level when under heating and cooling processes}
    \label{fig:working_principles}
\end{figure}

\subsubsection{Thermoplastic selection:} Thermoplastic, the core material enabling re-moldability, can be easily formed through the application of heat and mold. Thermoplastic typically has a lower glass transition temperature (Tg) range (-130$^{\circ}$C to 150$^{\circ}$C) compared to thermosets (150$^{\circ}$C to 300$^{\circ}$C), which is used for high strength and heat-endurance parts (See \Cref{fig:working_principle_transition}). Tg is the point at which a polymer transitions from a hard, glassy state to a softer, rubbery state. In this rubbery state, the thermoplastic become malleable, allowing the re-molding process to occur as shown in \Cref{fig:working_principle_smeffect}. Tm, on the other hand, is the glass melting temperature, which is not favored for our purpose as the thermoplastic has already transitioned into its liquid state at that point. In this work, we chose to work nylon monofilament with Tg ranging between 47°C to 57°C (approximately 117°F to 135°F) to ensure the temperature is low enough to prevent users from being harmed during the molding process while prioritizing textile integration and tool accessibility within the HCI community and the textile industry.

\subsubsection{Fabric selection:} The choice of fabric can contribute to the desired performance of ExoFabric, such as providing sufficient stiffness for shape-retainability and ease of geometrical formation. In our effort to identify the relevant fabric parameters for ExoFabric, we conducted a thoroughly examination of available market options, analyzing factors such as weight (e.g., GSM, or grams per square meter), stretchability (e.g., 25 percent stretch), stretch direction (e.g., two-way vs all-way), and material composition (e.g., 25\% cotton and 75\% polyester). In this work, we focus on high GSM fabrics to create ExoFabric because the property of these types of fabric are more likely to meet our requirements of stiffness or shape-retainability. 

\subsection{Fabrication tools} 

Thermoplastics exist in a wide variety of form factors (e.g., filaments, threads, sheets, films, pellets, powders) and each form factor has its associated fabrication process and machinery. We identified three fabrication methods that can incorporate thermoplastic into textiles that are commonly used in hobbyist and/or commercial settings, these include:  textile machinery, 3D printing, and lamination. 

\begin{figure}[htbp]
    \centering
    \begin{subfigure}[t]{0.3\linewidth}
        \centering
        \includegraphics[width=\textwidth]{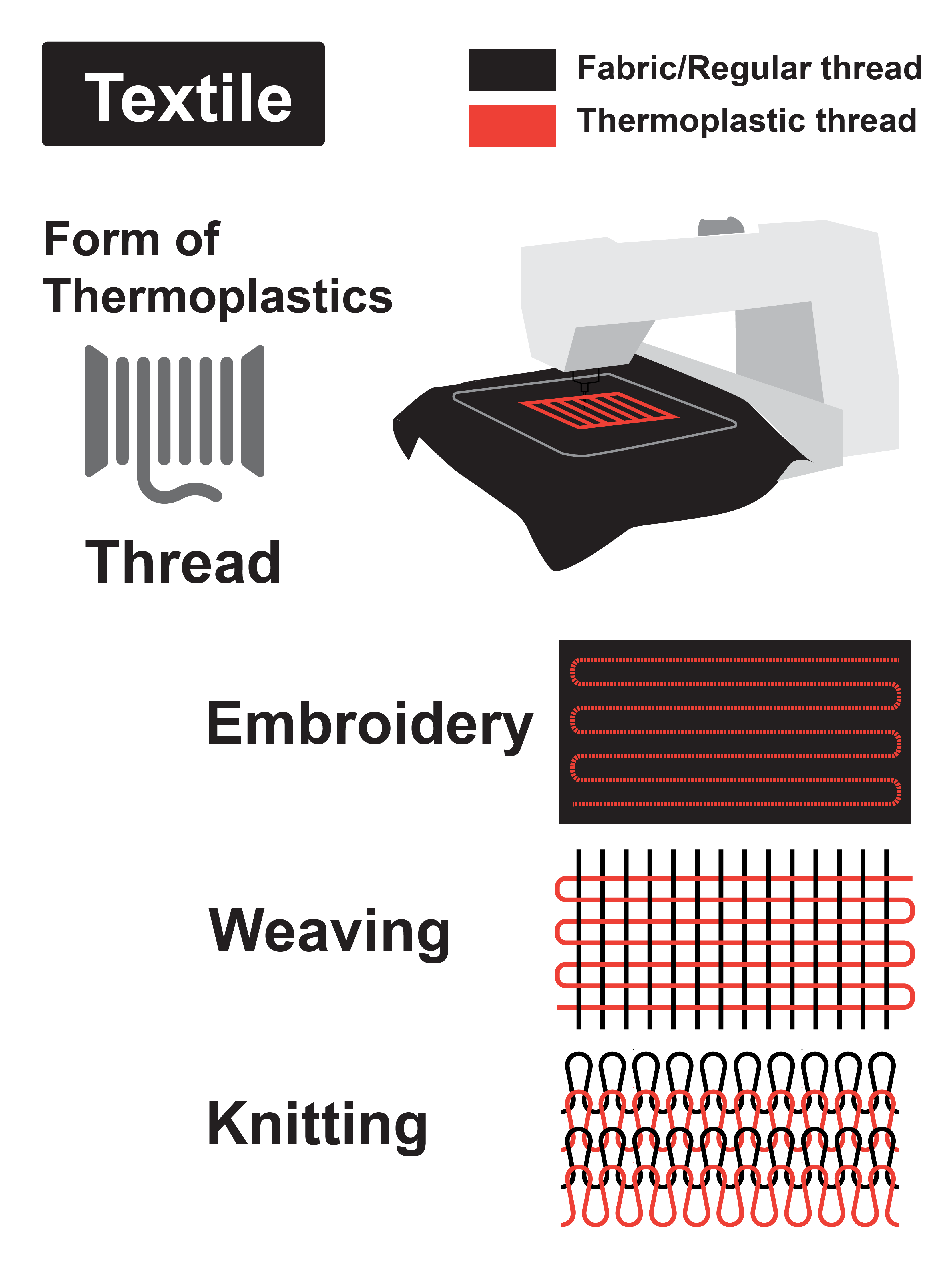}
        \caption{Textile}
        \label{fig:fab_method_textile}
    \end{subfigure}
    \hfill
    \begin{subfigure}[t]{0.3\linewidth}
        \centering
        \includegraphics[width=\textwidth]{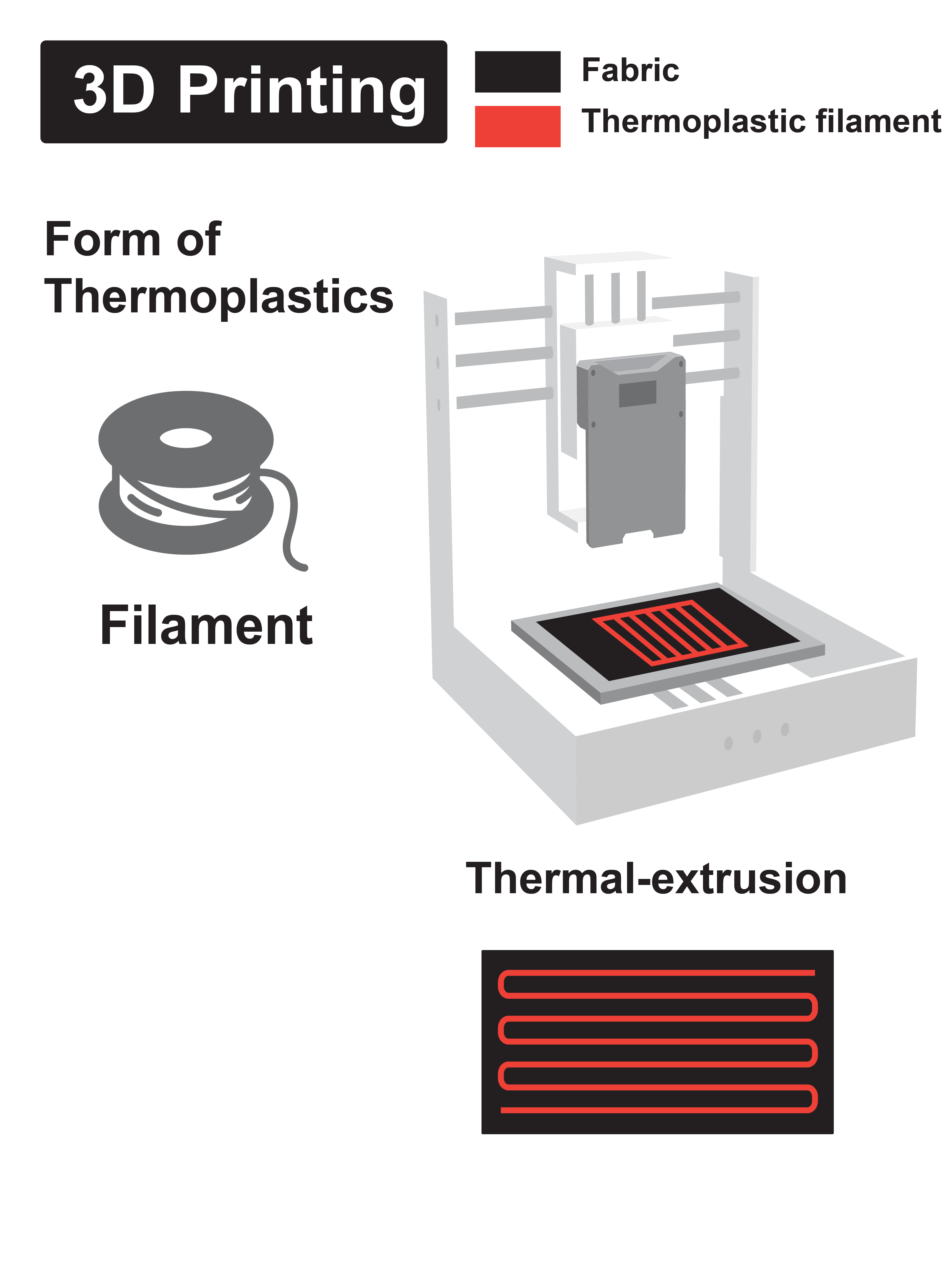}
        \caption{3D Printing}
        \label{fig:fab_method_3dprint}
    \end{subfigure}
    \hfill
    \begin{subfigure}[t]{0.3\linewidth}
        \centering
        \includegraphics[width=\textwidth]{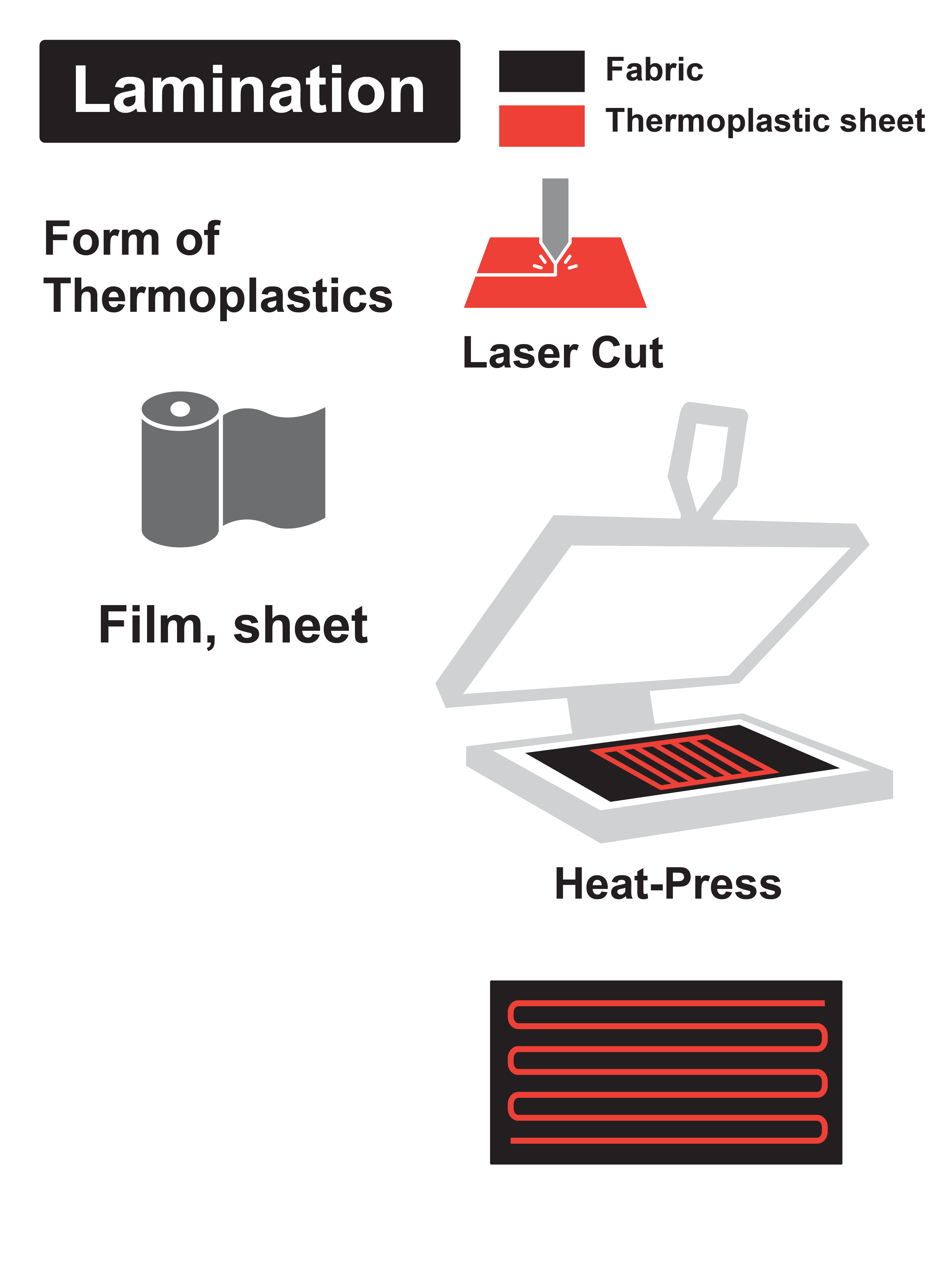}
        \caption{Lamination}
        \label{fig:fab_method_lamination}
    \end{subfigure}   
    \caption{Three methods for integrating thermoplastics into ExoFabric: (a) Textile methods use thermoplastic thread for embroidery, weaving, and knitting; (b) 3D Printing utilizes thermoplastic filament and thermal extrusion; (c) Lamination employs thermoplastic films or sheets, processed through laser cutting and heat pressing.}
    \label{fig:design_space_fab_methods}
\end{figure}

\subsubsection{Textile machinery:}
To allow for better compatibility with standard textile machinery and fabrication processes (e.g., embroidery, knitting, weaving), we chose to work with commercially available thermoplastic monofilament which is a softer, thinner, and more robust fiber (See \Cref{fig:fab_method_textile}). We use a desktop computerized embroidery machine to work with the thermoplastic thread because it allows for flexible thread placement at any direction and location via two-dimensional CNC-axis movement. It is affordable and it is easy to swap the fabric and thread. Automated or computerized knitting or weaving machines are more expensive, less accessible, and require a more complex process that demands proprietary knowledge and extensive training. Beyond the fabrication benefits, using the fiber first approach described here is more likely to provide users with comfortable daily wear and easy maintenance (e.g., washing, drying).

\subsubsection{3D printing}
Similar to \Cref{fig:fab_method_3dprint}, 3D printing with thermoplastics like PLA or TPU is also popular in the HCI community due to its low cost, accessibility, and established design and fabrication software. Thermoplastics have been used as scaffolds in prosthesis that require form-fitting and structural support. However, clothing printed with thermoplastic may lack the comfort and softness needed for daily wear, and the layer-by-layer process could impose adhesion issues affecting durability. When it comes to washability, a 3D printed piece is more likely to be damaged or deformed in the laundering process due to its stiffness and brittleness. 

\subsubsection{Lamination}
On the other hand, thermoplastics in the form of sheets or film, combined with lamination and heat-press processes, are commonly used as a backing material in commercial goods like shoes, clothing, or consumer electronics for providing structural support or shaping (See \Cref{fig:fab_method_lamination}). For example, shoes or socks with a stiffening pad installed around the toe can reduce penetration damage, thus increasing durability. However, additional processes like cutting and positioning are required before lamination, which may hinder the overall streamlining of the process.


\subsection{Fabrication parameters and ExoFabric affordances}  

As mentioned earlier, the desired characteristics of ExoFabric include re-moldability, geometrical formability, stiffness, and, depending on the application, stretchability. We further break down each affordance of ExoFabric and connect them to the relevant fabrication parameters, providing design guidelines for creating ExoFabric. These parameters include the material properties of thermoplastics, the quantity and direction of thermoplastics, the types of fabric, and the molding tools used. The correlations between these parameters and ExoFabric affordances are summarized in \Cref{tab:correlations_fab_params_and_affordance}.

\subsubsection{Quantity of Thermoplastics:}
\label{sec:quantity_of_thermoplastics}
The quantity of thermoplastics primarily contributes to the stiffness of ExoFabric, with a greater amount of material leading to increased stiffness. Stiffness can be enhanced by adjusting two key parameters: line spacing and stitch spacing. \textbf{Line spacing} refers to the distance between parallel rows of stitches, and reducing this distance increases the density of thermoplastic material. Similarly, \textbf{stitch spacing}, which is the distance between individual stitch points, can be reduced to create a denser pattern, further contributing to greater stiffness.

\subsubsection{Directionality of Thermoplastics:}
The directionality of thermoplastics influences the stiffness, geometric formation, and stretchability of ExoFabric. This is due to the angle relationship between the thermoplastic threads and the direction of molding and stretching. Stiffness can be increased when the continuous thermoplastic threads are aligned parallel to the molding direction. Expanding on this principle, the thread layout differs between singly-curved and doubly-curved geometries. Singly-curved surfaces are better formed with continuous, straight lines parallel to the molding direction, while doubly-curved surfaces benefit from a radial thread layout with curved lines rather than concentric circles with straight lines. This configuration allows for a more stretchable area to form a dome shape. Similarly, stretchability can be enhanced by placing thermoplastic threads perpendicular to the stretching direction, allowing for maximum flexibility.

\subsubsection{Material property of Thermoplastics:}
As mentioned in 3.1.1, thermoplastic is the core material to enable re-moldability due to its shape-memory effect under the application of heat (See \Cref{fig:working_principle_smeffect}). 

\subsubsection{Fabric Types:}
As previously mentioned in the material selection, different fabric types can influence the stiffness, geometric formation, and conformability of ExoFabric. For instance, heavier-weight fabrics contribute to increased stiffness. Non-stretch fabrics are preferred for creating singly-curved surfaces due to their crispness, while stretch fabrics are better suited for forming doubly-curved surfaces, as they allow for greater deformation. For applications requiring high conformability, we recommend using stretch fabric to achieve smoother formed surfaces, as non-stretch fabrics can lead to wrinkling.

\begin{table}[htbp]
\small
\centering
\caption{An overview of relationship between fabrication parameters and affordance in ExoFabric}
\label{tab:correlations_fab_params_and_affordance}
\begin{tabular}{|P{0.2\textwidth}|P{0.15\textwidth}|P{0.22\textwidth}|P{0.15\textwidth}|P{0.15\textwidth}|}
\hline
\rowcolor{linDarkGray} \cellcolor{white}&\multicolumn{4}{c}{\color{white}ExoFabric Affordance} \\ 
\rowcolor{linLightGray} \cellcolor{white} &\color{white}Stiffness& \color{white}Geometrical formability& \color{white}Stretchability & \color{white}Re-moldability \\ 
\rowcolor{linLightGray} \cellcolor{white} \multirow{-2}{*}{\color{black}Fabrication Parameters} & \includegraphics[width=0.14\textwidth]{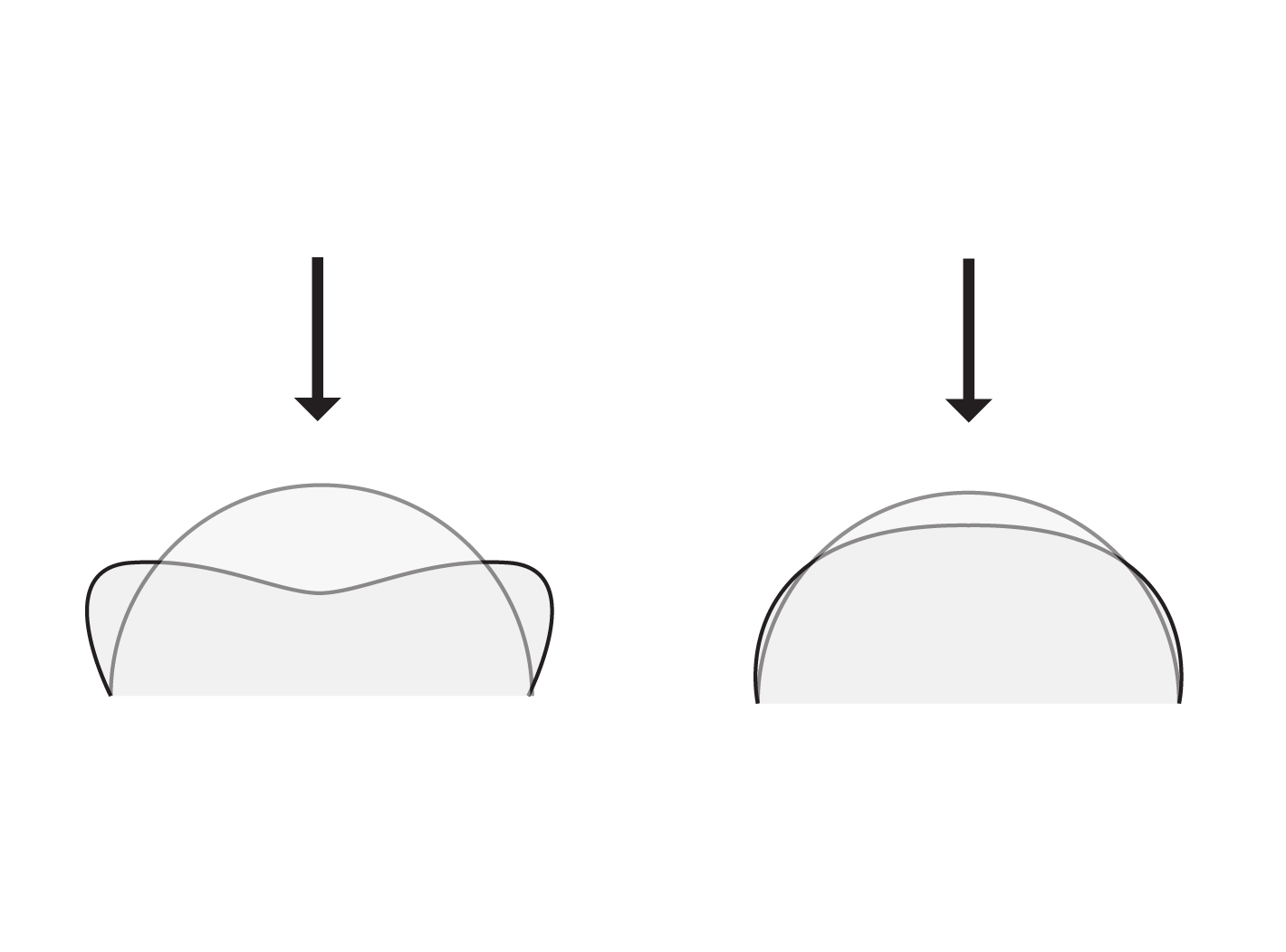} & \includegraphics[width=0.21\textwidth]{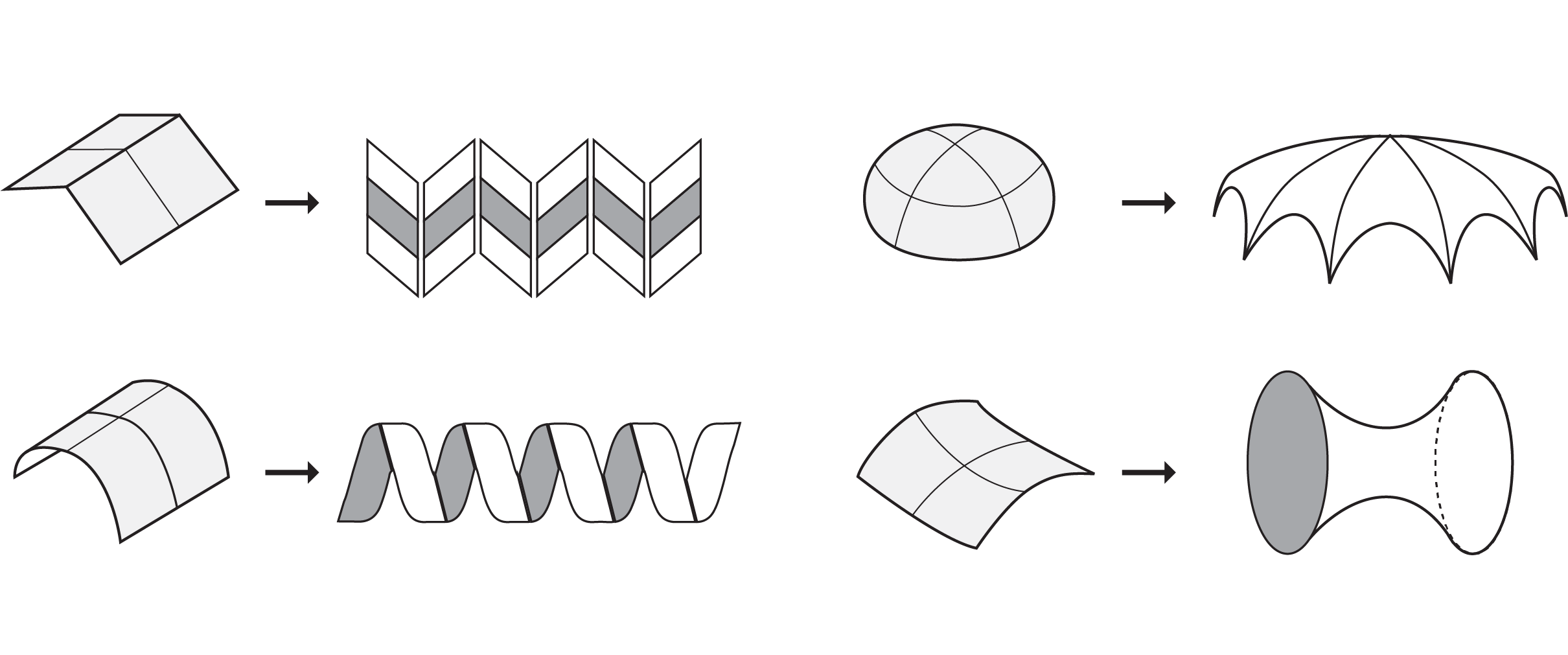} & \includegraphics[width=0.14\textwidth]{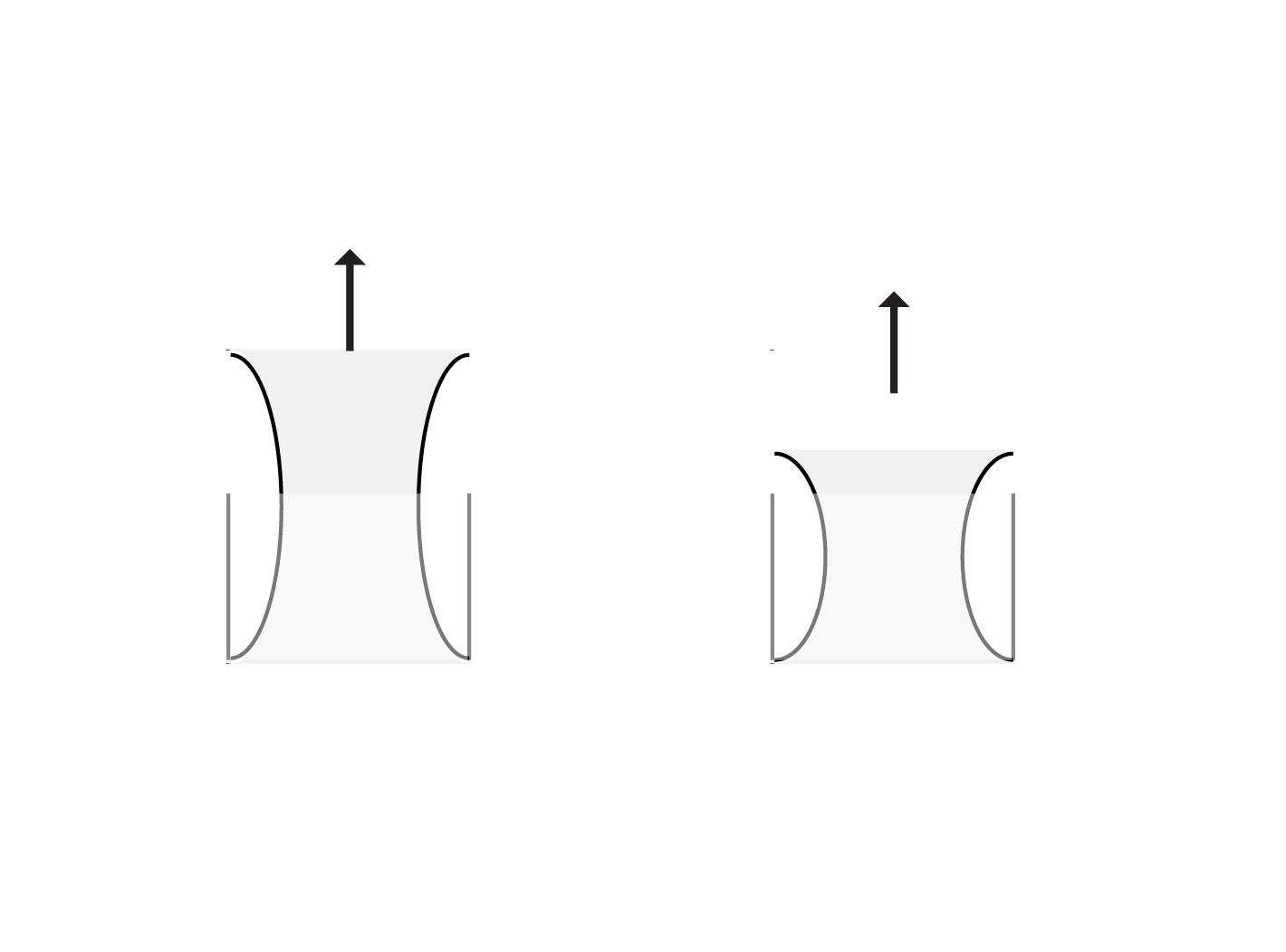} & \includegraphics[width=0.14\textwidth]{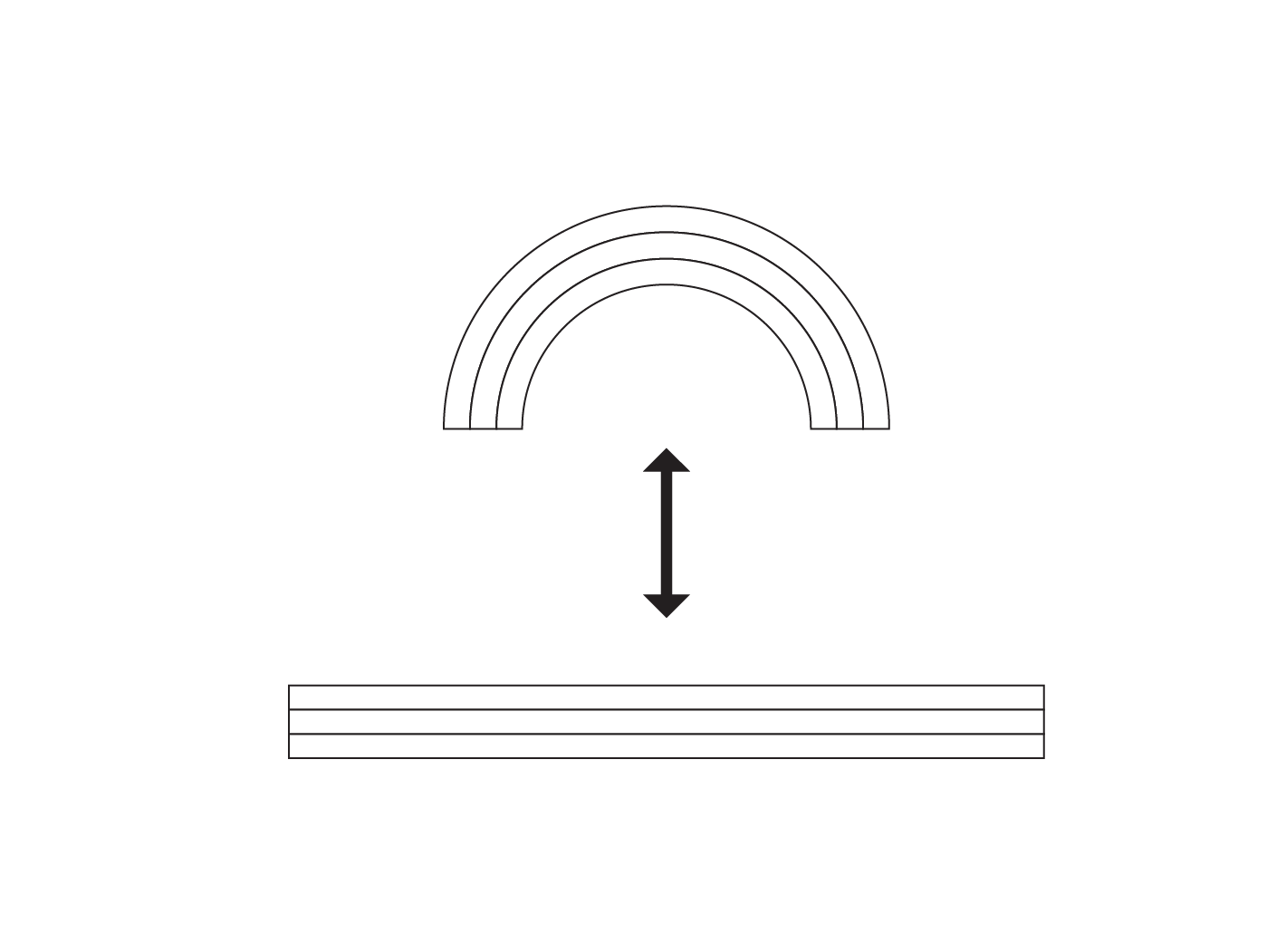}\\ \hline
Thermoplastic quantity  & & & & \\ 
\includegraphics[width=0.19\textwidth]{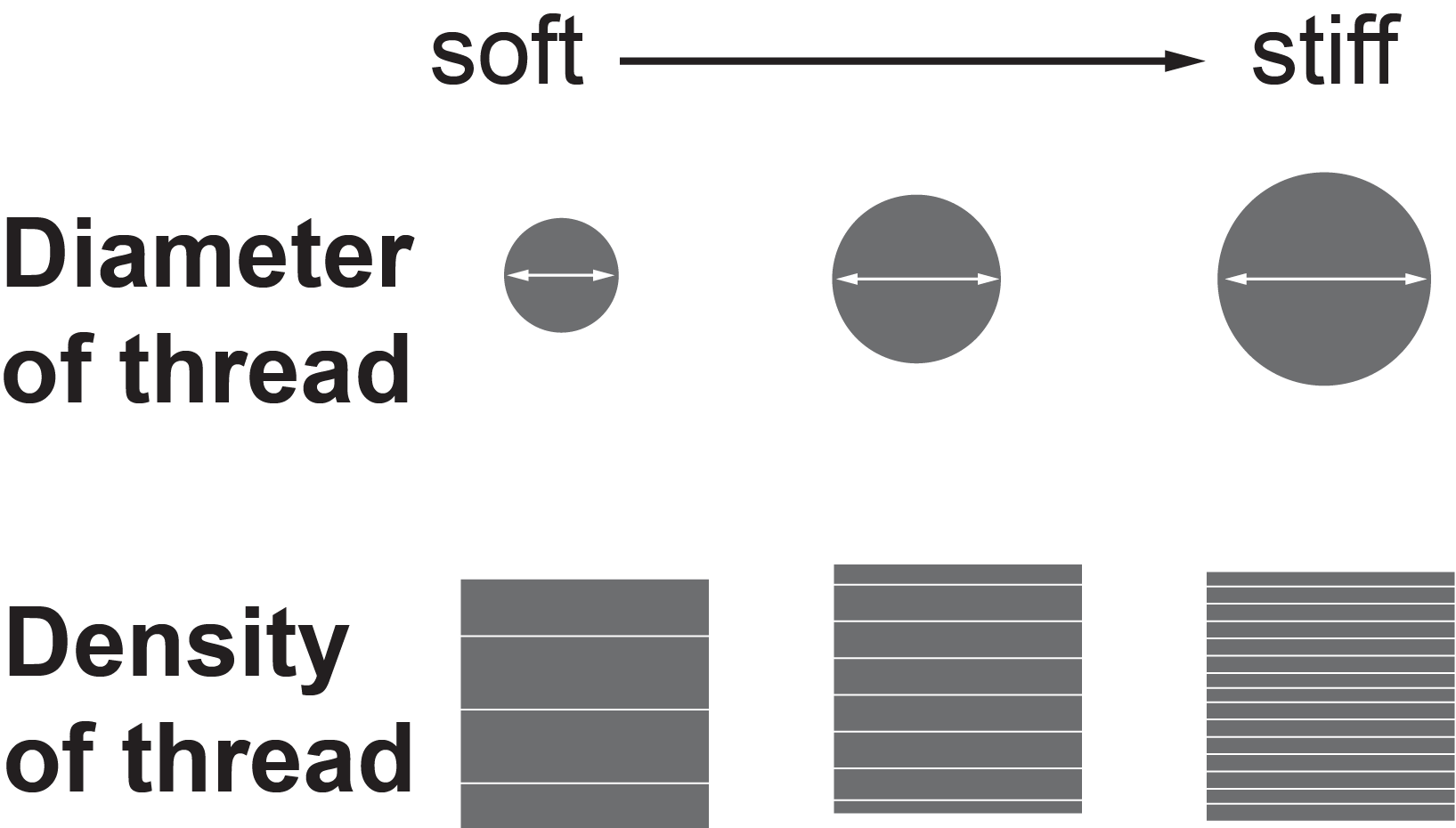} & \multirow{-1}{*}{\includegraphics[scale=0.3]{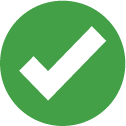}} & & &\\ \hline
Thermoplastic direction & & & &\\
\includegraphics[width=0.19\textwidth]{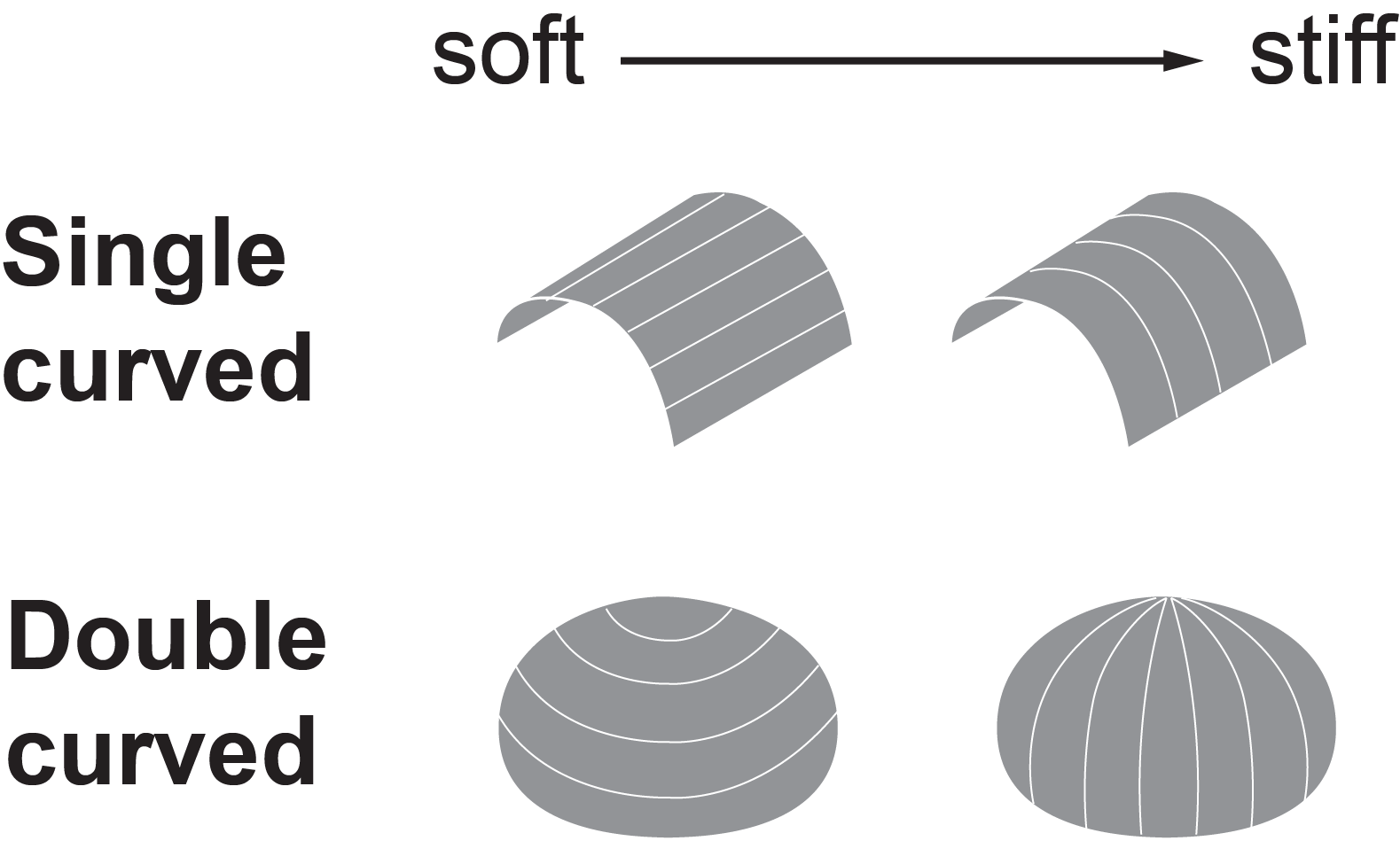} & \multirow{-1}{*}{\includegraphics[scale=0.3]{checkmark.png}} & \multirow{-1}{*}{\includegraphics[scale=0.3]{checkmark.png}} & \multirow{-1}{*}{\includegraphics[scale=0.3]{checkmark.png}} &\\ \hline
Thermoplastic property  & & & & \\
\includegraphics[width=0.19\textwidth]{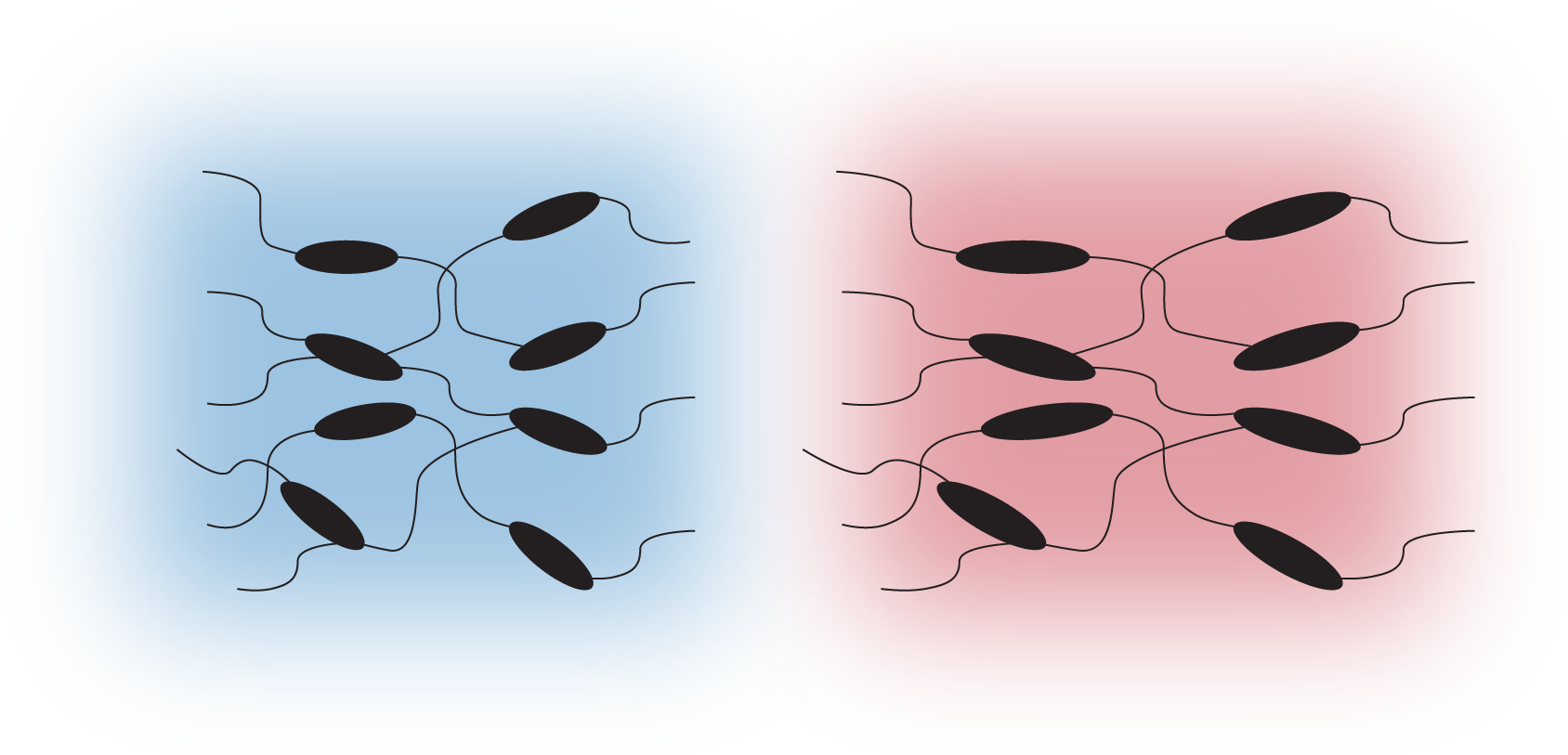} &  & & & \multirow{-1}{*}{\includegraphics[scale=0.3]{checkmark.png}} \\ \hline
Fabric types  & & & & \\ 
\includegraphics[width=0.19\textwidth]{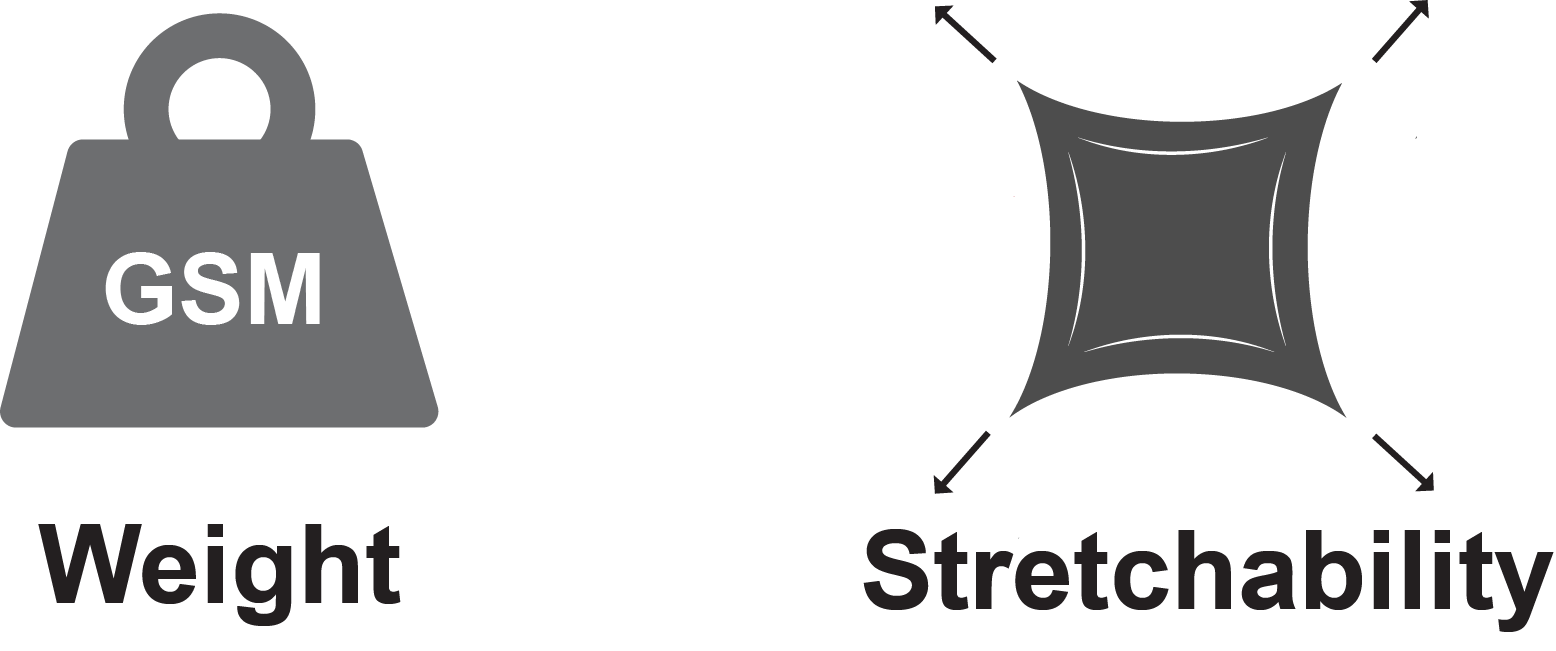} & \multirow{-1}{*}{\includegraphics[scale=0.3]{checkmark.png}} & \multirow{-1}{*}{\includegraphics[scale=0.3]{checkmark.png}} & \multirow{-1}{*}{\includegraphics[scale=0.3]{checkmark.png}} &\\ \hline
\end{tabular}
\end{table}

\subsubsection{Stiffness:} Stiffness refers to ExoFabric’s ability to resist deformation when subjected to force. It may vary between soft, easily deformable states and stiff, rigid states, depending on the force applied to the fabric. This property allows ExoFabric to switch between flexible and structured forms; thus makes it suitable for dynamic applications that require both flexibility and support.

\subsubsection{Geometrical Formability:} Geometrical formability describes ExoFabric’s capability to adopt various shapes, including single and double curves. It can be folded or bent to create simple curves or more complex shapes like domes and saddles. This feature could enhance ExoFabric’s versatility for creating intricate, three-dimensional structures.

\subsubsection{Stretchability:} Stretchability refers to the ability to elongate under force. It can be designed to either stretch or remain non-stretchable, with this programmable feature tailored to the specific application. This property is crucial for wearables and other applications in which adaptability to body movement or dynamic environments is required.

\subsubsection{Re-moldability:} Re-moldability highlights the ability to transition between a malleable state and a shaped state. After being shaped, it can be re-heated or re-manipulated to return to a flexible form. This property makes ExoFabric reusable and adjustable for various design needs that change over time.

\begin{figure}[htbp]
    \centering
    \begin{subfigure}[t]{0.48\linewidth}
        \centering
        \includegraphics[width=\textwidth]{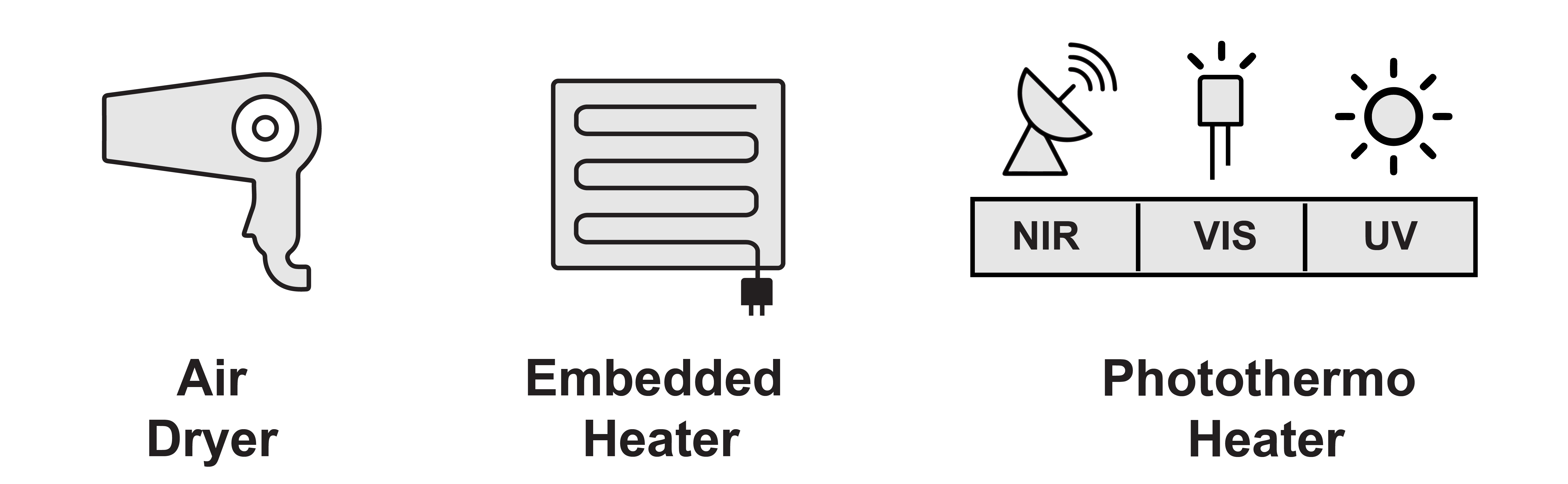}
        \caption{Activation Methods}
        \label{fig:fab_method_heating}
    \end{subfigure}
    \hfill
    \begin{subfigure}[t]{0.48\linewidth}
        \centering
        \includegraphics[width=\textwidth]{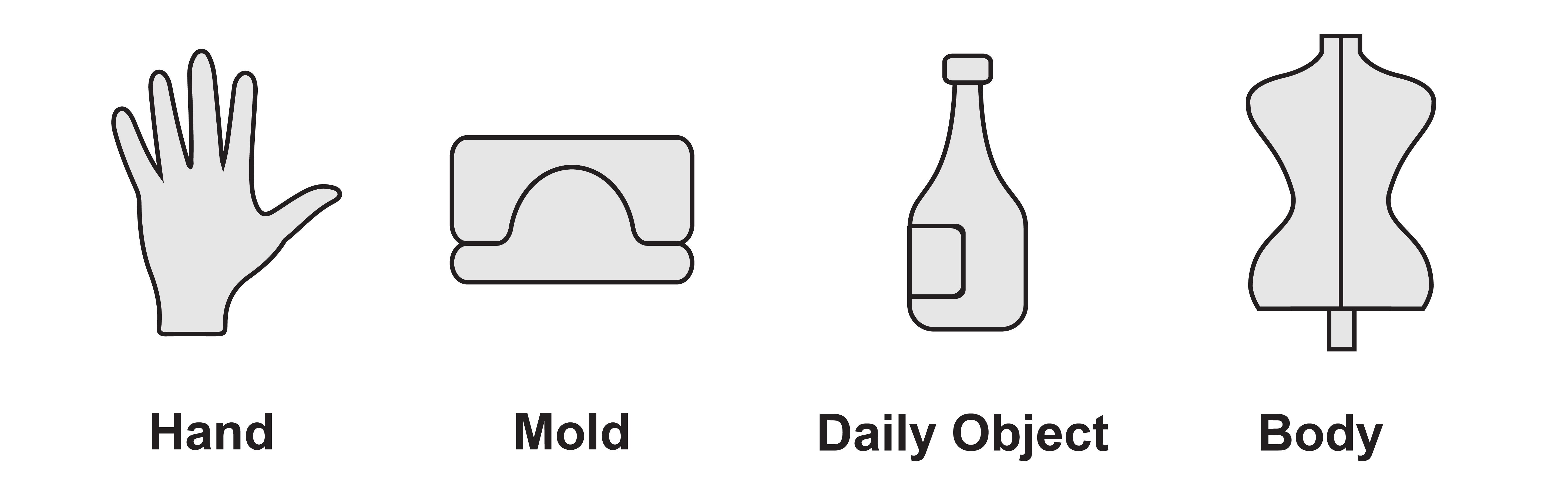}
        \caption{Molding Methods}
        \label{fig:fab_method_molding}
    \end{subfigure}   
    \caption{Two essential parts of ExoFabric fabrication: (a) activation methods include air dryer, embedded heater, and photothermal techniques such as NIR, visible light, and UV. (b) molding approaches range from hand shaping to the use of specialized apparatus, everyday objects, or the human body}
    \label{fig:design_space_heating_molding}
\end{figure}

\subsection{ExoFabric Molding} 

\subsubsection{Activation Methods:} To transition ExoFabric from a solid to a malleable state, heat must be applied to reach the thermoplastic's glass transition temperature. Users can achieve this through various methods, including common household heating devices like a blow dryer, an embedded system that utilizes joule heating within the textile, or a photothermal heating system that applies a near-infrared (NIR) absorbing coating to heat and actuate the ExoFabric (See \Cref{fig:fab_method_heating}). Household heating devices are convenient and easy to use, though they may lack precise control over the heated area and offer simpler construction. In contrast, an embedded system provides precise control over both the heating area and temperature but requires a digital control system to regulate current output, which adds complexity to the fabrication process. A photothermal system, while capable of being heated by light sources such as solar, infrared, or even light-emitting diode, presents challenges due to the potential toxicity of the coating, necessitating protective layers, and generally offers slower heating efficiency compared to the other two methods.

\subsubsection{Molding Methods:} Molding methods play a crucial role in determining the final shape. As shown in \Cref{fig:fab_method_molding}, molding can be achieved in various ways, including using well-designed molds for single-curved or doubly-curved forms, hands-on manipulation for creating versatile shapes, or utilizing any available object in the user's surroundings. The human body itself can also serve as a molding interface for custom-fit wearables.

\begin{figure}[htbp]
    \centering
       \centering
    \begin{subfigure}[t]{0.33\linewidth}
        \centering
        \includegraphics[width=\textwidth]{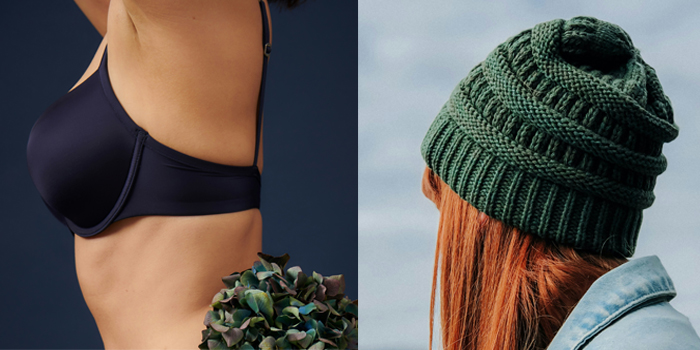}
        \caption{Form-fitting Wearables}
    \end{subfigure}
    \hfill
    \begin{subfigure}[t]{0.33\linewidth}
        \centering
        \includegraphics[width=\textwidth]{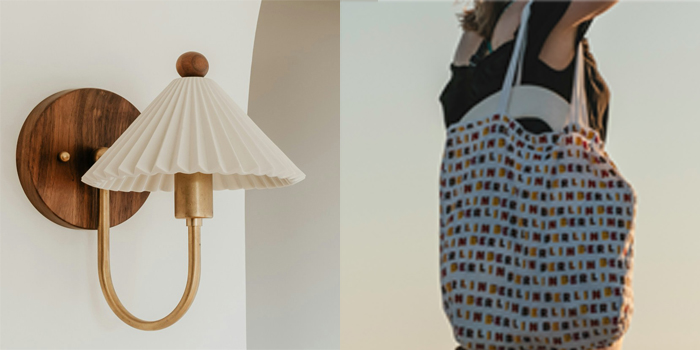}
        \caption{Reconfigurable Soft Goods}
    \end{subfigure}
    \hfill
    \begin{subfigure}[t]{0.33\linewidth}
        \centering
        \includegraphics[width=\textwidth]{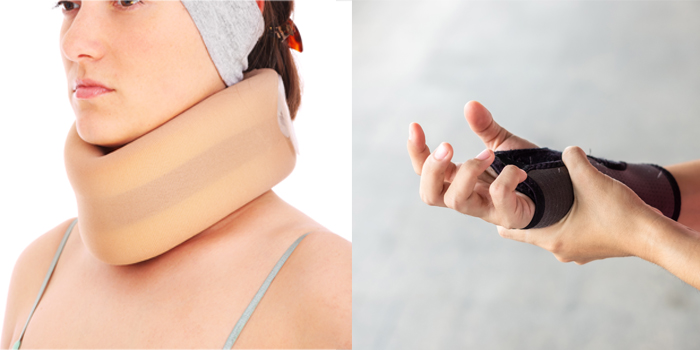}
        \caption{Structural Support}
    \end{subfigure}  
    \caption{Three domains of ExoFabric applications: (a) Form-fitting wearables for customized clothing, (b) Reconfigurable soft goods for flexible and adaptable designs, (c) Structural support for sports and rehabilitation to provide strength and stability in wearable devices.}
    \label{fig:design_space_apps}
\end{figure}

\subsection{Applications} 

Leveraging the understanding of the defined fabrication parameters and outcomes of ExoFbaric, we propose three application domains for ExoFabric as shown in \Cref{fig:design_space_apps}. The first application domain is form-fitting wearables. It is designed to snugly fit the contours of the wearer's body for intimate wear like bras, custom fit clothing, and health and fitness clothing that monitor biometrics. ExoFabric may provide an opportunity for fine-tuning clothing that fits an individual’s silhouette or body changes overtime without the need for mass-producing discrete sizes or paying a highly trained tailor to create a custom fit item. Our second proposed application domain is the structural support for sport and rehabilitation. As most of the commercial products focus heavily on functionality but do not much consider social perceptions or the aesthetics of matching with a daily outfit, we proposed a fabric-based approach to prosthesis that can turn a bulky prosthesis into a thinner and more personalized form factor. The last application space is reconfigurable soft goods, where the hands-on molding process may facilitate highly expressive forms of furniture, accessories, and costume design. The repeatable shaping allows for interchangeable functions in the short term, which can ultimately reduce the number of new purchases or material usage for a variety of custom builds, contributing to sustainability efforts over time.  

\section{Technical Characterization}

As outlined in \Cref{sec:exofabric_working_principle}, where we defined the materials, fabrication parameters, and desired outcomes for ExoFabric, this section provides a detailed evaluation of each parameter and its impact on the final outcome to identify the optimal combinations for specific applications. We begin with a preliminary characterization to establish a foundational workflow, followed by identifying the optimal embroidery configurations for stiffness, geometrical formability, and stretchability on a single layer of fabric. Finally, to enhance the performance of ExoFabric, we investigate the effect of multiple layers on its mechanical strength.

\subsection{Preliminary characterization} 
The objective of conducting preliminary characterizations is to establish a robust foundation for subsequent detailed characterization. This involves identifying optimal parameters for thread and fabric to ensure  desired material performance, fabrication stability, and finish quality. Once determined, these parameters are set as constants to serve as the foundational basis for subsequent experiments. 

\begin{figure}[htbp]
    \centering
    \begin{subfigure}[t]{\linewidth}
        \centering
        \includegraphics[width=\linewidth]{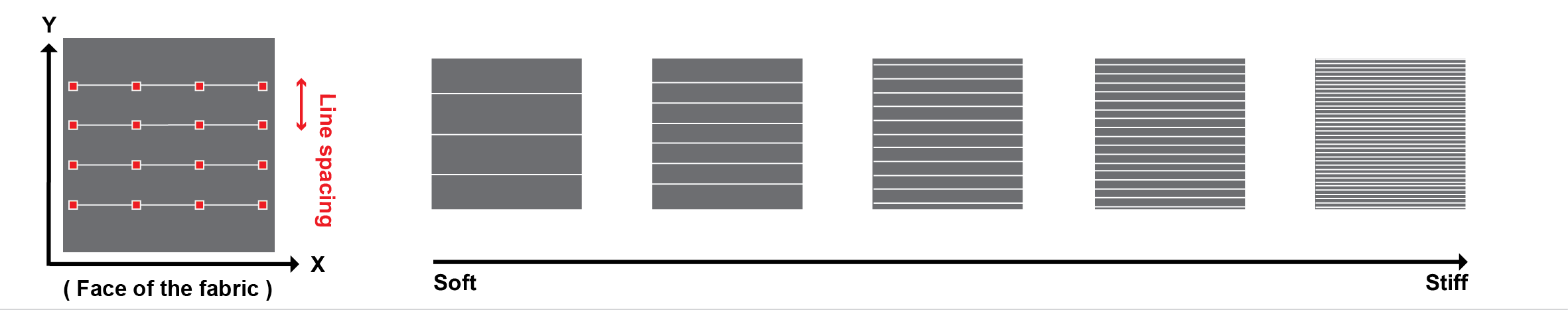}
        \caption{Line Spacing}
        \label{fig:embroidery_param_line_count}
    \end{subfigure}
    \hfill
    \begin{subfigure}[t]{\linewidth}
        \centering
        \includegraphics[width=\linewidth]{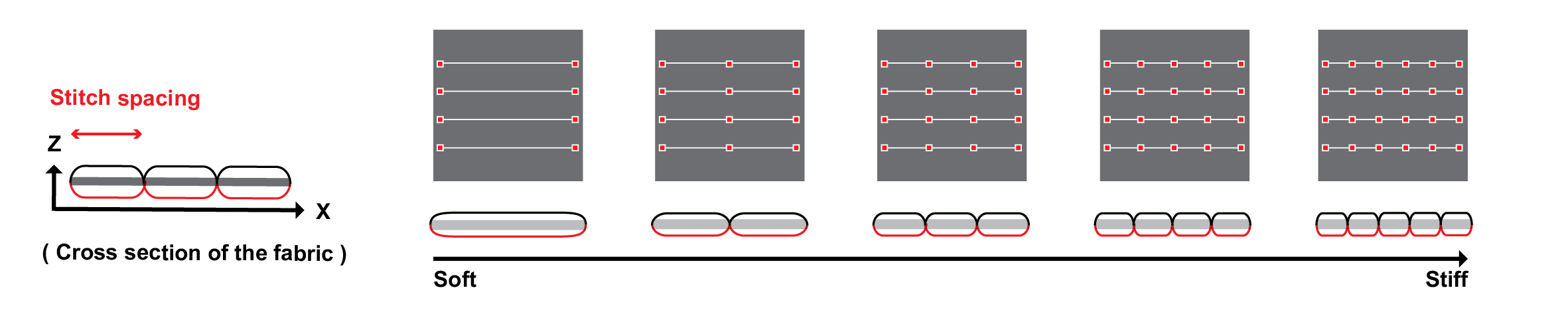}
        \caption{Stitch Spacing}
        \label{fig:embroidery_param_stitch_space}
    \end{subfigure}    
    \caption{Illustration of the effects of varying line spacing (a) and stitch spacing (b) on stitching outcomes.}
    \label{fig:embroidery_param}
\end{figure}

\subsubsection{Embroidery Parameters}
As discussed in \Cref{sec:quantity_of_thermoplastics}, we identified two embroidery parameters---\textbf{line spacing} and \textbf{stitch spacing} as they could potentially influence fabric stiffness. Reducing line spacing results in a greater concentration of thermoplastic threads in the X-Y plane, thereby increasing stiffness (\Cref{fig:embroidery_param_line_count}). Likewise, decreasing stitch spacing increases the density of thermoplastic threads in the cross-section of the fabric, which also enhances stiffness (\Cref{fig:embroidery_param_stitch_space}). 

\begin{figure}[htbp]
    \centering
    \begin{subfigure}[t]{0.48\linewidth}
        \centering
        \includegraphics[width=\textwidth]{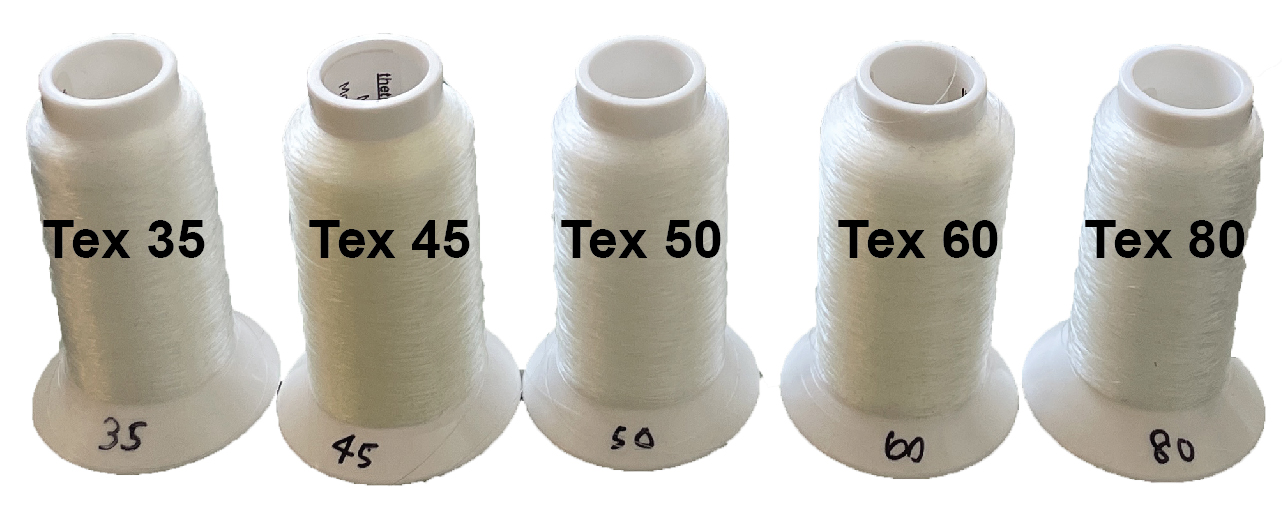}
        \caption{Thread Sizes}
        \label{fig:exp_thread_size_samples}
    \end{subfigure}
    \hfill
    \begin{subfigure}[t]{0.48\linewidth}
        \centering
        \includegraphics[width=\textwidth]{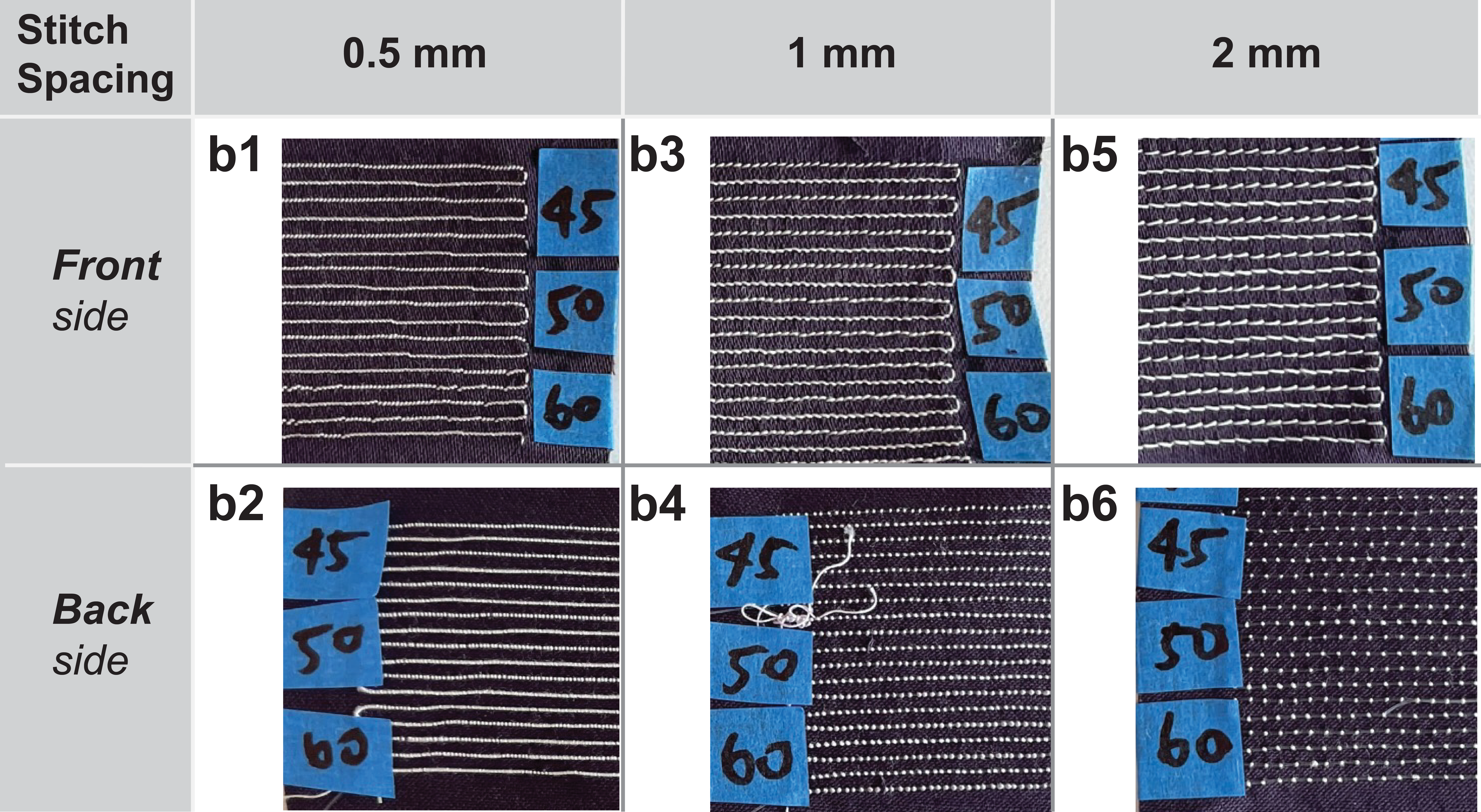}
        \caption{Fabrication Results}
        \label{fig:exp_thread_size_results}
    \end{subfigure}    
    \caption{Analysis of thread size impact on finish quality and machine stability. (a) Threads in sizes Tex 45, 50, 60. (b) Fabrication results across three thread sizes and three stitch spacings.}
    \label{fig:exp_thread_size2}
\end{figure}

\subsubsection{Thread Size} 
We identified the appropriate thermoplastic thread size (i.e., cross-sectional area) to achieve an optimal balance between stiffness enhancement and fabrication stability. While it is obvious that larger thread sizes contribute to increased stiffness due to the higher thermoplastic content, they are also subject to machine limitations. To assess the stability, we used two hobbyist embroidery machines (Janome MC500 and Brother SE600) to embroider single-sided thermoplastic threads with sizes of Tex 35, 45, 50, 60, 80 (\Cref{fig:exp_thread_size_samples}) on a non-stretch 336 GSM fabric. During several rounds of fabrication, we observed that the machines frequently jammed when using threads with Tex 80 as the embroidery machine is  designed for operating with thinner threads and soft materials. Tex 35, on the hand, is too soft to perform the desired mechanical strength, suggesting that there is no need to consider sized lower that this. 
Thus, we excluded Tex 35 and 80 from our testing candidates. The results in \Cref{fig:exp_thread_size_results} demonstrated promising outcomes across all tested thread sizes (Tex 45, 50, 60) and stitch spacings (0.5 $mm$, 1 $mm$, 2 $mm$). The front and back stitching showed good alignments, with only minor tension issues in Tex 50 at 1 mm spacing (b4). Overall, all three thread sizes performed well.We selected thread size with Tex 60 because it offers the greatest stiffness among the tested thread size while maintaining the fabrication stability and finish quality.

\subsubsection{Thread Side}
\begin{figure}[htbp]
    \centering
    \begin{subfigure}[t]{0.48\linewidth}
        \centering
        \includegraphics[width=\textwidth]{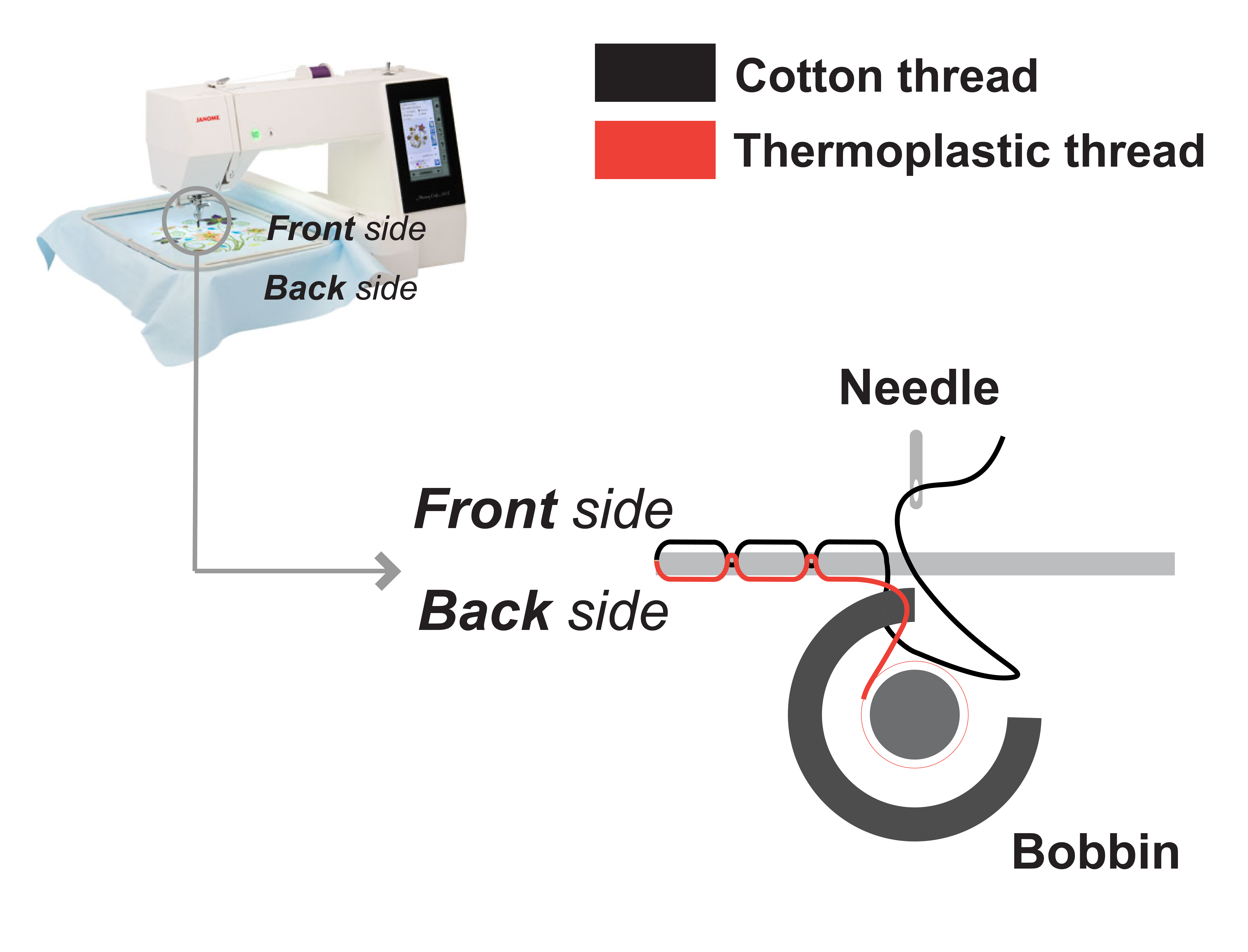}
        \caption{Thread Side}
        \label{fig:exp_thread_side_method}
    \end{subfigure}
    \hfill
    \begin{subfigure}[t]{0.48\linewidth}
        \centering
        \includegraphics[width=\textwidth]{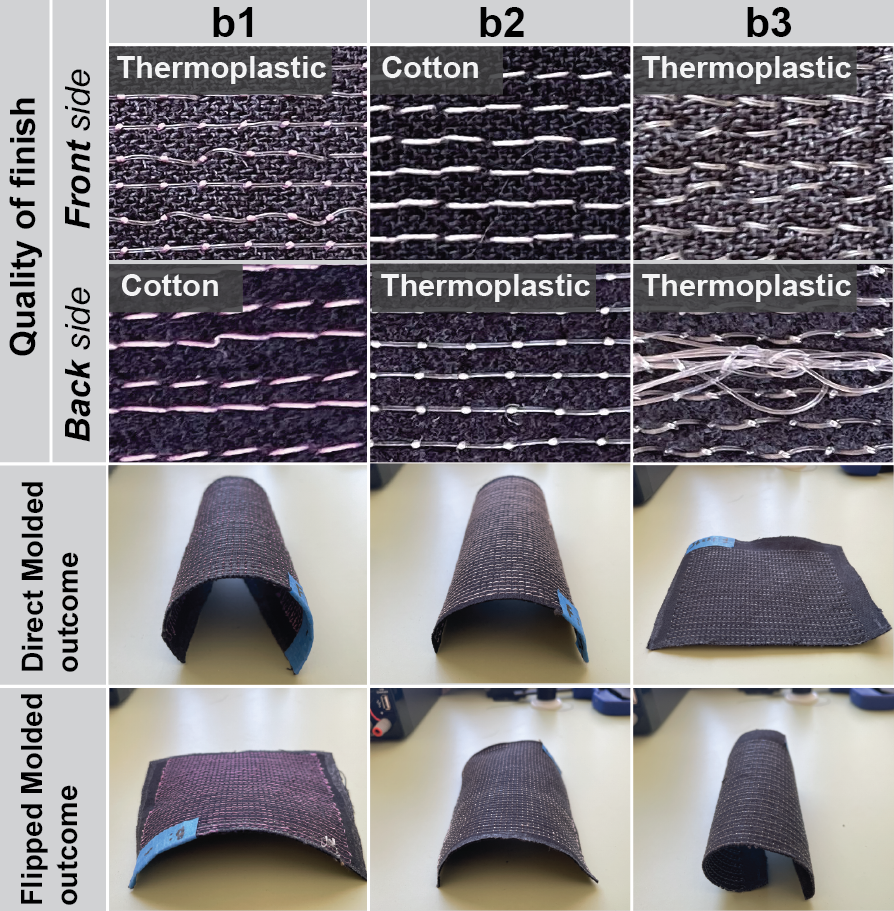}
        \caption{Fabrication Results}
        \label{fig:exp_thread_side_results}
    \end{subfigure}    
    \caption{Examination of thread size and its impact on fabric moldability, finish quality, and machine stability. (a) Embroidery machine mechanism: stitches are created by intertwining two threads from the fabric’s front and back. (b1) Thermoplastic on the front side and cotton on the back. (b2) Cotton on the front and thermoplastic on the back. (b3) Thermoplastic on both sides.}
    \label{fig:exp_thread_side}
\end{figure}

We further investigated the effects of embroidering thermoplastic threads on different sides of the fabric, including double-sided versus single-sided applications, as well as front versus back positioning. Double-sided embroidery involves stitching thermoplastic threads on both sides of the fabric, whereas single-sided embroidery has thermoplastic threads on one side and cotton threads on the other. We investigated three configurations: thermoplastic on the front side, cotton on the back side (\Cref{fig:exp_thread_side_results}1), cotton on the front side, thermoplastic on the back side (\Cref{fig:exp_thread_side_results}2), and thermoplastic on both sides (\Cref{fig:exp_thread_side_results}3). These configurations were molded using two approaches, with the threads on either the front or back side positioned on the elongated part of a bend structure. We observed a significant difference in the molded outcome when molded with the threads facing the opposite side ( \Cref{fig:exp_thread_side_results}1\&3). In  \Cref{fig:exp_thread_side_results}1, the sample formed as intended when the front-side threads faced upward, but not when molded with the opposite orientation. Similarly, the results in \Cref{fig:exp_thread_side_results}3 shows the the sample can only be formed well on one side. However, in \Cref{fig:exp_thread_side_results}2, the molded outcome perform relatively stable on both sides, leading us to select this configuration for further testing and prototype development. This suggests that the configuration in \Cref{fig:exp_thread_side_results}2 offer the most balanced tension, while the configurations in \Cref{fig:exp_thread_side_results}1\&3 perform the unbalanced tension between the sides. In addition to its shape-forming limitations, the double-sided configuration often results in lower finish quality, causing the fabric to become bumpy and the threads to tangle due to the heavy-duty nature of the thermoplastic thread (See the back side of \Cref{fig:exp_thread_side_results}3). In the single-sided configuration, front-side threading generally experiences higher tension than back-side threading given the threading mechanism of the embroidery machine (\Cref{fig:exp_thread_side_method}). When placing heavy-duty threads like thermoplastic on the front side of the fabric, it increases the risk of machine jamming (\Cref{fig:exp_thread_side_results}1). To minimize the risk of fabrication instability, we opted for single-sided threading, with the thermoplastic placed on the back side of the fabric (\Cref{fig:exp_thread_side_results}2). 

\subsubsection{Thread Direction and Geometry}
\begin{figure}[htbp]
    \centering
    \includegraphics[width=1\linewidth]{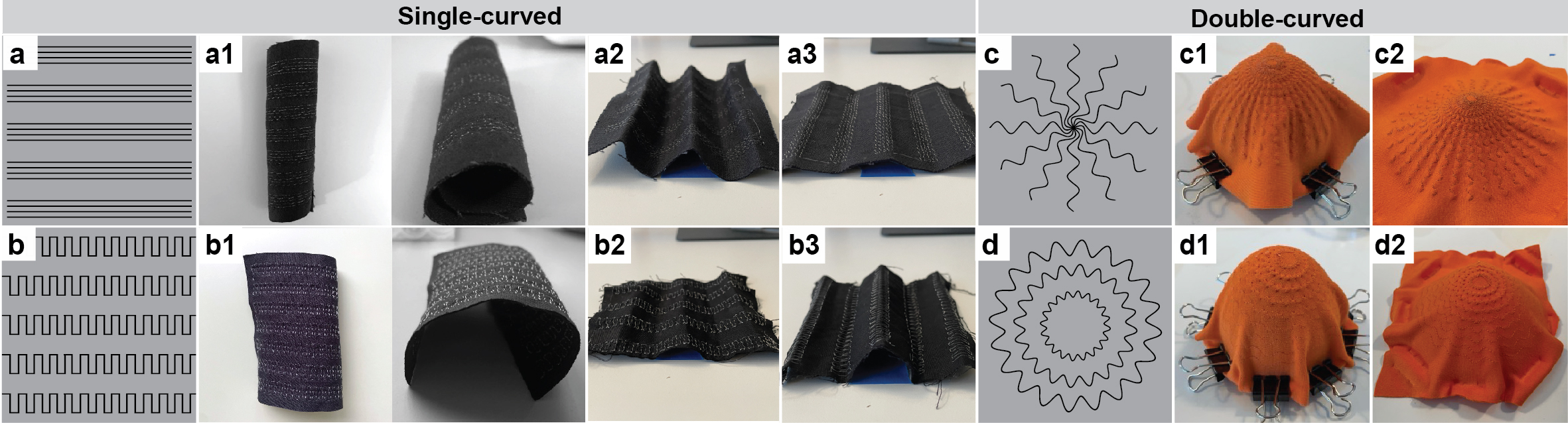}
    \caption{Examination of thread direction and its impact on single and doubly curved geometry formation. (a) Linear patterns parallel to the bending direction: (a1) sharp bending, (a2) sharp folding, (a3) less sharp folding. (b) Linear patterns perpendicular to the bending direction: (b1) less sharp bending, (b2) less sharp folding, (b3) sharp folding. (c) Radial pattern: (c1 \& c2)  sharp doubly-curved molding. (d) Concentric pattern: (d1 \& d2) sharp doubly-curved molding.}
    \label{fig:Thread direction}
\end{figure}

Thread direction is another crucial factor influencing stiffness. Since ExoFabric is an anisotropic material, the angle between the fiber direction and the molding direction is crucial for maintaining structural integrity. In a single-curved geometry, such as  configurations a and b in Figure \ref{fig:Thread direction}, we demonstrated that threads aligned with the bending direction (Figure \ref{fig:Thread direction}a1) produce a more defined cylindrical shape compared to those aligned perpendicular to the bending direction (Figure \ref{fig:Thread direction}b1). Similarly, the folding outcomes in \Cref{fig:Thread direction}a3 and b2 are less sharp than \Cref{fig:Thread direction}a2 and b3, as they do not effectively utilize the thread direction. For double-curved geometries, the continuous thread should be aligned with the molding direction, either in the radial or concentric pattern to allow the maximum deformation in these directions, hence, to ensure the success of the doubly-curved geometries formation. Unlike the straight lines used in single-curved geometries, the wavy curve design in double-curved geometries accommodated fabric stretching, enabling the desired shape formation as shown in \Cref{fig:Thread direction}c and d.

\subsubsection{Fabric Types}
\begin{figure}[htbp]
    \centering
    \begin{subfigure}[t]{0.18\linewidth}
        \centering
        \includegraphics[width=0.8\textwidth]{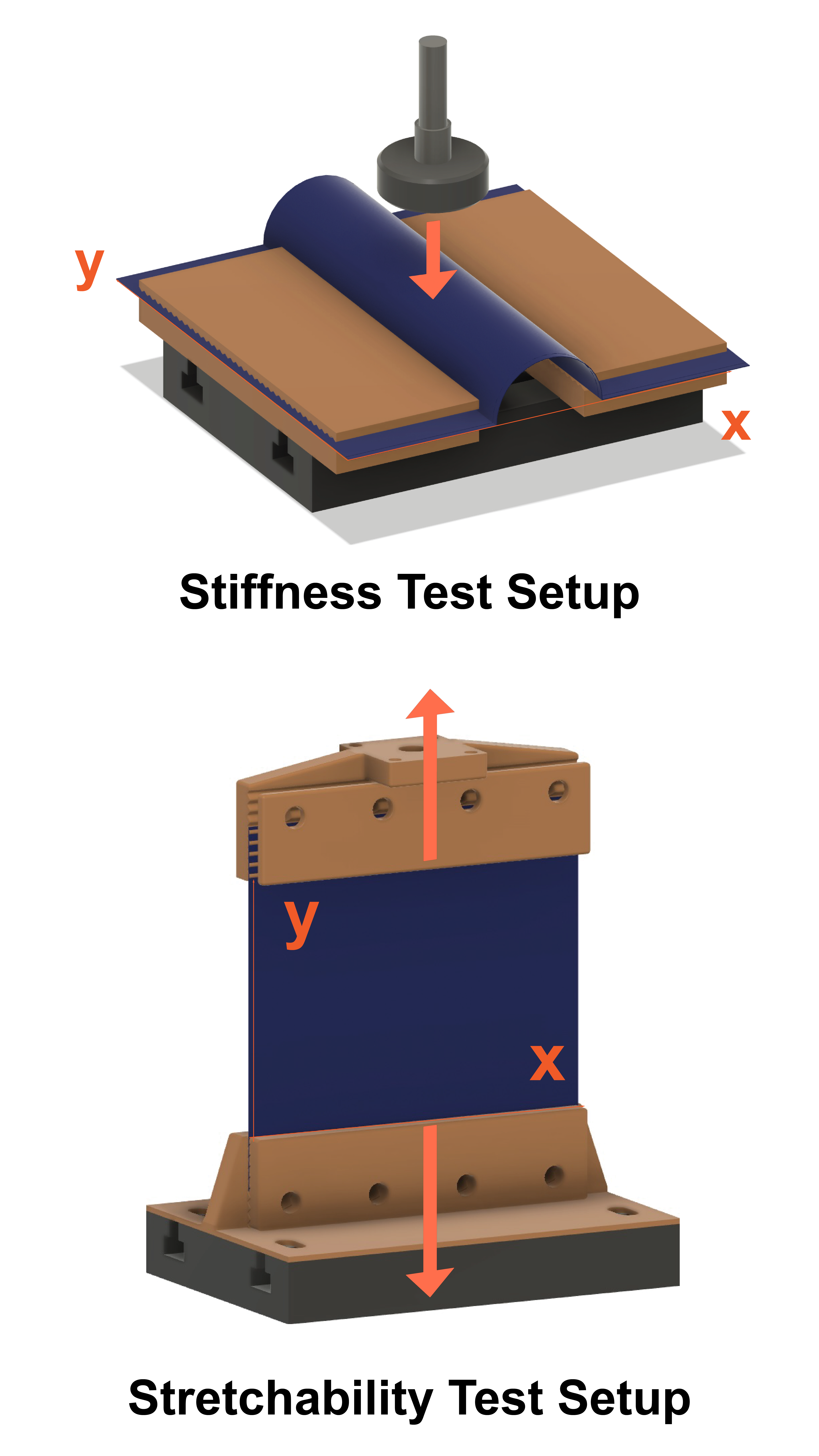}
        \caption{Fabric Test Setups}
        \label{fig:exp_fabric_test_setup}
    \end{subfigure}
    \hfill
    \begin{subfigure}[t]{0.4\linewidth}
        \centering
        \includegraphics[width=\textwidth]{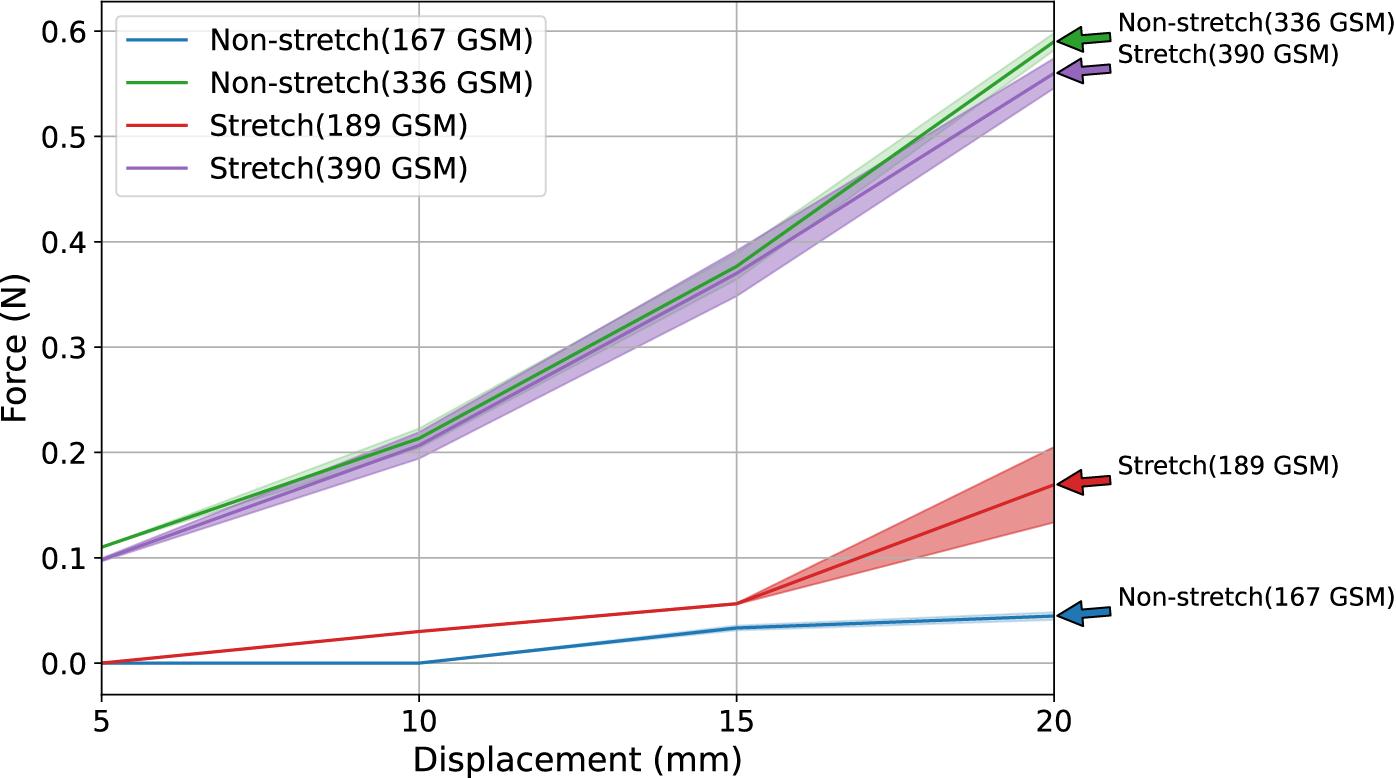}
        \caption{Stiffness Results}
        \label{fig:exp_fabric_test_stiffness}
    \end{subfigure}
    \hfill
    \begin{subfigure}[t]{0.38\linewidth}
        \centering
        \includegraphics[width=\textwidth]{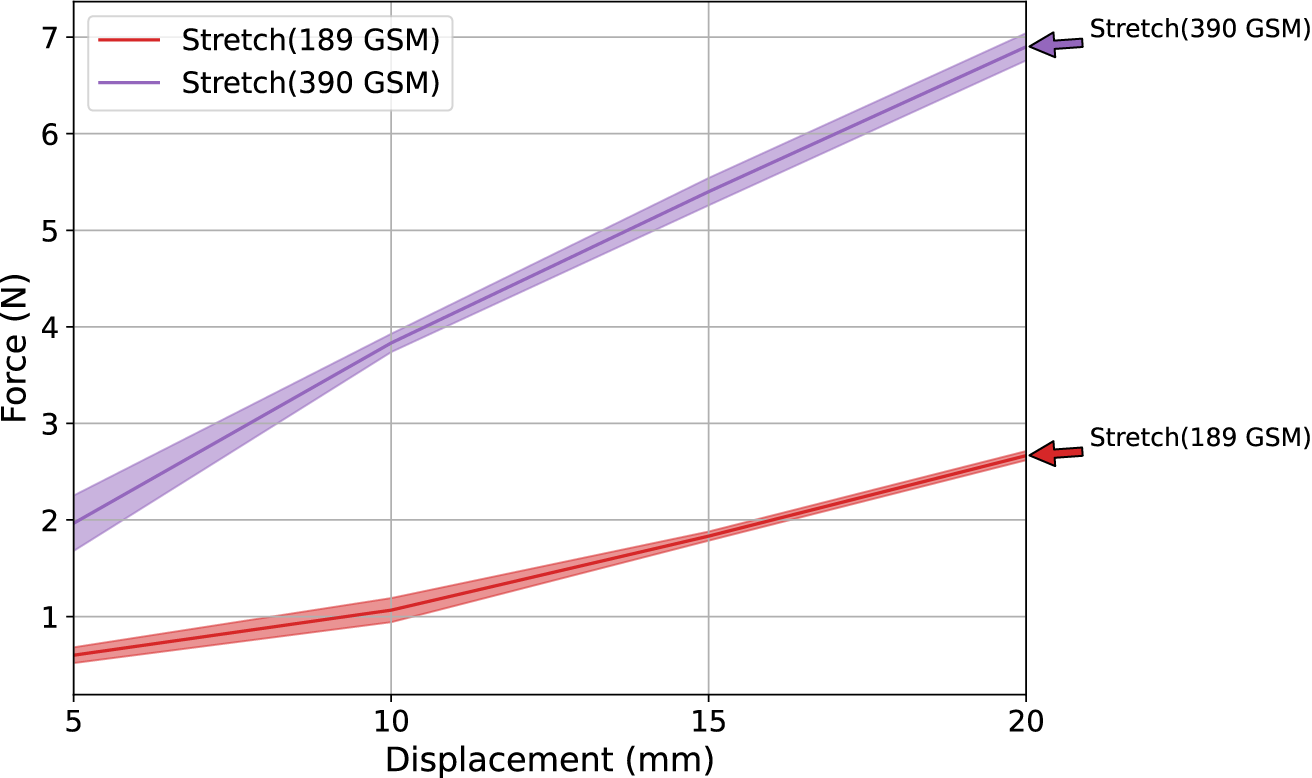}
        \caption{Stretchability Results}
        \label{fig:exp_fabric_test_stretchability}
    \end{subfigure}    
    \caption{Examination of the mechanical properties of four fabrics. (a) Test setups. (b) Stiffness test results for non-stretch 336 GSM, non-stretch 167 GSM, stretch 390 GSM, and stretch 189 GSM. (c) Stretchability test results for stretch fabrics: 390 GSM and 189 GSM.}
    \label{fig:fabric_type}
\end{figure}

We identified fabric weight and stretchability as two key parameters influencing the desired performance of ExoFabric. We hypothesize that fabric weight contributes to shape-holding capability, while stretchability provides conformability, allowing the formation of doubly-curved geometry . To test this hypothesis, we evaluated four fabric options: non-stretch 167 GSM, non-stretch 336 GSM, stretch 189 GSM, and stretch 390 GSM, using two test setups for each stiffness and stretchability test as shown in \Cref{fig:exp_fabric_test_setup}. The results indicated that heavier-weight fabrics offer better mechanical stability (See \Cref{fig:exp_fabric_test_stiffness}). Among the heavier fabrics, we selected the non-stretch 336 GSM fabric  (constructed with 98\% cotton, 2\% elastane, twill weaving) for applications requiring structural integrity and the stretchable 390 GSM fabric (constructed with 62\% rayon, 32\% nylon, 6\% spandex, knitting) for applications requiring conformability (See \Cref{fig:exp_fabric_test_stretchability}).

\subsection{Identifying the optimal embroidery configuration across stiffness, geometrical formability, and stretchability } 
\begin{figure}[htbp]
    \centering
    \begin{subfigure}[t]{0.28\linewidth}
        \centering
        \includegraphics[width=1\textwidth]{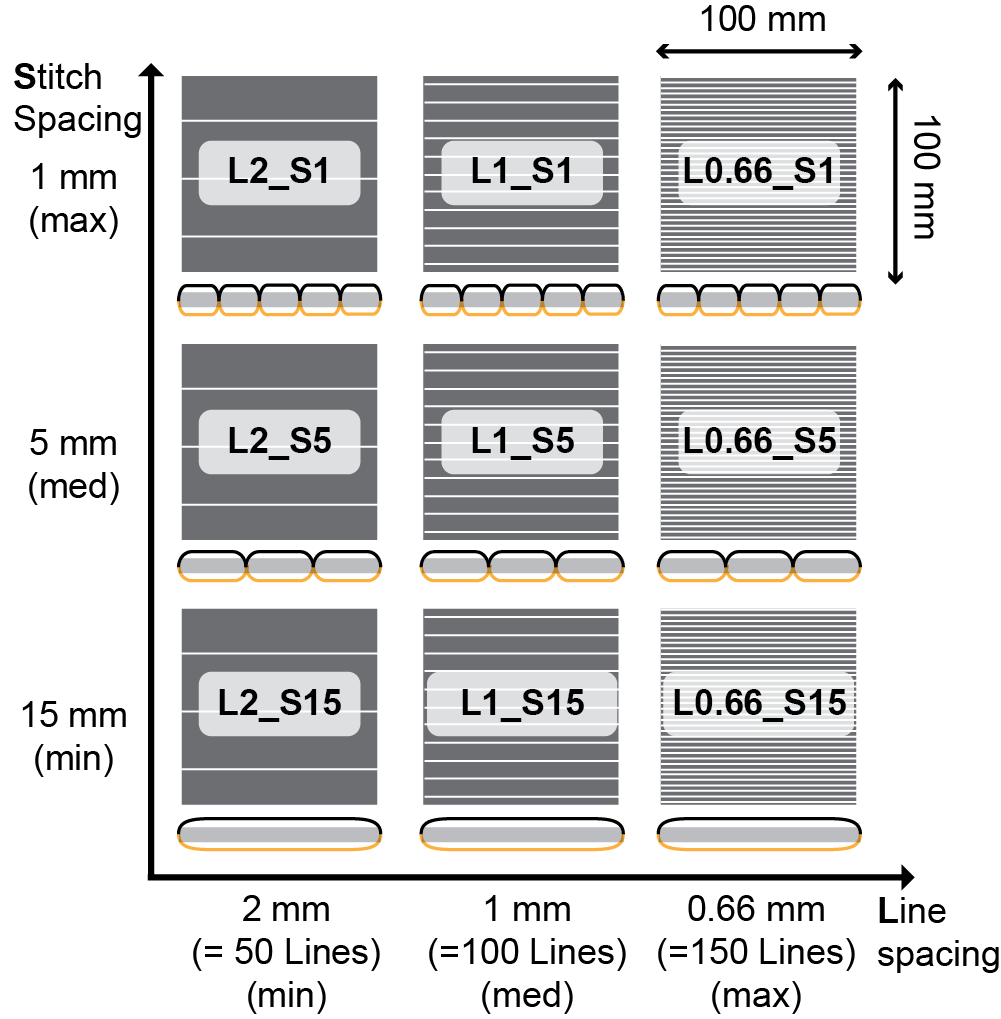}
        \caption{Embroidery Configurations}
        \label{fig:embroidery_config}
    \end{subfigure}
    \hfill
    \begin{subfigure}[t]{0.35\linewidth}
        \centering
        \includegraphics[width=\textwidth]{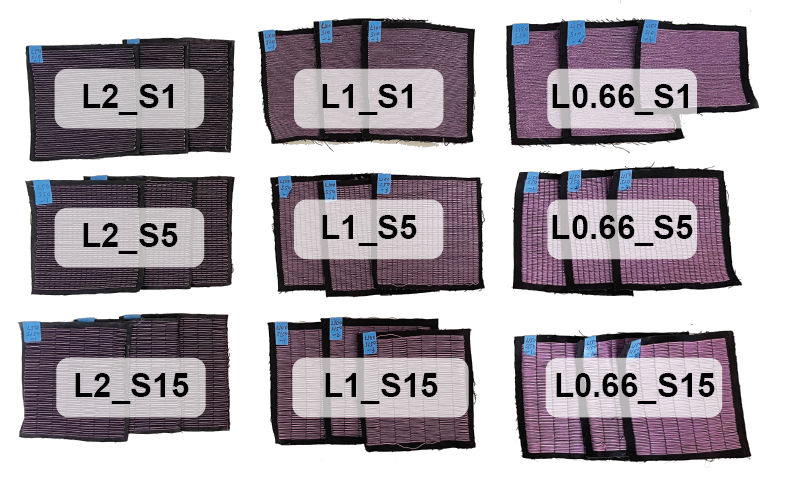}
        \caption{Non-Stretch Fabric Swatches}
        \label{fig:embroidery_config_non_stretch}
    \end{subfigure}
    \hfill
    \begin{subfigure}[t]{0.35\linewidth}
        \centering
        \includegraphics[width=\textwidth]{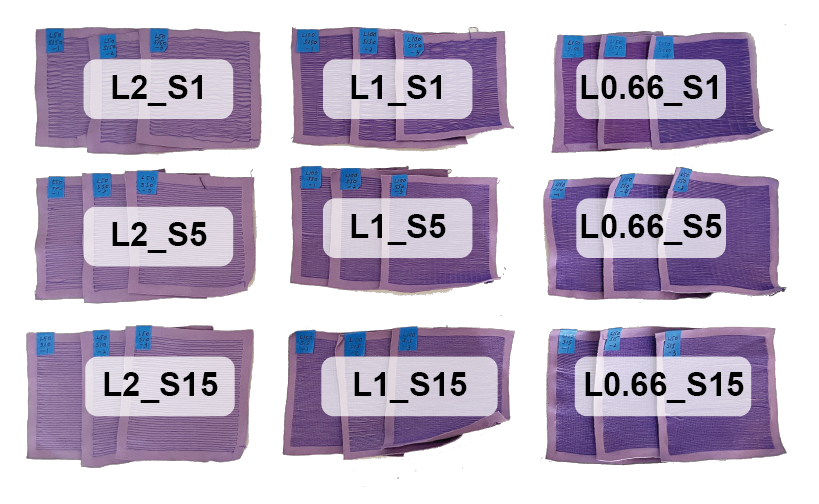}
        \caption{Stretch Fabric Swatches}
        \label{fig:embroidery_config_stretch}
    \end{subfigure}    
    \caption{Nine embroidery configurations with varying line spacing and stitch spacing parameters (a), along with fabricated swatches on non-stretch fabric (b) and stretch fabric (c).}
    \label{fig:embroidery_params}
\end{figure}

In this section, we conduct multivariate tests by selecting three configurations that best represent the minimum, medium, and maximum values for both line spacing and stitch spacing. This forms nine distinct embroidery configurations as shown in \Cref{fig:embroidery_params}. The line spacing (horizontal axis in \Cref{fig:embroidery_config}) ranges from 2 mm (50 lines within a 100 x 100 mm fabric), 1 mm (100 lines), to 0.66 mm (150 lines), representing the minimum, medium, and maximum densities of thermoplastic threads. The stitch spacing (vertical axis  in \Cref{fig:embroidery_config}) varies from 1 mm (maximum stitch density), 5 mm (medium), to 15 mm (minimum stitch density). Each combination of these parameters (e.g., L2\_S1, L1\_S5, L0.66\_S15) results in a unique configuration, which serves as the foundation for our multivariate tests. These tests aim to assess the effects of three variables---line spacing, stitch spacing, and fabric type (non-stretch and stretch)---on the stiffness, geometrical formability, and stretchability of ExoFabric. Each configuration was fabricated using two types of fabric, with three swatches per fabric type, to ensure statistical confidence in subsequent technical experiments (\Cref{fig:embroidery_config_non_stretch,fig:embroidery_config_stretch}).
By analyzing these effect, we identify the optimal embroidery configurations that balance stiffness and stretchability for applications like comfortable wear. The subsequent experiments includes (1) stiffness via compression tests, (2) geometrical formability via molded shape assessments, and (3) stretchability via tensile tests.

\begin{figure}[htbp]
    \centering
    \begin{subfigure}[t]{0.55\linewidth}
        \centering
        \includegraphics[width=\textwidth]{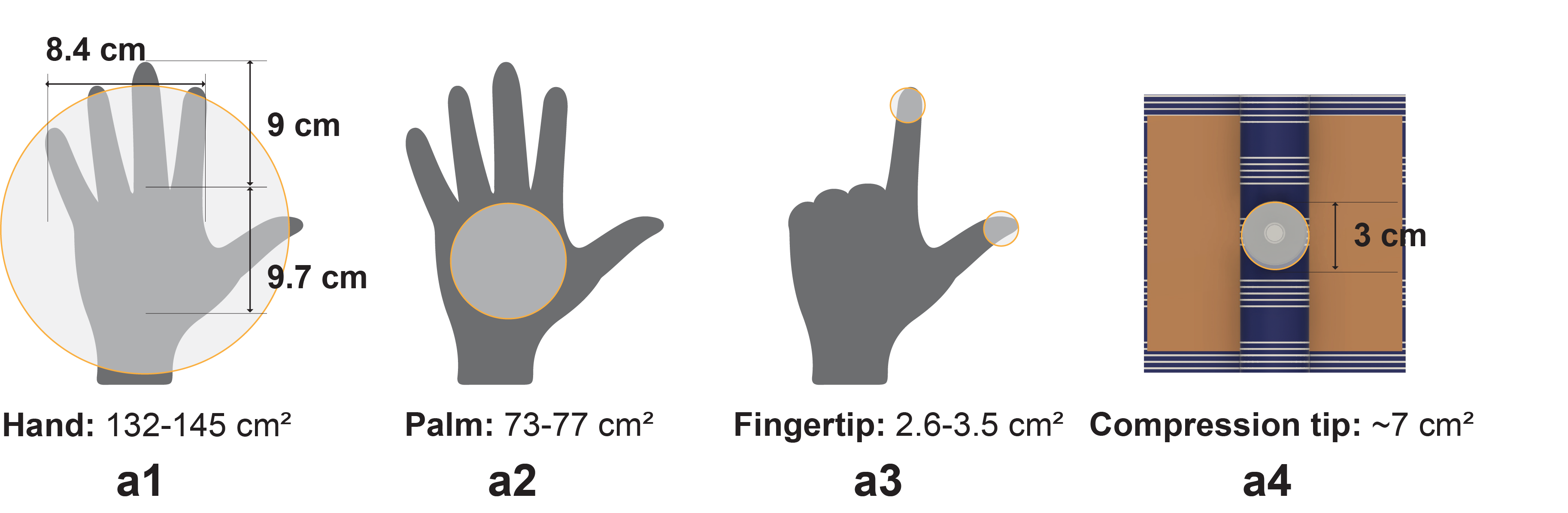}
        \caption{Tip Sizes}
        \label{fig:compression_test_tip_size}
    \end{subfigure}
    \hfill
    \begin{subfigure}[t]{0.22\linewidth}
        \centering
        \includegraphics[width=\textwidth]{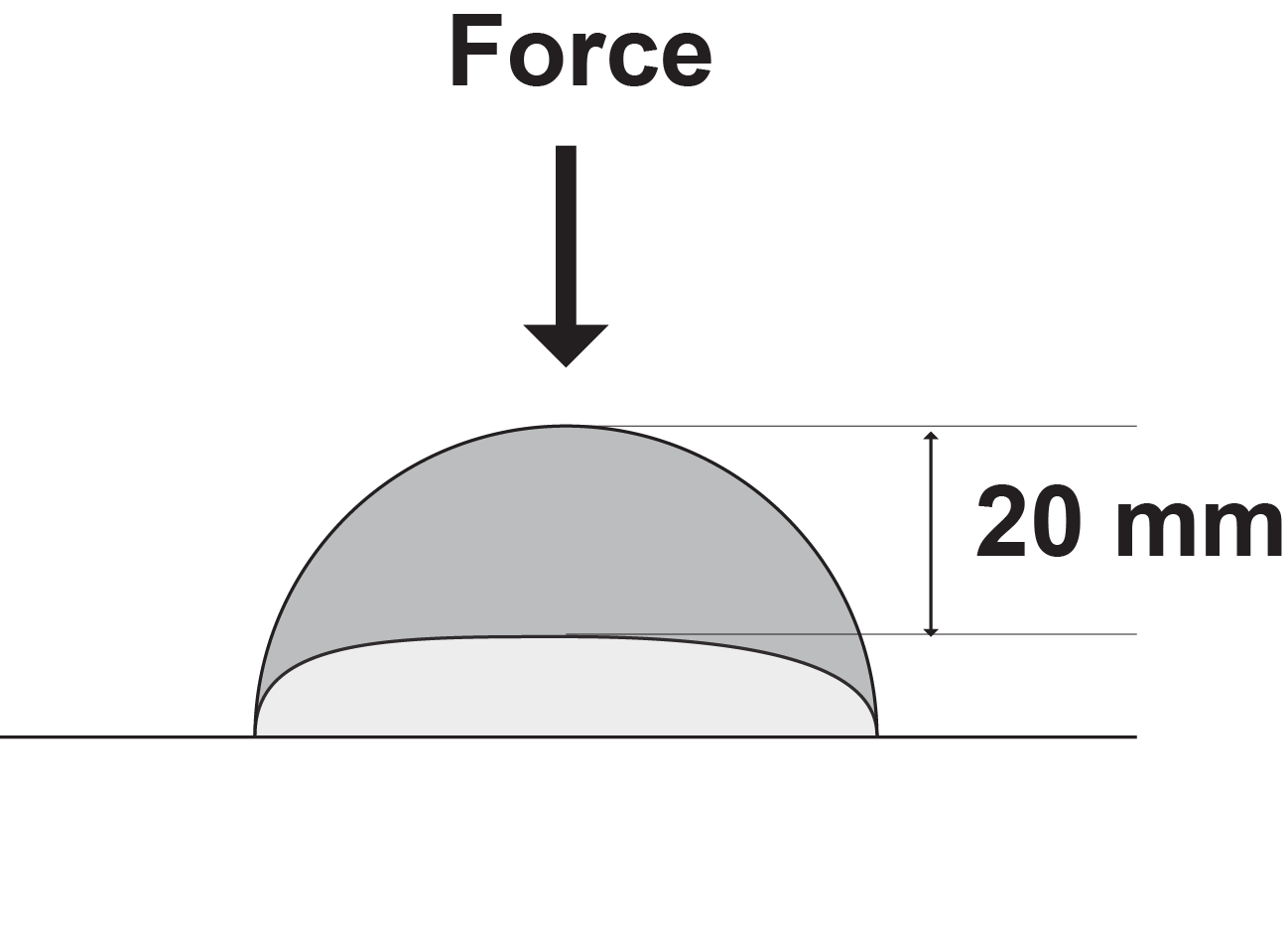}
        \caption{Deformation Range}
        \label{fig:compression_test_deform_range}
    \end{subfigure}
    \hfill
    \begin{subfigure}[t]{0.20\linewidth}
        \centering
        \includegraphics[width=\textwidth]{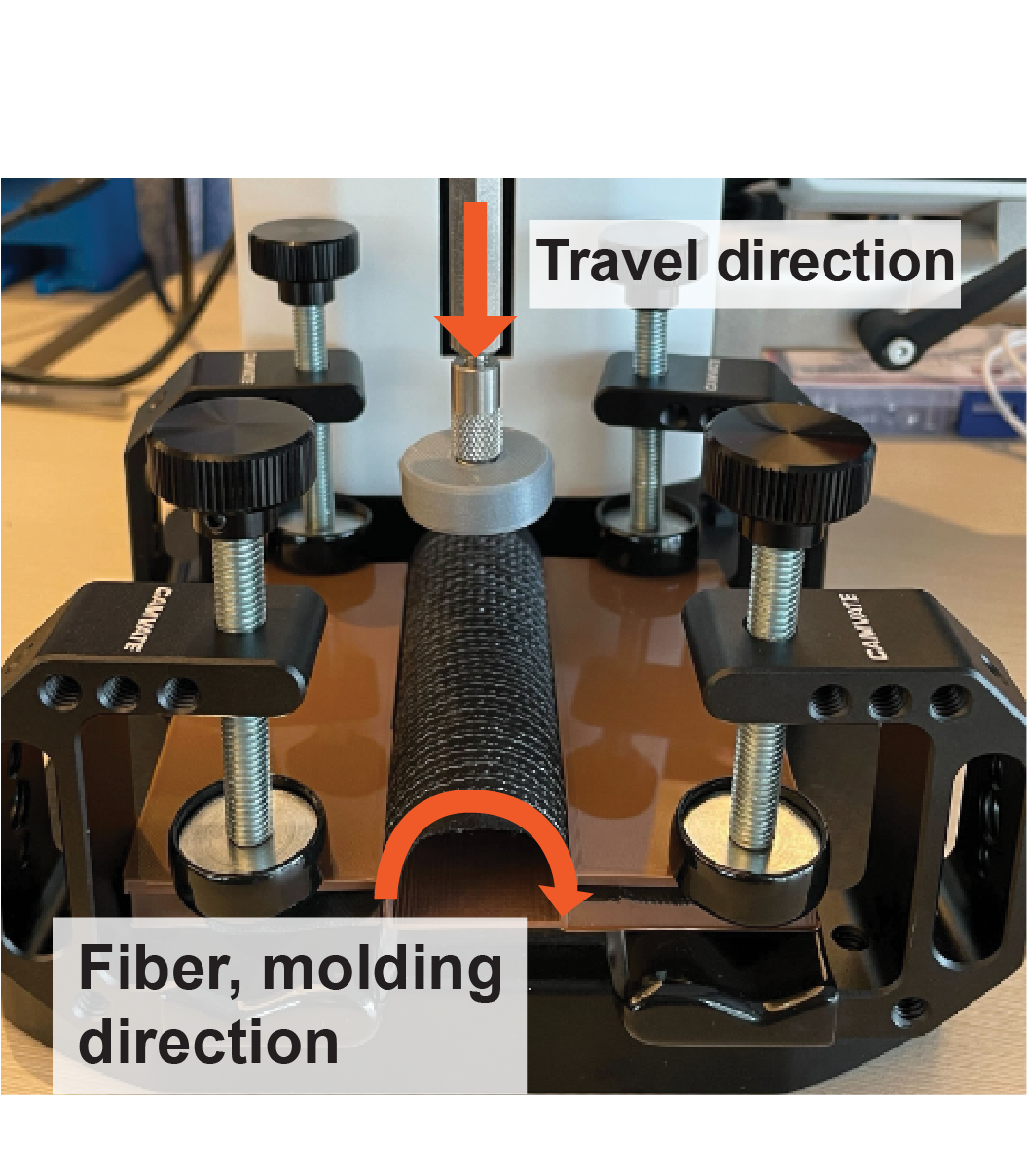}
        \caption{Actual Setup}
        \label{fig:compression_test_actual_setup}
    \end{subfigure}
    \begin{subfigure}[t]{1\linewidth}
        \centering
        \includegraphics[width=0.95\textwidth]{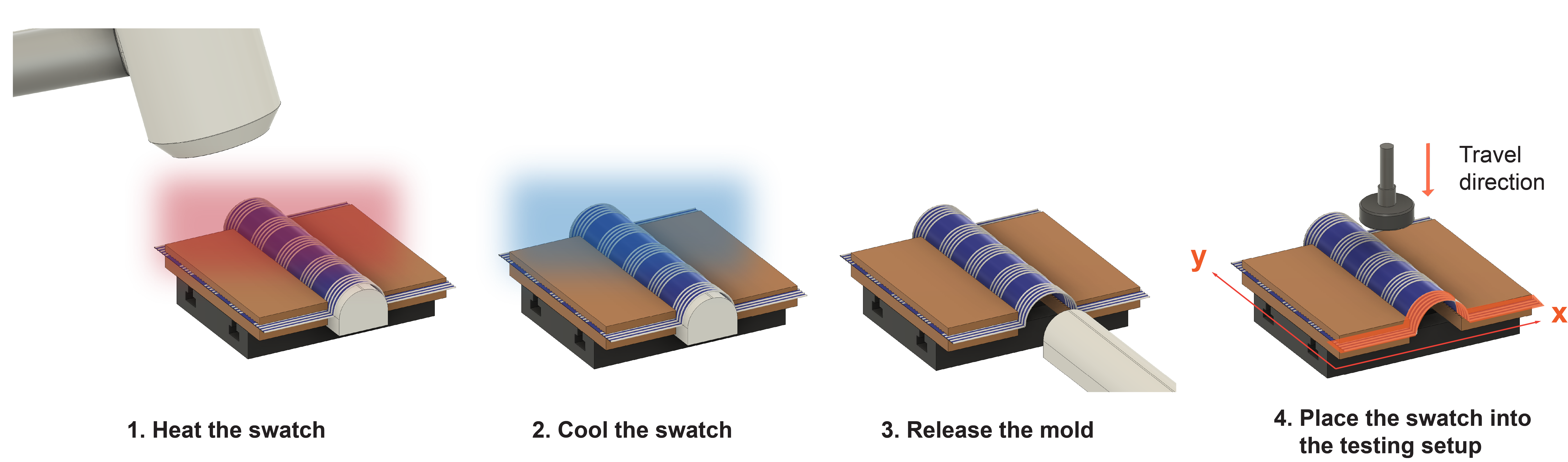}
        \caption{Molding Process}
        \label{fig:compression_test_molding}
    \end{subfigure}  
    \caption{Compression test setup, showing (a) compression tip, (b) predetermined travel distance, (c) actual test setup, and (d) molding process.}
    \label{fig:compression_test}
\end{figure}

\subsubsection{Stiffness Evaluation via Compression Testing}
\label{sec:stiffness_eval_via_compress_testing}
We performed compression tests on the nine embroidery configurations across two fabric types to identify the optimal ExoFabric combinations that can maintain their desired shape under applied load. The test setup included a compression tip, a fixture to secure the swatch, a predetermined travel distance, and the test swatches (See \Cref{fig:compression_test_actual_setup}). The dimensions of the hand, palm, and fingertips were used as guidelines for setting the compression tip’s contact area as shown in \Cref{fig:compression_test_tip_size}. Each swatch was clamped onto a 30 mm cylindrical mold within the fixture, aligning the thermoplastic threads along the bending curve to maximize stiffness. The swatch was heated with a heat gun for 10 second, reaching 70$^{\circ}$C, then cooled for 20 seconds until the temperature dropped to 22$^{\circ}$C. After cooling, the cylindrical mold was removed, and the swatch was placed in the testing setup (see \Cref{fig:compression_test_molding} for an overview of the molding process). For each swatch, force data was recorded 10 times using an HP-500 digital force gauge at 5 mm intervals, measured with an electronic micrometer caliper, until the total travel distance reached 20 mm.

\begin{figure}[htbp]
    \centering
    \begin{subfigure}[t]{0.48\linewidth}
        \centering
        \includegraphics[width=\textwidth]{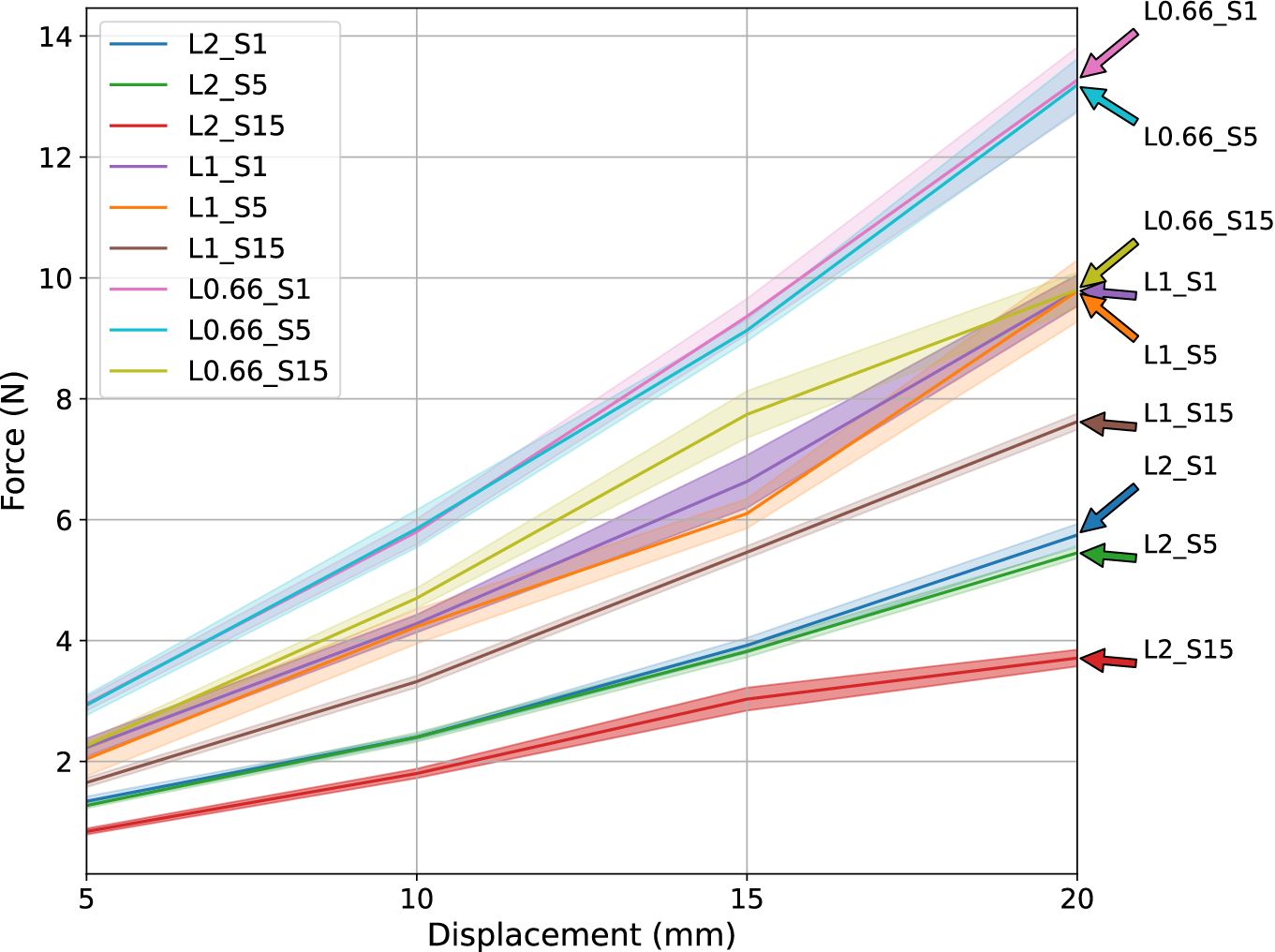}
        \caption{Non-Stretch Fabric}
        \label{fig:comp_test_1_layer_non_strech}
    \end{subfigure}
    \hfill
    \begin{subfigure}[t]{0.48\linewidth}
        \centering
        \includegraphics[width=\textwidth]{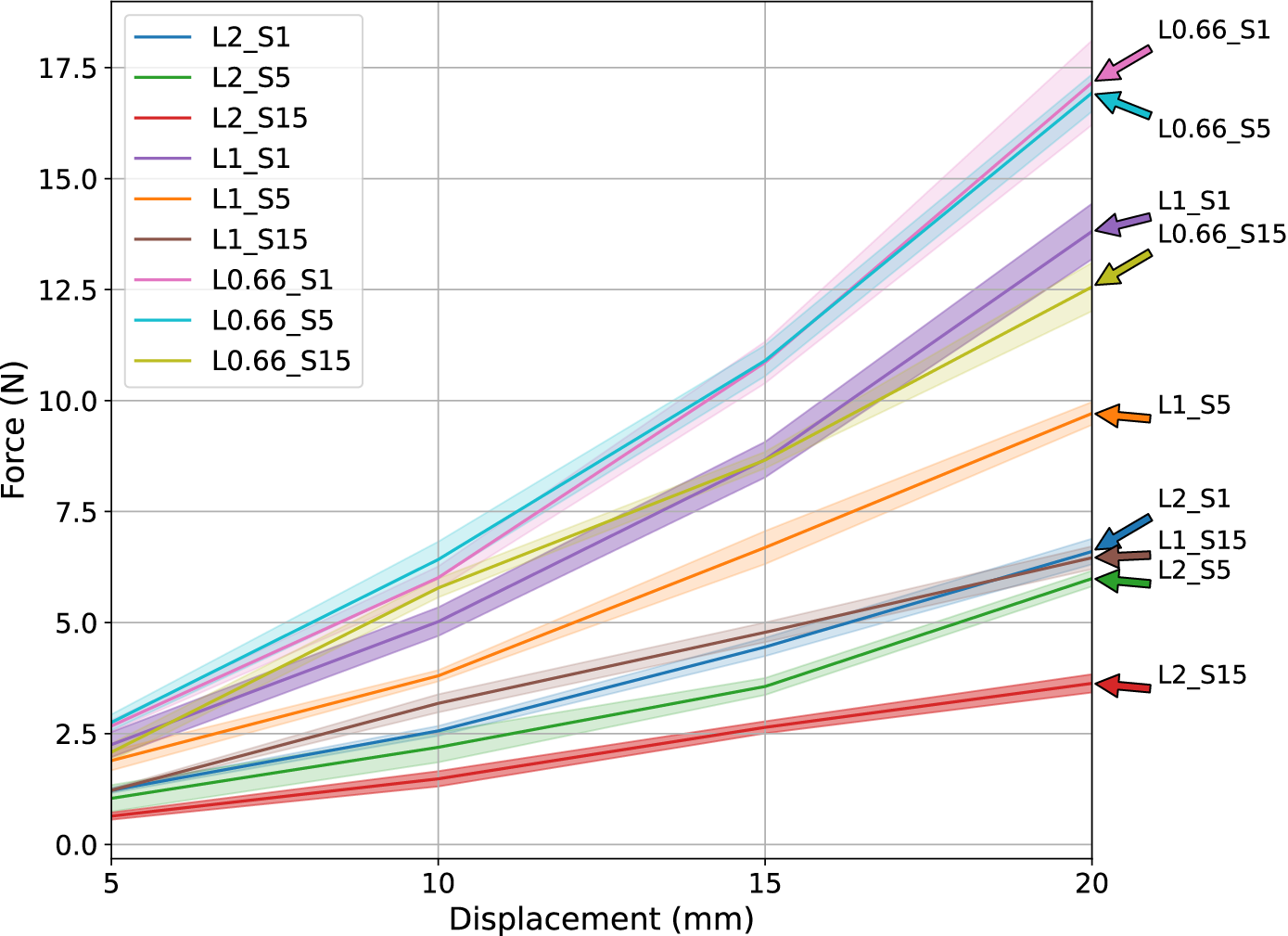}
        \caption{Stretch Fabric}
        \label{fig:comp_test_1_layer_strech}
    \end{subfigure}    
    \caption{Compression test results for three line spacing (L2, L1, L0.66) and three stitch spacings (1 mm, 5 mm, 15 mm) in (a) non-stretch fabric and (b) stretch fabric. Note that the transparent areas represent the standard deviation in force measurements.}
    \label{fig:comp_test_result_1_layer}
\end{figure}

As shown in \Cref{fig:comp_test_1_layer_non_strech}, the results demonstrated that for non-stretch fabrics, an decrease in line spacing combined with a decrease in stitch spacing leads to a significant increase in stiffness, which aligns with our initial hypothesis. The L0.66\_S1 (i.e., line spacing 0.66 mm, stitch spacing 1 mm) configuration exhibited the highest stiffness, while the L2\_S15 configuration showed the lowest stiffness. Some configurations demonstrate a combined effect between line spacing and stitch spacing; for example, swatches L0.66\_S15 and L1\_S1 exhibited similar stiffness. We observed that swatches can exhibit similar stiffness despite variations in stitch spacing. For example, configurations L2\_S1 and L2\_S5 demonstrated comparable stiffness across the displacement range, suggesting that increasing stitch spacing does not significantly impact stiffness when the line spacing is at a higher range. However, the effect of stitch spacing becomes more pronounced at smaller line spacing, as seen with L0.66\_S5 and L0.66\_S15, where reducing the stitch spacing leads to greater stiffness. These results indicate that the relationship between stitch spacing and stiffness is more sensitive at lower line spacing in non-stretch fabrics. In terms of line spacing, we observed a notable rise in the force required to achieve the same displacement as the line spacing decreased. For instance, for L2\_S5, L1\_S5, and L0.66\_S5, the average force needed to compress the swatches at 10 mm was 2.4 N, 4.2 N, and 5.9 N, respectively. This indicates that lower line spacing lead to greater stiffness, as the increased material density results in more resistance to compression.

In the case of stretch fabric, the results generally followed a similar trend to non-stretch fabric, but required more force to achieve the same displacement in higher stiffness configurations, as shown in \Cref{fig:comp_test_1_layer_strech}. For instance, in the highest stiffness configuration (L0.66\_S1), stretch fabric required an average of 17.2 N to compress at 20 mm, compared to 13.3 N for non-stretch fabric. The results suggest that for applications requiring high stiffness, a lower line spacing combined with smaller stitch spacing is optimal, with additional considerations for stretch fabrics due to their increased resistance. Other factors such as fabrication efficiency, fabric quality, and aesthetics should also be considered as key decision points when selecting the optimal configuration. For example, the fabrication time for L0.66\_S1 is about 20 minutes while L2\_S15 is about 5 minutes. Denser stitch spacing and line spacing can cause damage to the fabric due to it being over punched. These elements play a crucial role in balancing performance with practical implementation and overall design appeal.

\begin{figure}[htbp]
    \centering
    \includegraphics[width=0.7\linewidth]{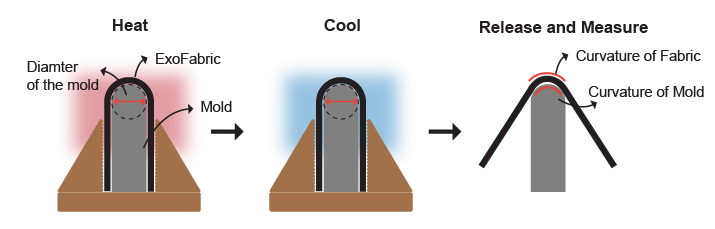}
    \caption{Molding workflow for evaluating geometrical formability.}
    \label{fig:formability_setup_workflow}
\end{figure}

\subsubsection{Geometrical formability evaluation}
Formability, or the ability to conform to the shape of intended molds, is another critical factor in our evaluation as it facilitates the molding of ExoFabric into the desired shape. We assessed formability by observing the shape discrepancy between the molded fabric and the mold, and by evaluating the durability of the fabric's deformation over time without external force. Each swatch was fixed tightly on a cylindrical mold with diameter of 30 mm, 20 mm, and 10 mm, heated for 10 seconds, cooled for 20 seconds, released from the mold and placed in room temperature for an 48 hours to ensure the durability of the formed samples (See \Cref{fig:formability_setup_workflow}). We then photographed the result and conducted visual inspections for the shape differences between the mold and molded swatches. We also compared the formed results across nine configurations and two fabric types.

\begin{figure}[htbp]
    \centering
    \includegraphics[width=1\linewidth]{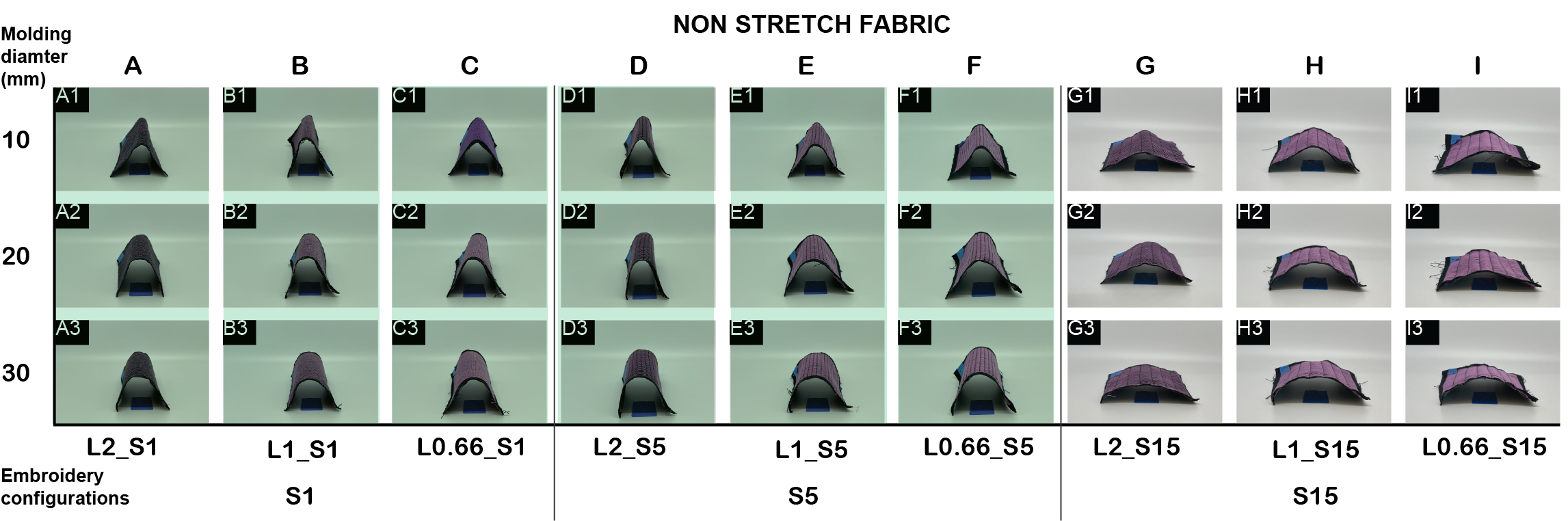}
    \caption{Geometrical formability test results for non-stretch fabric across nine embroidery configurations}
    \label{fig:formability_test_results_ns}
\end{figure}

With the non-stretch fabric, better-shaped swatches correlated with denser stitch spacing, such as S1 and S5 (\Cref{fig:formability_test_results_ns} A-F), regardless of their line spacing. Conversely, swatches with sparser stitch spacing, like S15 (\Cref{fig:formability_test_results_ns} G-I), failed to maintain shape effectively. Our hypothesis is that large stitch spacing, with longer thermoplastic threads (rod) and less stitch points (joint), may constraint the overall flexibility of the linkage structure presented in thread form that is necessary to form a proper curvature. We compared formability and stiffness outcomes across nine embroidery configurations and two fabric types. Of the top six best-formed swatches (\Cref{fig:formability_test_results_ns} A-F), four were among the stiffest (\Cref{fig:formability_test_results_ns} B, C, E, F ), while the other two were on the softer end. This variation suggests that the formability is primarily influenced by stitch spacing as swatches with larger stitch spacing (e.g., S15) did not form as well as those with smaller stitch spacing (e.g., S5 and S1). In contrast, the stiffness is determined by the combined effect of line spacing and stitch spacing. Based on the combined results of formability and stiffness, we recommend using denser stitch spacing, such as S1 and S5, with higher line spacing at L2 (\Cref{fig:formability_test_results_ns} A, D) for applications where formability is the primary focus, as this combination also enhances fabrication efficiency. For applications requiring both high stiffness and improved formability, we suggest decreasing line spacing to L1 or L0.66 while maintaining stitch spacing at S1 or S5 (\Cref{fig:formability_test_results_ns} B, C, E, F).

\begin{figure}[htbp]
    \centering
    \includegraphics[width=1\linewidth]{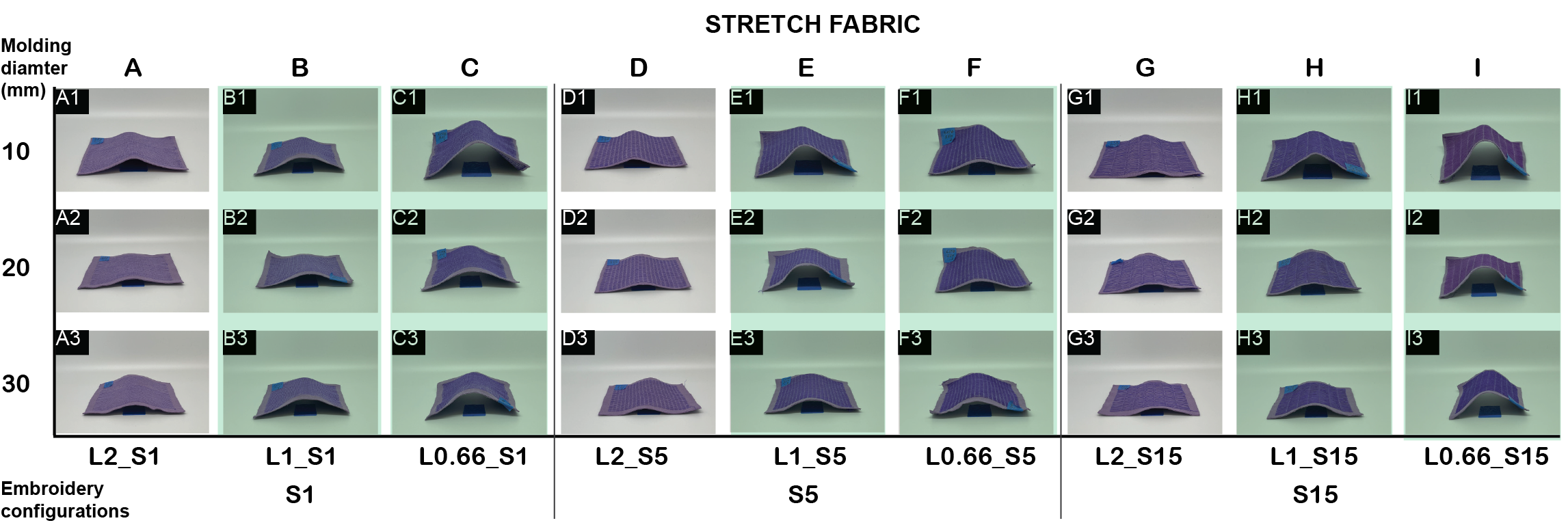}
    \caption{Geometrical formability test results for stretch fabric accross nine embroidery configurations}
    \label{fig:formability_test_results_s}
\end{figure}

In stretch fabrics, a contrasting pattern emerged compared to non-stretch fabric: better-shaped swatches were associated with smaller line spacings, such as L1 and L0.66 (\Cref{fig:formability_test_results_s} B, C, E, F, H, I), regardless of stitch spacing variations. Conversely, swatches with higher line spacing, such as L2 (\Cref{fig:formability_test_results_s} A, D, G), struggled to maintain shape. This indicates that line spacing is the primary factor influencing formability likely because stretch fabric has less structural integrity, and lower line spacing help to add rigidity. We also examined the correlation between formability and stiffness. Among the top six best-formed configurations, five were also among the stiffest (\Cref{fig:formability_test_results_s} B, C, E, F, I). For applications that focus on formability alone, we recommend using sparser stitch spacing, like S15, with smaller line spacings, such as L1 or L0.66 (\Cref{fig:formability_test_results_s} H, I), to achieve good formability while maintaining fabrication efficiency. For applications that require both formability and stiffness, we suggest adjusting the stitch spacing to S1 or S5 to enhance stiffness (\Cref{fig:formability_test_results_s} B, C, E, F). 

\begin{figure}[htbp]
    \centering
    \begin{subfigure}[t]{0.3\linewidth}
        \centering
        \includegraphics[width=0.9\textwidth]{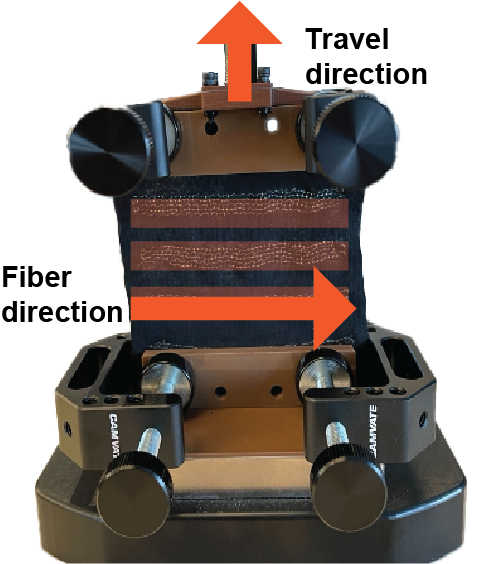}
        \caption{Tensile Test Setup}
        \label{fig:tensile_test_1_layer_test_setup}
    \end{subfigure}
    \hfill
    \begin{subfigure}[t]{0.68\linewidth}
        \centering
        \includegraphics[width=0.9\textwidth]{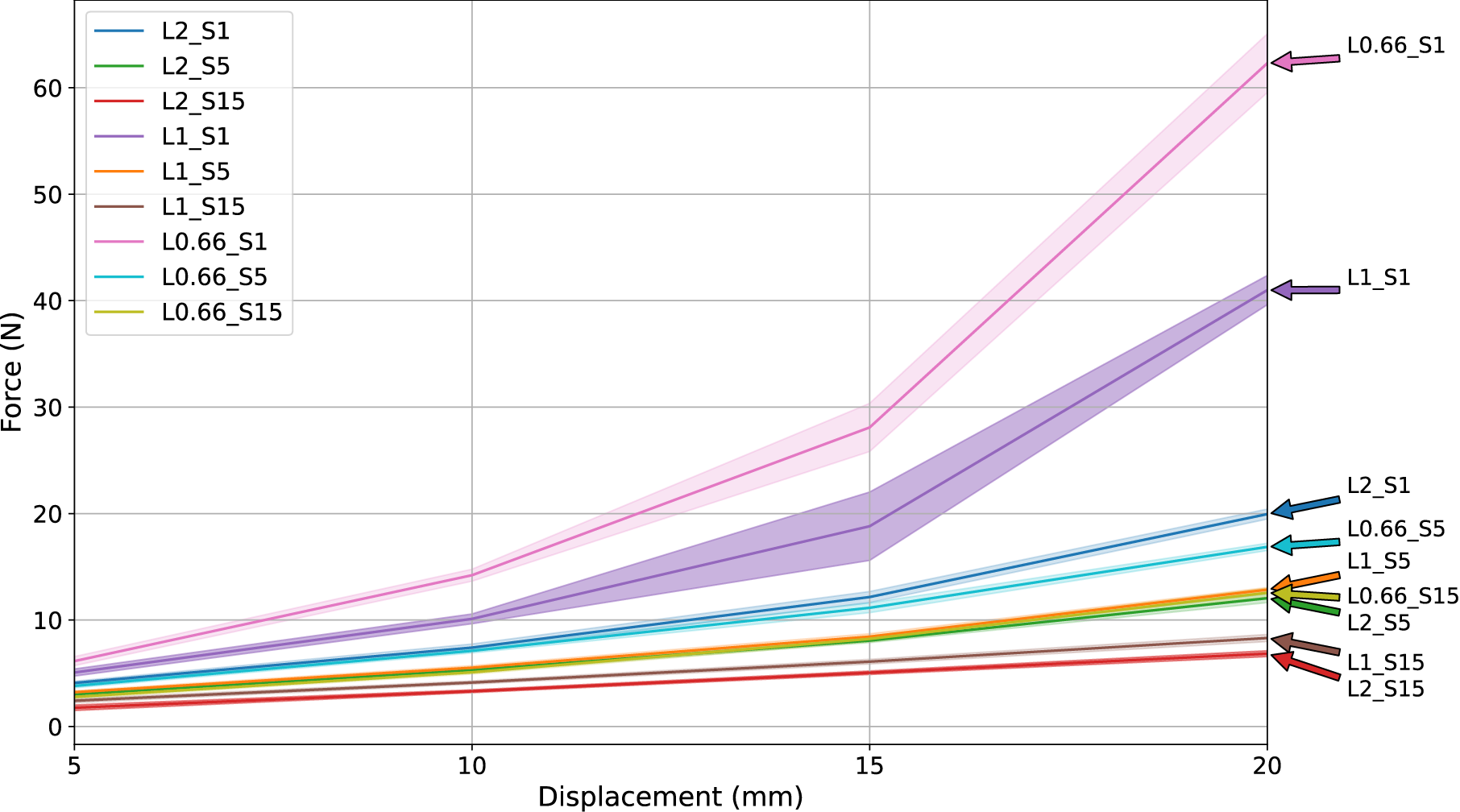}
        \caption{Test Results}
        \label{fig:tensile_test_1_layer_strech}
    \end{subfigure}    
    \caption{Tensile test results for three line spacings (L2, L1, L0.66) and three stitch spacings (1 mm, 5 mm, 15 mm) using (a) the test setup in (b) stretch fabric. Note that the fiber direction is perpendicular to the tensile direction in the test setup to maximize stretchability.}
    \label{fig:tensile_test_result_1_layer}
\end{figure}

\subsubsection{Stretchability evaluation via tensile testing}
In the stretchability evaluation, we tested only swatches made from stretchable fabrics to determine which embroidery patterns provide less or more stretchability while also enhancing stiffness. In the test setup, we oriented the tensile direction perpendicular to the fiber layout to maximize the stretchable area and prevent premature exhaustion of the swatches (\Cref{fig:tensile_test_1_layer_test_setup}).

\Cref{fig:tensile_test_1_layer_strech} demonstrates that both line spacing and stitch spacing significantly impact the stretchability. Lower line spacings (L1, L0.66) result in less stretchability, as evidenced by the higher forces required for the same displacement. For example, L0.66\_S1 exhibits the lowest stretchability, requiring an average force of 62.3N of force at 20 mm displacement, while L1\_S1 follows with an average of 41.0 N. The effect of stitch spacing, though less pronounced than line spacing, also plays a role. Smaller stitch spacing (S1) generally lead to lower  stretchability compared to larger stitch spacing (S15). For instance, L0.66\_S1 is less stretchable than L0.66\_S15. Larger line spacings (L2) combined with larger stitch spacing (S15) produce the most flexible swatches, with L2\_S15 requiring the least force for displacement, remaining below 7.0 N at 20 mm. Overall, the results suggest that both line spacing and stitch spacing work together to influence the stretchability of the fabric. Larger line spacings combined with larger stitch spacing results in the most flexible configurations, while smaller line spacings and smaller stitch spacing leading to reduced stretchability and stiffer swatches.

\subsection{Layering ExoFabric} 
Based on the evaluation of single-layer samples, we selected the embroidery configuration that best performs in the stiffness and formability tests for both non-stretch and stretch fabrics (i.e., L0.66\_S1). We then investigated multi-layer configurations as an additional parameter to assess how adding layers could further mechanical strength, particularly when embroidery parameters (i.e., line spacing, stitch spacing) reach their limits within the given fabric size.

\begin{figure}[htbp]
    \centering
    \begin{subfigure}[t]{0.33\linewidth}
        \centering
        \includegraphics[width=\textwidth]{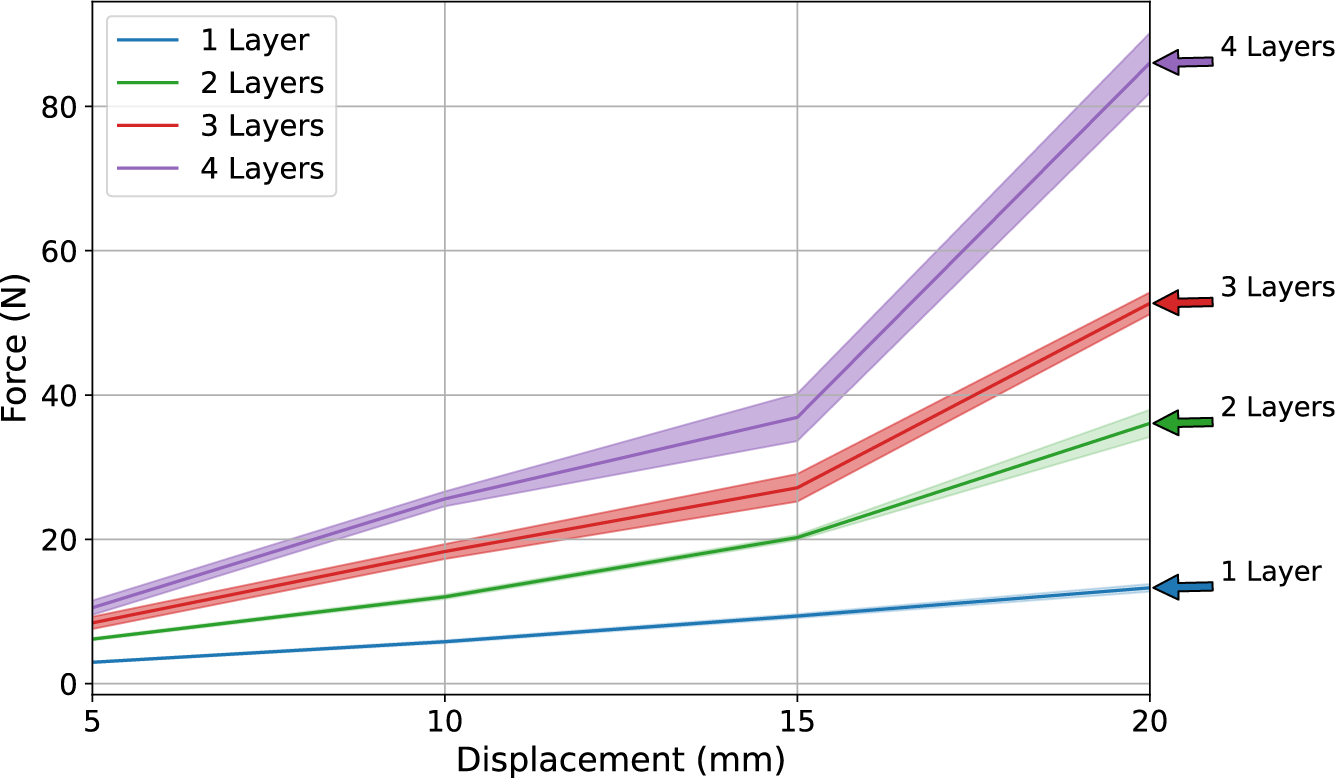}
        \caption{Stiffness Results for Non-Stretch Fabric}
        \label{fig:multilayer_comp_ns_result}
    \end{subfigure} 
    \hfill
    \begin{subfigure}[t]{0.33\linewidth}
        \centering
        \includegraphics[width=\textwidth]{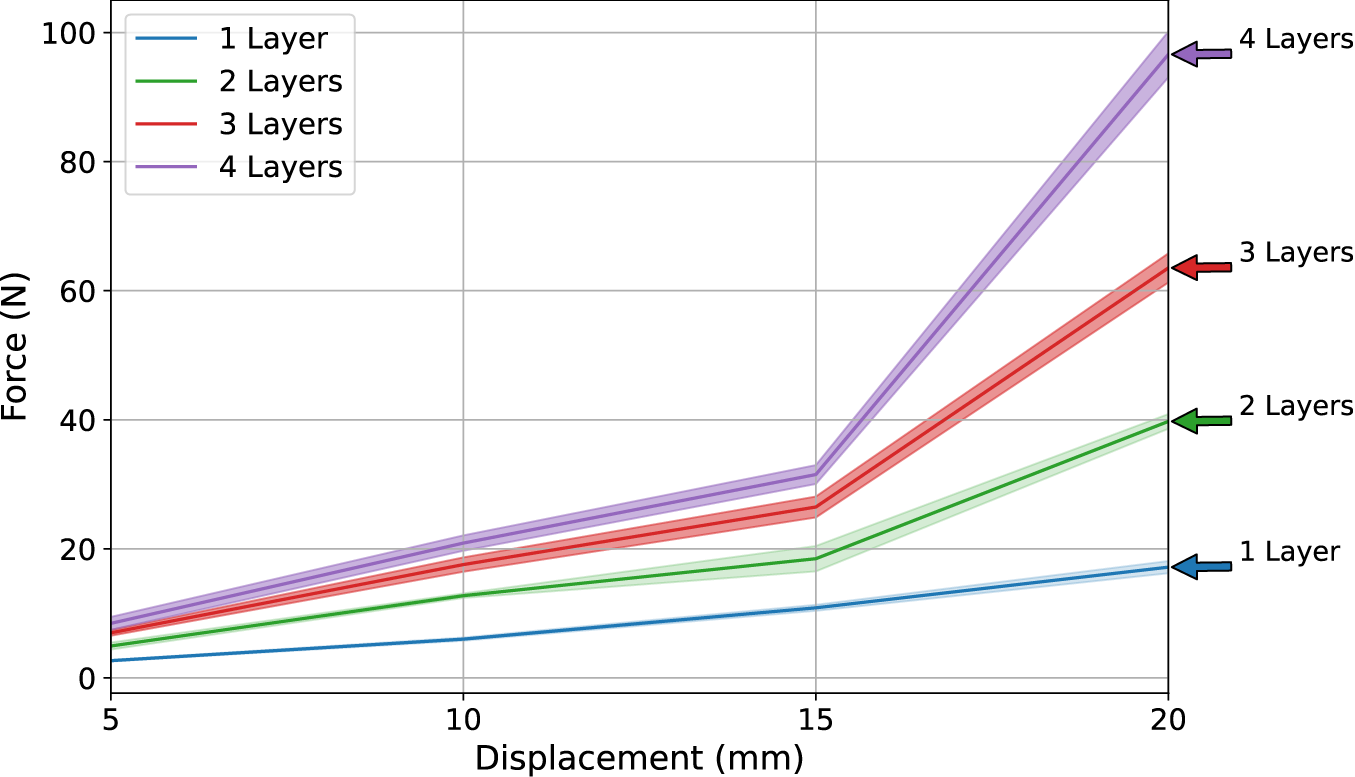}
        \caption{Stiffness Results for Stretch Fabric}
        \label{fig:multilayer_com_s_result}
    \end{subfigure} 
    \hfill
    \begin{subfigure}[t]{0.33\linewidth}
        \centering
        \includegraphics[width=\textwidth]{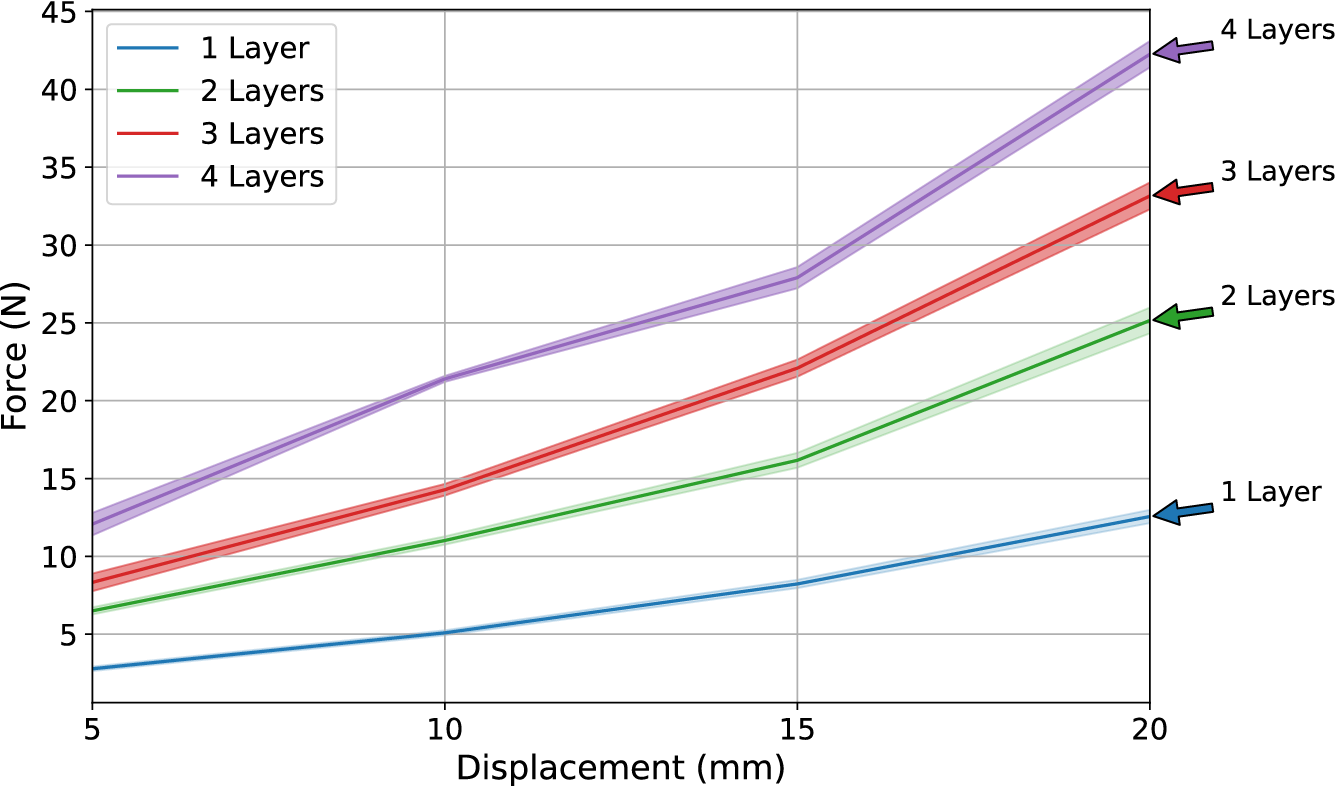}
        \caption{Tensile Results for Stretch Fabric}
        \label{fig:multilayer_tens_s_result}
    \end{subfigure} 
    \hfill
    \hfill
    \hfill
    \begin{subfigure}[t]{0.49\linewidth}
        \centering
        \includegraphics[width=\textwidth]{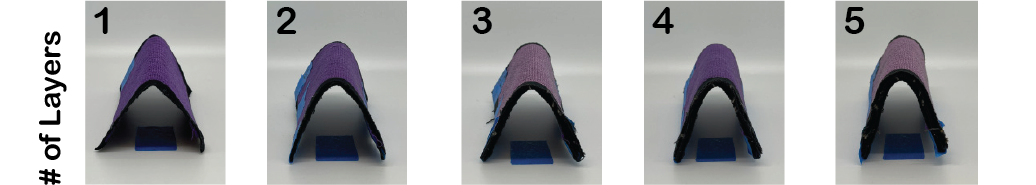}
        \caption{Results of Formability Tests on Non-Stretch Fabric}
        \label{fig:multilayer_ns_vis_result}
    \end{subfigure}
    \hfill
    \begin{subfigure}[t]{0.49\linewidth}
        \centering
        \includegraphics[width=\textwidth]{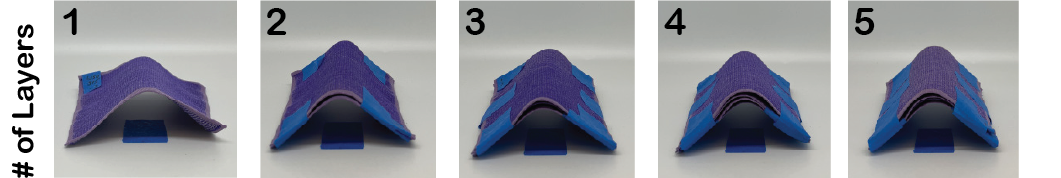}
        \caption{Results of Formability Tests on Stretch Fabric}
        \label{fig:multilayer_s_vis_result}
    \end{subfigure}
    \caption{Multi-layering effects on ExoFabric. (a) Stiffness results for non-stretch fabric and (b) stretch fabric during compression tests using the L0.66\_S1 configuration. (c) Tensile results for stretch fabric using the L0.66\_S15 configuration. (d) Formability test results on non-stretch fabric and (e) stretch fabric. }
    \label{fig:multilayer_effect}
\end{figure}

In the stiffness evaluation, for both non-stretch and stretch fabrics, stiffness significantly increases as the number of layers increases (\Cref{fig:multilayer_comp_ns_result,fig:multilayer_com_s_result}). For example, with non-stretch fabric, the 4-layer swatch requires an average of 86.0 N of force at 20 mm displacement, compared to an average of 13.3 N for a single layer. Similarly, with stretch fabric, the 4-layer swatch requires an average of 96.6 N of force at 20 mm displacement, compared to an average of 17.2 N for a single layer. Overall, the increase in stiffness is proportional to the number of layers added. The non-linear changes observed at around 15 mm displacement are likely due to the concave shape formed when the tip contacts the fabric. This geometric change induces some non-linear behavior, which may be mitigated if the fabric size is sufficiently large.

In the formability evaluation, we observed similar outcomes to the single-layer tests, with enhanced stability for both non-stretch and stretch fabrics. However, the multi-layer configurations did not improve the precision of the shape conformed to the mold due to increased stiffness, which limited the fabric’s ability to fully conform to complex contours (\Cref{fig:multilayer_ns_vis_result,fig:multilayer_s_vis_result}). 

The tensile test results for the L0.66\_S15 configuration on stretch fabric highlight the significant impact of multi-layering on stretchability (\Cref{fig:multilayer_tens_s_result}). As additional layers are introduced, the force required to achieve a given displacement increases, indicating a clear trade-off between stretchability and stiffness. As layers are added, the force required to reach the same displacement grows, with four layers requiring more than three times the force of a single layer. This near-linear increase in force required to stretch the fabric suggests that multi-layering is an effective strategy for enhancing the mechanical strength. However, it is also evident that increased stiffness comes at the cost of reduced flexibility and stretchability, which may not be ideal for applications that require greater freedom of movement or adaptability. 

Overall, these evaluations help us create design guidelines for a wide range of application requirements based on their stiffness, formability, and stretchability by selecting the optimal combinations of fiber, fabric, embroidery configuration, and layering. For example, for wearable applications that prioritize comfort, we recommend using stretch fabrics and choosing embroidery configurations that maximize formability and stretchability. For structural applications, such as protective wear, we suggest prioritizing stiffness to ensure the material can withstand higher external loads effectively. This balance between flexibility and mechanical strength is crucial in tailoring the fabric’s performance to meet the needs of diverse use cases.

\section{Applications}
With the mechanical and geometrical properties of ExoFabric building blocks being thoroughly characterized, along with their corresponding fabrication parameters, we can leverage these insights to design ExoFabric-based applications. By adjusting key parameters such as line spacing, stitch spacing, and layering, we showcased ExoFabric’s adaptability to a variety of use cases. In this section, we present three applications—an adjustable finger splint, a customizable bra, and a shape-changeable lampshade that highlight ExoFabric’s capability to provide structural support, comfort, and resilience in both wearable technologies and everyday objects. These examples illustrate how the dynamic tunability of ExoFabric’s stiffness, geometrical formability, and stretchability can address real-world challenges.

\begin{figure}[htbp]
    \centering
    \begin{subfigure}[t]{0.19\linewidth}
        \centering
        \includegraphics[width=\linewidth]{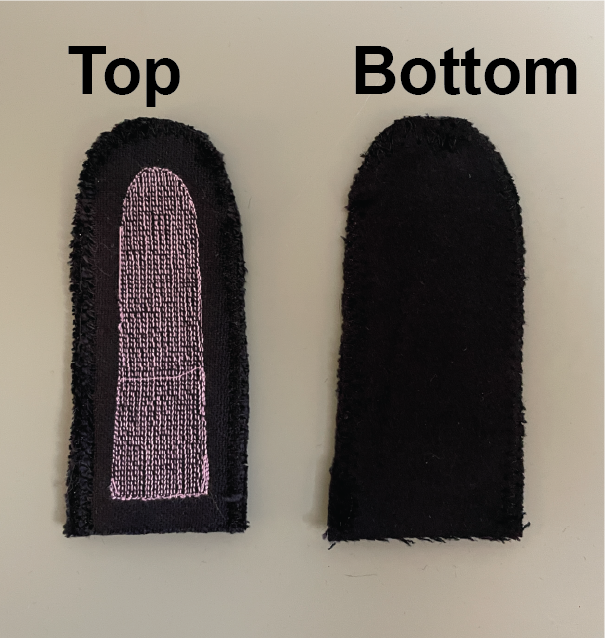}
        \caption{Top and Bottom View}
        \label{fig:finger_splint_top_bottom}
    \end{subfigure}
    \hfill
    \begin{subfigure}[t]{0.19\linewidth}
        \centering
        \includegraphics[width=\linewidth]{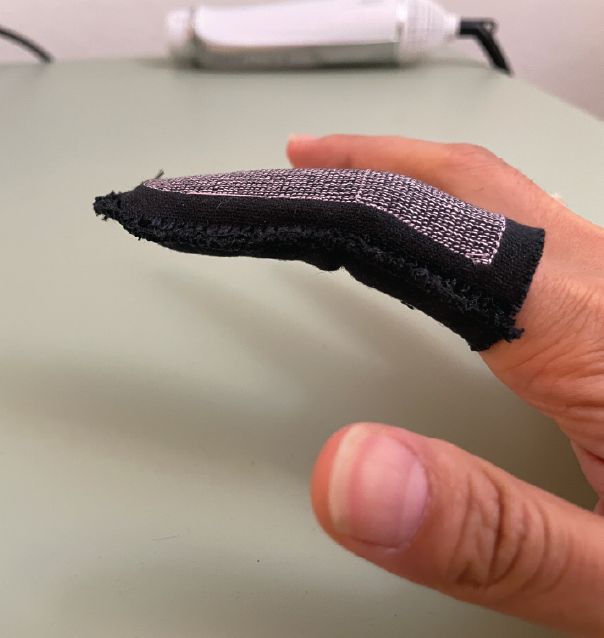}
        \caption{2 Layers of ExoFabric}
        \label{fig:finger_splint_2_layers}
    \end{subfigure}
    \hfill
    \begin{subfigure}[t]{0.19\linewidth}
        \centering
        \includegraphics[width=\linewidth]{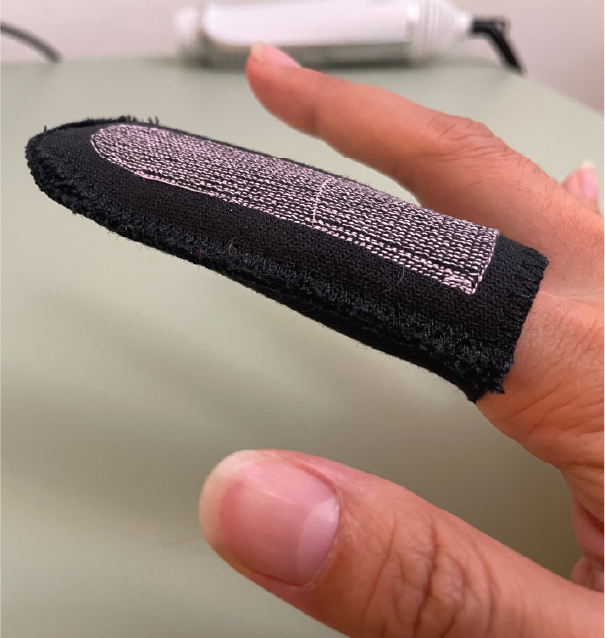}
        \caption{4 Layers of ExoFabric}
        \label{fig:finger_splint_4_layers}
    \end{subfigure}
    \hfill
    \begin{subfigure}[t]{0.19\linewidth}
        \centering
        \includegraphics[width=\linewidth]{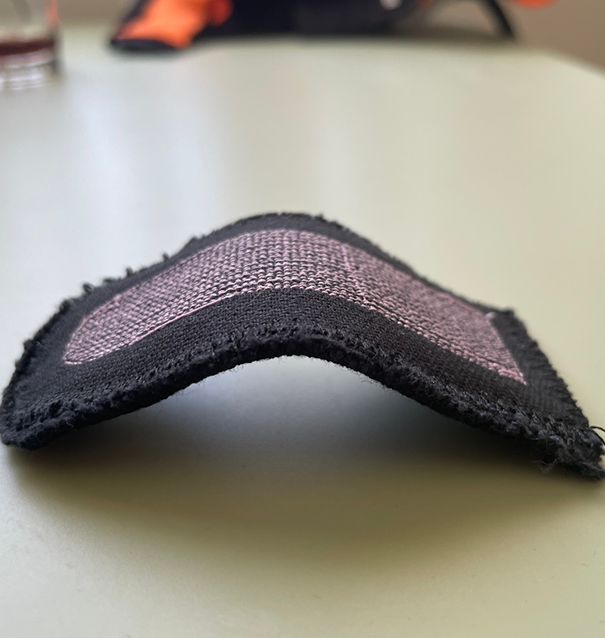}
        \caption{Molding Result}
        \label{fig:finger_splint_molded}
    \end{subfigure}
    \hfill
    \begin{subfigure}[t]{0.19\linewidth}
        \centering
        \includegraphics[width=\linewidth]{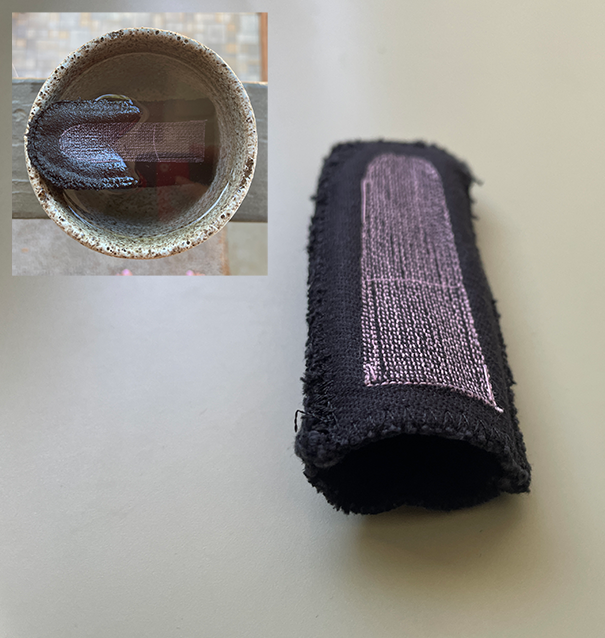}
        \caption{Washability Result}
        \label{fig:finger_splint_wash_result}
    \end{subfigure}
    \caption{ExoFabric-based finger splint. (a) Top and bottom view of the splint. (b) The top layer consists of two layers of the L0.66\_S1 configuration ExoFabric, with the bottom made of pure non-stretch fabric. (c) The top layer is composed of four layers of ExoFabric. (d) The molding test demonstrated that the 4-layer configuration can be fixed at a desired angle using the molding method described in \Cref{sec:stiffness_eval_via_compress_testing}. (e) After hand washing and complete drying, the proposed splint maintained its shape and performance.}
    \label{fig:finger_splint}
\end{figure}

\subsection{Adjustable Finger Splint}
\label{sec:app_finger_splint}
Stack or Zimmer finger splints have traditionally been used to stabilize injured or weakened fingers, but these devices present several drawbacks in terms of materials and customization methods \cite{teng_adjustable_2024}. These splints are typically worn for extended periods, raising hygiene concerns due to the difficulty in cleaning and drying them. Additionally, adjusting these splints as treatment progresses is challenging, often leading to discomfort and improper fit, which can result in treatment failure rates as high as 50\% \cite{teng_adjustable_2024}. Researchers have explored 3D-printed splints as alternatives \cite{zolfagharian_patient-specific_2024}, but ExoFabric-based splints could also offer a promising solution to address these issues. 

To investigate its potential, we applied ExoFabric to create a customized splint. We selected the stiffest parameters, L0.66\_S1, to ensure sufficient rigidity for restricting movement and providing protection. Furthermore, we used non-stretch fabric and applied the chosen embroidery configuration to fabricate an ExoFabric splint designed to fit a typical finger as shown in \Cref{fig:finger_splint_top_bottom}. Note that the thermoplastic embedded fabric was located on the top side of the finger while the pure non-stretch fabric is placed on the back side of the finger to provide the maximum restricting force. The fiber orientation followed the finger’s bending direction to ensure it withstands forces from bent finger parts. The force specifications for a commercial finger splint typically requires it to withstand 13.7 N when flexed at 45$^{\circ}$C or more, and 6.4 N at full extension \cite{noauthor_spring_nodate}. We initially selected the stiffness configuration from \Cref{sec:stiffness_eval_via_compress_testing}; however, the data (as shown in \Cref{fig:comp_test_1_layer_non_strech}) is based on 100 x 100 mm fabric, meaning the results may not be directly applicable to different fabric sizes. 

To meet these force requirements, we tested a single layer of ExoFabric and progressively increased the number of layers up to four. The four-layer configuration successfully met the force requirements, withstanding 7.8 N at 5 mm displacement when placed flat in our bending test setup, while three layers withstood 6.3 N, and two layers withstood 2.6 N at 5 mm displacement, respectively. Additionally, we conducted a heuristic evaluation with a member of the research team (female, height 163 cm, weight 53 kg). Similar to the force-displacement measurements, she reported difficulty moving her index finger with the four-layer splint, while the two-layer splint was easier to bend (\Cref{fig:finger_splint_2_layers,fig:finger_splint_4_layers}). For washability, we hand-washed and fully dried the 4-layer splint, then retested the force-displacement and wearability.  The 4-layer splint withstood a force of 7.5 N, and no noticeable changes were observed in performance or fit, as shown in \Cref{fig:finger_splint_wash_result}.

\begin{figure}[htbp]
    \centering
    \begin{subfigure}[t]{0.42\linewidth}
        \centering
        \includegraphics[width=\linewidth]{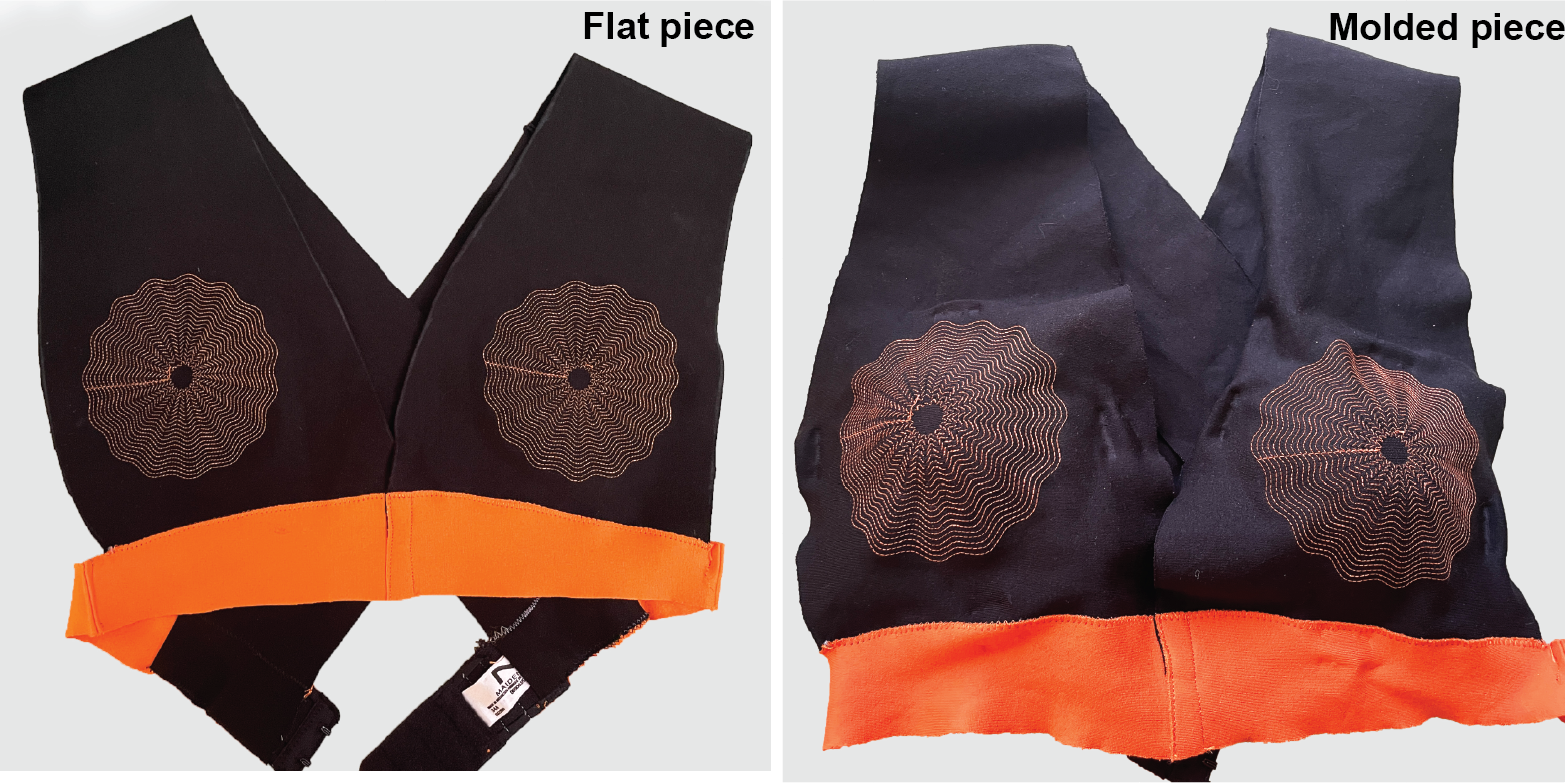}
        \caption{Flat and Molded Status}
        \label{fig:bra_flat_molded}
    \end{subfigure}
    \hfill
    \begin{subfigure}[t]{0.27\linewidth}
        \centering
        \includegraphics[width=\linewidth]{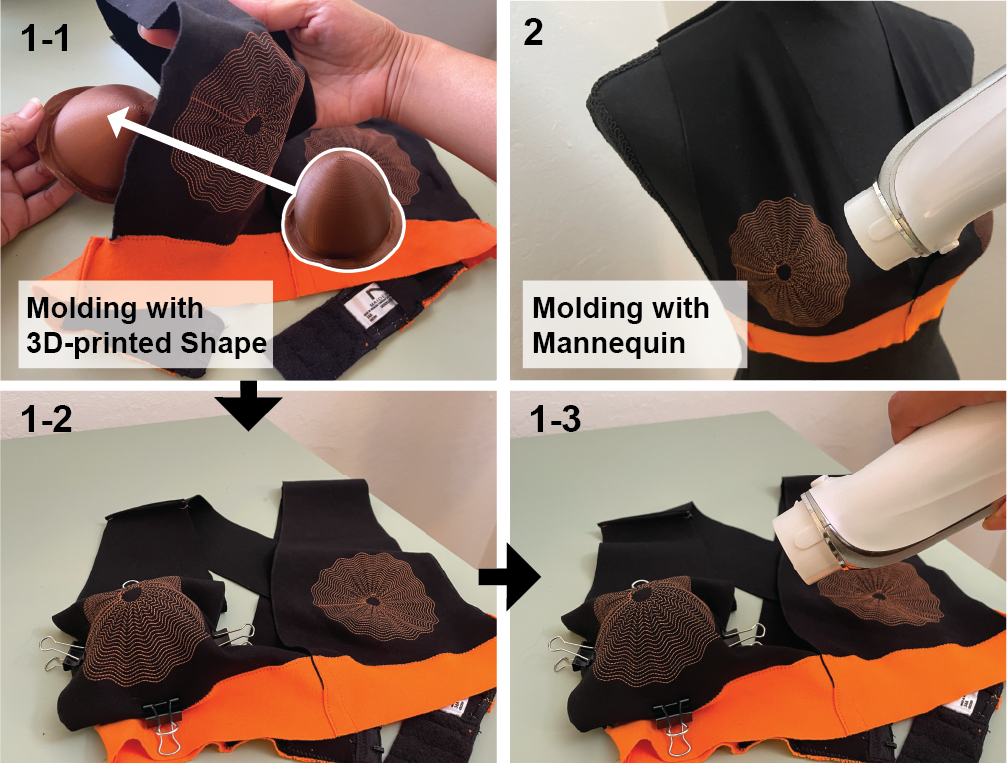}
        \caption{Molding Process}
        \label{fig:bra_molding_process}
    \end{subfigure}
    \hfill
    \begin{subfigure}[t]{0.27\linewidth}
        \centering
        \includegraphics[width=\linewidth]{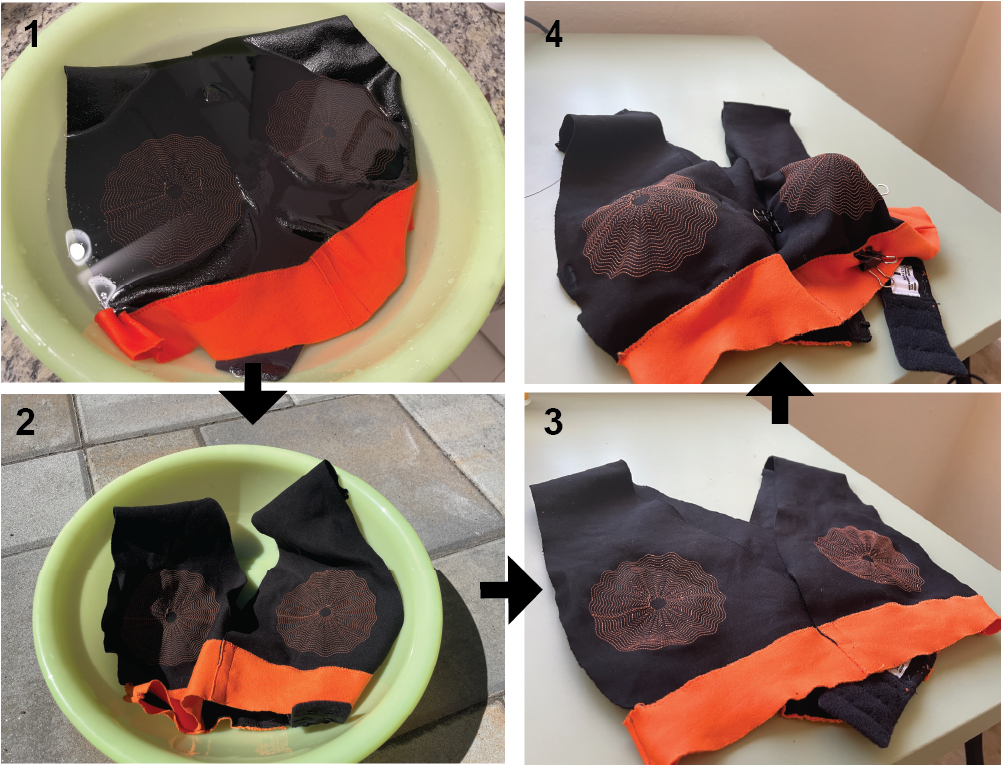}
        \caption{Washability Test }
        \label{fig:bra_washability}
    \end{subfigure}
    \caption{ExoFabric-based customizable bra. (a) Initial flat form and final outcome after molding. (b) Two different molding methods for the bra application: Method 1 uses a 3D-printed mold, while Method 2 uses a mannequin. (c) The process and outcome of the washability test, demonstrating that the bra maintained its shape after washing.}
    \label{fig:exofabric_bra}
\end{figure}

\subsection{Customizable Bra}
Secondly, we explored the potential of ExoFabric for creating customizable bras. Research shows that many women wear incorrectly sized bras due to the limitations of current sizing systems, which often results in discomfort \cite{baidak_custom_2023}. This is particularly important for adolescent girls, as bras need to adjust as they grow to provide proper chest support and posture correction \cite{luh_girls_2020}. Additional considerations, such as asymmetrical breasts or aesthetic preferences, also lead women to seek customizable options \cite{lazaro_vasquez_auto-adjustable_2019}. ExoFabric offers a promising foundation for such applications due to its tunable mechanical properties, including customizable stiffness, controlled stretchability, and precise geometrical formability. 

Similar to the design process in \Cref{sec:app_finger_splint}, we selected parameters that best demonstrate formability and stretchability to allow for conformability by choosing stretch fabric with concentric thread layouts (\Cref{fig:Thread direction}d) to facilitate the formation of a dome shape in \Cref{fig:bra_flat_molded}. The activation process starts from a flat bra, followed by molding and heating to form the intended shape as shown in \Cref{fig:bra_molding_process}. \Cref{fig:bra_flat_molded}-1 presents the results of molding using a pre-made mold, while \Cref{fig:bra_flat_molded}-2 shows the outcome of molding directly on a body silhouette.  

We conducted the washability test to validate the prototype can perform the intended function after soaking in water and air drying for a few hours as \Cref{fig:bra_washability}-1\&2. After washing and drying, we were able to re-mold the bra from its flat state (\Cref{fig:bra_washability}-3) back to the intended shape (\Cref{fig:bra_washability}-4). While bras do not typically require high force resistance, we examined the force-displacement behavior to ensure that the ExoFabric-based bra could maintain its shape in typical use scenarios. The compression test recorded forces of 0.6 N at 10 mm, 0.9 N at 15 mm, and 1.8 N at 19 mm. According to a prior study on bra-breast forces \cite{mcghee_brabreast_2013}, 1.8 N represents the minimum force the bra should support, indicating that this application meets the requirement. For larger or heavier breasts, additional force support would be necessary, which can be achieved by minimizing line spacing, reducing stitch spacing, and/or adding more layers in the ExoFabric design to enhance structural support.

\begin{figure}[htbp]
    \centering
    \begin{subfigure}[t]{0.16\linewidth}
        \centering
        \includegraphics[width=\linewidth]{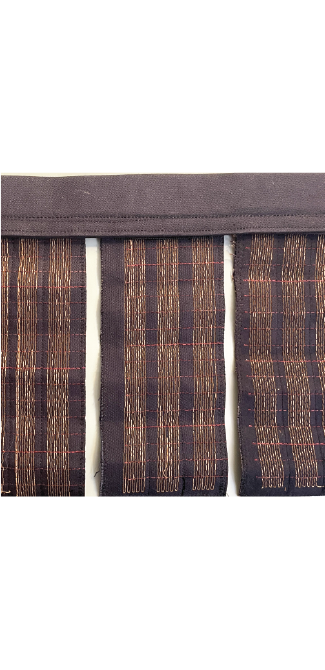}
        \caption{Base Shape}
        \label{fig:lampshape_base}
    \end{subfigure}
    \hfill
    \begin{subfigure}[t]{0.30\linewidth}
        \centering
        \includegraphics[width=\linewidth]{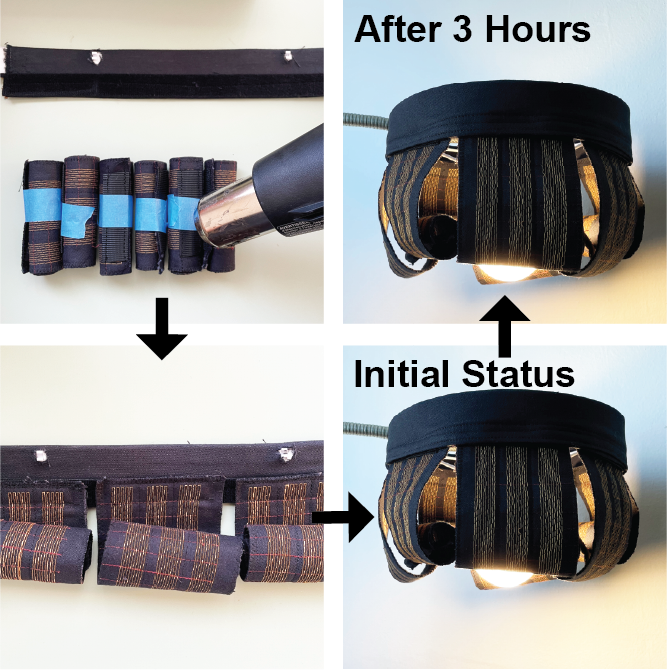}
        \caption{Design \#1}
        \label{fig:lampshape_design_01}
    \end{subfigure}
    \hfill
    \begin{subfigure}[t]{0.30\linewidth}
        \centering
        \includegraphics[width=\linewidth]{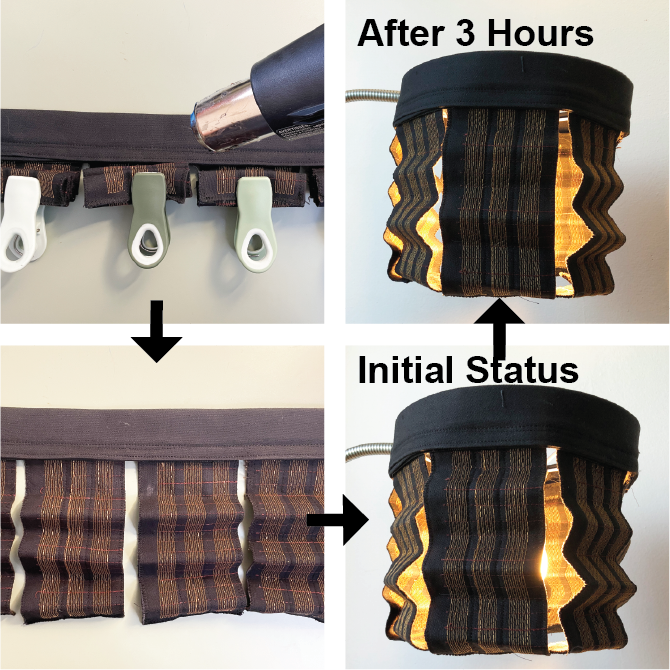}
        \caption{Design \#2}
        \label{fig:lampshape_design_02}
    \end{subfigure}
    \begin{subfigure}[t]{0.15\linewidth}
        \centering
        \includegraphics[width=0.95\linewidth]{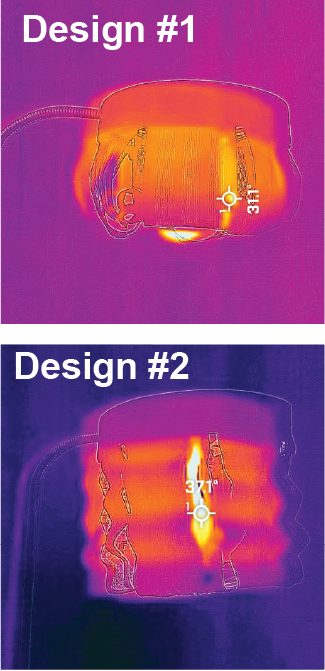}
        \caption{Heat Maps}
        \label{fig:lampshape_heatmaps}
    \end{subfigure}
    \caption{ExoFabric-based shape-changeable lamp shade. (a) The base design used for both Designs \#1 and \#2. (b) The molding process of the curled lamp shade (Design \#1). After 3 hours of use with a 60W light bulb, no noticeable shape change was observed. (c) After testing Design \#1, the lamp shade was reheated, reverted to the base shape, and remolded into a zigzag pattern (Design \#2). No visible changes were observed after 3 hours of testing. (d) Heat maps around the light bulb with the ExoFabric-based lamp shade were captured for both designs.}
    \label{fig:lampshape}
\end{figure}

\subsection{Shape-changeable lamp shade}
As the final application, we sought to evaluate the potential of ExoFabric in everyday objects and selected a lamp shade as a test case. This experiment aimed to evaluate whether the heat generated by the lamp could activate the thermoplastic properties of ExoFabric, potentially causing unintended shape deformations. This application allowed us to determine the feasibility of using ExoFabric in mild heat-generating devices, while ensuring its stability under such conditions. Additionally, we aimed to assess how easily the lamp shade’s shape could be altered, as it may be adjusted according to the ambiance or to personal preference. 

In the design process, we prioritized formability over stiffness to improve fabrication efficiency, selecting the L1\_S5 embroidery configuration with non-stretch fabric. Since a lampshade is primarily a decorative element and not typically subject to external forces, this configuration was ideal. Using this setup, we created a base shape, as shown in \Cref{fig:lampshape_base}, which was applied to two designs: curled and origami. For the first design, we rolled the flat base shape into a cylinder and fixed it while heating, as demonstrated in the first three steps of \Cref{fig:lampshape_design_01}. The diameter of the cylinder can influence the degree of curl. After installing the lampshade on a lamp with a 60W light bulb (emitting heat around 31-37$^{\circ}$C, See \Cref{fig:lampshape_heatmaps}), it was exposed to mild heat for three hours, during which no noticeable shape change occurred. The lampshade was then reheated, reverted to the base shape (\Cref{fig:lampshape_base}), and remolded into a zigzag pattern, as shown in \Cref{fig:lampshape_design_02}. The number of zigzag turns determines whether the origami design is looser or denser. This remolded lampshade was tested under the same conditions with the same light bulb, and no significant deformation was observed.

\section{Discussion}
We proposed ExoFabric as a novel textile platform to enable customizable mechanical properties, including stiffness, geometrical formability, and stretchability, through the integration of thermoplastic threads into traditional fabric structures. ExoFabric allows for dynamic tuning of its mechanical characteristics by adjusting parameters such as thread orientation, line spacing, stitch spacing, fabric types, and the number of fabric layers. Our study confirmed that these parameters play a critical role in determining the performance of ExoFabric, especially when balancing stiffness and stretchability. For instance, smaller line spacings combined with smaller stitch spacings resulted in significantly greater stiffness, while larger stitch spacings promoted flexibility, particularly in stretch fabrics. Multi-layer configurations further enhanced mechanical properties, providing a means to increase stiffness when single-layer designs reached their limits.

Integrating thermoplastic into fabric might initially seem like a straightforward method to enhance stiffness and reduce flexibility, but one of the key findings of this research is the complex trade-off between stiffness, geometrical formability, and stretchability. Additionally, the correlation and interplay between embroidery configurations and fabric types provide deeper insights into optimizing these mechanical properties. For example, in the stiffness evaluation, both non-stretch and stretch fabrics exhibit a similar trend where denser stitch and line spacing lead to greater stiffness. However, during geometrical formability evaluations, the configuration with the densest line spacing and loosest stitch spacing (i.e., L0.66\_S15 ) does not perform as intended in non-stretch fabric, whereas the same combination allows for proper formation in stretch fabric. In stretchability evaluations, it is evident that configurations with the highest stiffness, such as L0.66\_S1, can limit the fabric’s stretchability. Nevertheless, better geometrical formability is attainable with both stiffer configurations like L0.66\_S1 and L1\_S1, as well as more stretchable configurations, such as L1\_S15 and L0.55\_S15.

To evaluate the significance of the implications on stiffness, geometrical formability and stretchability, we incorporated them into the design process of three practical applications. In the bra application, for instance, the ability to fine tune the geometrical formability allows for the creation of form-fitting garments tailored to individual's needs. We conducted a washability test to evaluate ExoFabric’s ability to retain its re-moldable properties after exposure to typical washing and drying processes. This test was designed to assess not only the material’s structural integrity and performance post-wash but also its ability to revert to a flexible state and be reshaped as needed. By simulating real-world conditions, such as hand washing and air drying, we ensured that ExoFabric could endure repeated use and washing cycles while maintaining its re-moldability. Similarly, in the case of a finger splint that requires significant rigidity to constrain movement, multiple layers of the stiffest configuration (e.g., L0.66\_S1) are needed to provide sufficient stiffness. In the lamp shade example, we emphasize the geometrical formability and the simplicity of the molding process while enabling the creation of versatile designs by everyday users. This is demonstrated through the development of two distinct design variations. The durability of the formed shape was evaluated based on the usage scenario, specifically its ability to maintain the intended shape while being exposed to a mild heat source (i.e., light bulb) for three hours. 

When it comes to the fabrication tool, we primarily focused on using embroidery machines due to their versatility in thread placement and ease of material swapping. However, limitations such as thread size and working area arise. For example, Tex 80 was excluded due to frequent machine jamming. Dense stitching in a small area can also compromise fabric integrity, and the limited working area of embroidery machines also restricts the production of larger-scale items and requires multiple pieces to be assembled afterward. These challenges could be addressed with industrial embroidery machines, or roll-to-roll based textile machinery like knitting or weaving as well as alternative methods like 3D printing or lamination. In evaluating ExoFabric-based applications, our assessments effectively demonstrated the material’s feasibility and versatility across multiple domains. Future research could explore application-specific aspects in greater detail, such as the comfort of a customizable bra, taking into account other fabric qualities and garment geometries, or the efficacy of a finger splint, with its shape adjusted based on rehabilitation progress. 

Additionally, future research could explore additional materials and fabrication techniques to further optimize ExoFabric’s mechanical performance and fabrication efficiency. Investigating advanced thermoplastics or hybrid materials could offer new opportunities to enhance both durability and safety by choosing materials with a lower glass transition temperature. While plastics are often viewed as harmful to the environment, more sustainable alternatives such as bio-plastics or eco-friendly plastics could also be applied to ExoFabric, providing a sustainable alternative to material selection. Additionally, alternative heating sources, such as embedded Joule heating, could provide more efficient control over the material’s formability and stiffness in the targeted area. Combining ExoFabric with other functional threads, such as conductive threads, could also introduce new functionality, allowing for the tuning of stiffness, geometrical formability, and stretchability in smart textiles. Long-term durability and wearability testing under various environmental conditions would also provide insights into ExoFabric’s real-world performance. Moreover, exploring more intricate threading patterns may lead to new innovations in more complex geometries formation formability and mechanical strength. The insights gained from this study establish a foundation for future development of ExoFabric in diverse industries, ranging from fashion and healthcare to industrial design and soft goods. To enhance applicability across diverse domains, future research could focus on developing a comprehensive design tool for ExoFabric while incorporating mathematical models based on thoroughly characterized mechanical properties and their associated fabrication parameters and fabric types.

\section{Conclusion}
We introduced ExoFabric, a re-moldable textile system that customizes mechanical properties like stiffness, geometrical formability, and stretchability through the integration of thermoplastic threads via embroidery. Our characterization revealed that line spacing, stitch spacing, and layering significantly influence the fabric’s mechanical behavior, providing dynamic tunability for various applications. Smaller line and stitch spacings increase stiffness, while larger spacings enhance flexibility and formability, particularly in stretch fabrics. Multi-layer configurations further improved mechanical strength, achieving a balance between rigidity and flexibility, critical for applications requiring both support and comfort. Three applications—an adjustable finger splint, customizable bra, and shape-changing lamp shade—demonstrated ExoFabric’s versatility in providing protection, tailored support, and heat-resistant integrity. These examples underscore its potential in wearables and everyday objects, where stability and adaptability are essential. Further exploration of hybrid materials, advanced techniques, and functional thread integration could enhance performance and broaden its applications across healthcare, fashion, and industrial design. This study positions ExoFabric as a promising stepping stone for creating adaptive, reconfigurable textiles for evolving user needs.

\bibliographystyle{unsrtnat}
\bibliography{references}  






\end{document}